\documentclass[default]{aastex63}
\usepackage{abbrv}

\graphicspath{{./}{paper-figures/}{paper-tables/}}

\shorttitle{Multiphase Galactic Outflows}
\shortauthors{Kim et al.}

\begin{document}

\title{First results from SMAUG:
Characterization of Multiphase Galactic Outflows from a Suite of Local Star-Forming Galactic Disk Simulations
}

\author[0000-0003-2896-3725]{Chang-Goo Kim}
\affiliation{Department of Astrophysical Sciences, Princeton University, Princeton, NJ 08544, USA}
\affiliation{Center for Computational Astrophysics, Flatiron Institute, New York, NY 10010, USA}

\author[0000-0002-0509-9113]{Eve C. Ostriker}
\affiliation{Department of Astrophysical Sciences, Princeton University, Princeton, NJ 08544, USA}

\author[0000-0003-2835-8533]{Rachel S. Somerville}
\affiliation{Center for Computational Astrophysics, Flatiron Institute, New York, NY 10010, USA}
\affiliation{Department of Physics and Astronomy, Rutgers University, 136 Frelinghuysen Rd, Piscataway, NJ 08854, USA}

\author[0000-0003-2630-9228]{Greg L. Bryan}
\affiliation{Department of Astronomy, Columbia University, 550 West 120th Street, New York, NY 10027, USA}
\affiliation{Center for Computational Astrophysics, Flatiron Institute, New York, NY 10010, USA}

\author[0000-0003-3806-8548]{Drummond B. Fielding}
\affiliation{Center for Computational Astrophysics, Flatiron Institute, New York, NY 10010, USA}

\author[0000-0002-1975-4449]{John C. Forbes}
\affiliation{Center for Computational Astrophysics, Flatiron Institute, New York, NY 10010, USA}

\author[0000-0003-4073-3236]{Christopher C. Hayward}
\affiliation{Center for Computational Astrophysics, Flatiron Institute, New York, NY 10010, USA}

\author[0000-0001-6950-1629]{Lars Hernquist}
\affiliation{Harvard-Smithsonian Center for Astrophysics, 60 Garden Street, Cambridge, MA 02138, USA}

\author[0000-0002-2499-9205]{Viraj Pandya}
\affiliation{UCO/Lick Observatory, Department of Astronomy and Astrophysics, University of California, Santa Cruz, CA 95064, USA}
\affiliation{Center for Computational Astrophysics, Flatiron Institute, New York, NY 10010, USA}
\email{cgkim@astro.princeton.edu}

\begin{abstract}
Large scale outflows in star-forming galaxies are observed to be ubiquitous, and are a key aspect of theoretical modeling of galactic evolution, the focus of the SMAUG (Simulating Multiscale Astrophysics to Understand Galaxies) project. Gas blown out from galactic disks, similar to gas within galaxies, consists of multiple phases with large contrasts of density, temperature, and other properties. To study multiphase outflows as emergent phenomena, we run a suite of $\sim\pc$-resolution local galactic disk simulations using the TIGRESS framework. Explicit modeling of the interstellar medium (ISM), including star formation and self-consistent radiative heating plus supernova feedback, regulates ISM properties and drives the outflow. We investigate the scaling of outflow mass, momentum, energy, and metal loading factors with  galactic disk properties, including star formation rate (SFR) surface density ($\Sigma_{\rm SFR}\sim 10^{-4}-1{\rm\,M_\odot\,kpc^{-2}\,yr^{-1}}$), gas surface density ($\Sigma_{\rm gas}\sim1-100{\rm\,M_\odot\,pc^{-2}}$), and total midplane pressure (or weight; $P_{\rm mid}\approx \mathcal{W}\sim 10^3-10^6\,k_B{\rm\,cm^{-3}\,K}$). The main components of outflowing gas are mass-delivering cool gas ($T\sim10^4{\rm\,K}$) and energy/metal-delivering hot gas ($T\simgt 10^6{\rm\,K}$). Cool mass outflow rates measured at outflow launch points (one or two scale heights $\sim300\pc-1\kpc$) are 1--100 times the SFR (decreasing with $\Sigma_{\rm SFR}$), although in massive galaxies most mass falls back due to insufficient outflow velocity. The hot galactic outflow carries mass comparable to 10\% of the SFR, together with 10-20\% of the energy and 30-60\% of the metal mass injected by SN feedback. Importantly, our analysis demonstrates that in any physically-motivated cosmological wind model, it is crucial to include at least two distinct thermal wind components.
\end{abstract}


\section{Introduction}\label{sec:intro}
In current theories of galaxy formation and evolution, galactic winds are an important element, counteracting cosmic accretion to limit stellar mass growth of galaxies. Even in the earliest theoretical models of galaxy formation in dark matter halos, the issue of overproduction of stellar mass was recognized, necessitating mass and energy flows out of galaxies  \citep[e.g.,][]{1978MNRAS.183..341W,1986ApJ...303...39D,1991ApJ...379...52W}. Recent cosmological hydrodynamical simulations and semi-analytic models that successfully \emph{match} the observed galaxy statistics, including stellar mass-halo mass relations, all require ejection of a significant fraction of the gas mass accreted in the form of galactic-scale winds \citep[see reviews of][and references therein]{2015ARA&A..53...51S,2017ARA&A..55...59N}. With the enormous spatial and temporal domains required for cosmological-scale modeling, however, it is not possible to simultaneously represent the detailed properties of the star-forming interstellar medium (ISM) that lead to the production of galactic winds. Absent a means to directly model the physics within galactic disks, the usual practice is to adopt parameterized scaling relations (for both star formation rates and wind mass-loss rates), calibrating free parameters by reference to observations \citep[e.g.,][]{2013MNRAS.436.3031V,2015MNRAS.450.1937C,2018MNRAS.473.4077P}. The approach of empirically-constrained parameterization, while heretofore unavoidable, has been a major source of uncertainty in modern galaxy formation theory.
The SMAUG\footnote{Simulating Multiscale Astrophysics to Understand Galaxies; \url{https://www.simonsfoundation.org/flatiron/center-for-computational-astrophysics/galaxy-formation/smaug/}} project was initiated to address the need for developing and implementing subgrid treatments for cosmological models that are \textit{derived and calibrated from simulations that explicitly model and resolve key physical processes}.

In addition to the important role of winds in the theory of galaxy formation, galactic outflows are prevalent in observations of nearby dwarf starbursts and luminous/ultraluminous infrared galaxies (LIRGs/ULIRGs) \citep[e.g.,][]{1990ApJS...74..833H,1999ApJ...513..156M,2000ApJS..129..493H,2005ApJ...621..227M,2015ApJ...809..147H,2015ApJ...811..149C}. 
Winds appear to be even more ubiquitous in both AGN-host and star-forming galaxies at high-redshift  \citep[e.g.,][]{2001ApJ...554..981P,2003ApJ...588...65S,2007ApJ...663L..77T,2010ApJ...717..289S,2012ApJ...759...26E,2019ApJ...875...21F}, although the limited spatial resolution of these observations makes interpretation more  difficult.

Galactic outflows driven by star formation are the result of the feedback that is produced by populations of young stars. This feedback -- primarily associated with core-collapse supernovae (SNe) from massive stars, but with some contribution from stellar winds and radiation -- returns metal-enriched gas at extremely high velocity to the surrounding ISM. 
As a result of complex interactions driven by SN shocks (and potentially involving cosmic rays as an intermediary), a portion of the ISM gas is accelerated sufficiently to emerge as a galactic  wind, delivering mass, momentum, energy, and metals to the circumgalactic/intergalactic medium (CGM/IGM).
Because the massive stars responsible for feedback are buried deep within the ISM, properties of galactic outflows are not simply set by the immediate deposition at feedback sites.
Instead, localized energy injection events build up expanding bubbles (or superbubbles for correlated feedback events) with more and more momentum as they sweep up surrounding gas 
(see e.g., \citealt{1959sdmm.book.....S,1950RSPSA.201..159T,1972ApJ...178..159C,1977ApJ...218..148M} for a single SN and e.g., \citealt{1977ApJ...218..377W,1987ApJ...317..190M,1992ApJ...388...93K} for stellar winds or clustered SNe). 
The  hot ISM is produced by stellar-wind and SN shocks, and fills the interior of each bubble.  
Cooling of the shocked ISM when bubble expansion slows to $\simlt 200 \kms$ limits the momentum injection for the case of a single SN \citep[e.g.,][]{1988ApJ...334..252C,1998ApJ...500...95T}, while mixing of hot diffuse gas with dense gas at the bubble-shell boundary drains energy from superbubbles and reduces their dynamical impact \citep[e.g.][]{2019MNRAS.490.1961E}.  
When extreme star formation/feedback events occur, superbubble breakout from the ISM before the onset of cooling alters the dynamics significantly \citep[e.g.,][]{1986PASJ...38..697T,1989ApJ...337..141M,2008ApJ...674..157C} and enables delivery of a large fraction of pristine metals and original feedback energy to the CGM \citep[e.g.,][]{2017ApJ...834...25K,2018MNRAS.481.3325F}. 

Based on focused high-resolution numerical simulations with an inhomogeneous ISM  \citep[e.g.,][]{2015ApJ...802...99K,2015A&A...576A..95I,2015MNRAS.450..504M,2015MNRAS.451.2757W}, the net terminal momentum injection from single SNe has been shown to be quite insensitive to the background medium's average density and detailed structure (this insensitivity is because the onset of cooling by metal lines at postshock temperature $T\sim 10^6\Kel$ is insensitive to the density).
A practical application of this result to galactic simulations is the numerical approach of injecting the previously calibrated terminal momentum if the energy-conserving stage of a SN remnant (SNR) expansion is unresolved \citep[e.g.,][]{2014ApJ...788..121K,2017ApJ...846..133K,2014MNRAS.445..581H,2018MNRAS.477.1578H}. 
This ``momentum feedback'' approach captures the dynamical impact of SN feedback on the warm-cold ISM phases (essentially all of the ISM's mass) reasonably well, especially for driving  turbulence and therefore self-regulating the star formation rate (SFR), even at relatively low numerical resolution \citep[e.g.,][]{2011ApJ...743...25K,2013ApJ...776....1K,2012ApJ...754....2S,2015ApJ...815...67K,2017ApJ...846..133K,2014MNRAS.445..581H,2015MNRAS.451.2900K}. 
However, how much hot gas is created in the ISM by expanding SNe-driven bubbles, and how much is retained to vent from the ISM into the CGM, depends sensitively on the details of micro/macro physics and conditions of the vertically-stratified ISM.  

Physical elements that affect momentum injection and hot gas yield include turbulence, inhomogeneity, magnetization, thermal conduction, as well as temporal and spatial correlations of feedback \citep[e.g.,][]{2015ApJ...802...99K,2017ApJ...834...25K,2018MNRAS.481.3325F,2017MNRAS.465.2471G,2019MNRAS.483.3647G,2019MNRAS.490.1961E}. Although in principle hot gas generation can be implemented via deposition of ``residual'' thermal energy \citep[e.g.,][]{2015MNRAS.450..504M}, this will be immediately lost if resolution is too low and hot diffuse gas is not spatially separated from warm fast gas \citep{2019MNRAS.483.3363H}.  
In general, outflow properties are much more sensitive to resolution than SFRs \citep[e.g.,][]{2017MNRAS.466...11R,2017ApJ...846..133K,2018ApJ...853..173K,2018MNRAS.478..302S}; because proper hot gas generation and evolution is crucial, outflow properties will be incorrect if most SNe are realized in the form of momentum feedback \citep[e.g.,][]{2018ApJ...853..173K,2019MNRAS.483.3363H}. 

A key characteristic of galactic outflows, which is often overlooked in theoretical modeling, is their multiphase nature. Galactic outflows in observations are often detected in spectra of neutral and ionized gas tracers (e.g., \ion{Na}{1}, H$\alpha$, \ion{Si}{2}, \ion{Si}{4}; \citealt{1990ApJS...74..833H,1998ApJ...506..222M,2005ApJ...621..227M,2005ApJS..160..115R,2015ApJ...811..149C,2015ApJ...809..147H}) that trace gas at $T\sim10^{4-5}\Kel$, but there has also  been direct detection of kinematically confirmed hot winds ($T\sim 10^{6-7}\Kel$) via diffuse X-rays \citep[e.g.,][]{1997MNRAS.286..626R,1999ApJ...523..575L,2007ApJ...658..258S}, as well as cold atomic and molecular outflows \citep[e.g.,][]{2011ApJ...733L..16S,2013Natur.499..450B,2015ApJ...814...83L,2018ApJ...856...61M}. 
Due to the low density and hence low emissivity of the hot gas, quantitative characterization of full multiphase outflows from observations have been limited to a few best case examples \citep[e.g.,][]{2007ApJ...658..258S,2015ApJ...814...83L}; significant advances will require next-generation X-ray observatories (e.g., AXIS, ATHENA, and Lynx).

To date, systematic theoretical studies of outflow properties for different thermal phases have also been limited. Utilizing pc-resolution local, kpc-patches of galactic disks, resolved multiphase ISM simulations with SN feedback (and additional feedback processes) have been conducted by several groups. However, due to the complexity and expense of modeling full star-forming ISM physics with high resolution, many previous simulations studying galactic winds have adopted prescribed SN rates and positions \citep[e.g.,][]{2013MNRAS.429.1922C,2016MNRAS.456.3432G,2016ApJ...816L..19G,2016MNRAS.459.2311M,2017ApJ...841..101L,2018MNRAS.479.3042G} and run only for a short period of time, with a limited range of ISM conditions \citep[e.g.,][]{2017MNRAS.466.1903G,2020MNRAS.491.2088K}. To understand galactic outflows as emergent phenomena produced by the star-forming ISM, lack of self-consistency is a concern because the reported characteristics could be sensitive to the adopted feedback rates and SN locations. Previous controlled simulations with SNe imposed ``by  hand'' have shown that the resulting ISM and outflow properties change dramatically when SNe are located only in dense gas or randomly \citep[e.g.,][]{2016MNRAS.456.3432G}, or when clustering of SN is varied \citep[e.g.,][]{2018MNRAS.481.3325F}.
Simulations with a short duration are problematic because results may be strongly affected by imposed ISM initial conditions and numerical startup transients. 

A different approach  from high-resolution ``local patch'' simulations is global isolated galaxy and cosmological zoom simulations. In the case of cosmological zooms \citep[e.g.,][]{2015MNRAS.454.2691M,2016ApJ...824...57C,2017MNRAS.470.4698A,2019MNRAS.485.2511T}, cosmic accretion and merging/interaction of galaxies is included, which provides a ``natural'' CGM environment with which winds may interact \citep{2020arXiv200616316F}.  
For studying wind acceleration, global/zoom models also have an advantage over local models in that the effect of global geometry and quasi-conical wind expansion and acceleration is naturally captured \citep[e.g.,][]{1985Natur.317...44C}. However, for studying wind creation, zoom simulations are at a disadvantage compared to local models in that the ISM physics including star formation and feedback is at best only marginally resolved.
The adopted mass resolution ($\sim10^{3-5}\Msun$) is still insufficient to resolve the Sedov-Taylor stage of SNR evolution: since the remnant mass at the time of shell formation ($\sim10^{3}\Msun$) must be resolved by several elements, mass resolution of $\lesssim 100\Msun$ is needed \citep[][]{2015ApJ...802...99K}.
This is critical for accurately modeling hot gas production and the multiphase interactions inherent to wind launching \citep{2019MNRAS.483.3363H}.
The derived wind properties from zoom-in simulations are compromised by approximate treatments of SN feedback: artificially-delayed cooling \citep{2016ApJ...824...57C,2019MNRAS.485.2511T} or momentum feedback \citep{2015MNRAS.454.2691M,2017MNRAS.468.4170M,2017MNRAS.470.4698A}. 

Even in isolated galaxy simulations, achieving high enough resolution for resolving individual SNe as well as self-consistent modeling of star formation with self-gravity is challenging; currently such simulations are done only for very low mass galaxies (total gas mass $\sim 10^7\Msun$; e.g., \citealt{2018ApJ...865L..22E,2019MNRAS.483.3363H}). For more massive galaxies, prescribed rates and positions of SNe are still adopted \citep[e.g.,][]{2017MNRAS.470L..39F,2018ApJ...860..135S}. 
Note that in their study of superwinds, \citet{2020arXiv200210468S} consider an isolated global galaxy with comparable resolution to local simulations (up to 5~pc) but with a much smaller set of physics: neither self gravity nor magnetic fields, no cold ISM (cooling is truncated at $10^4\Kel$), and prescribed SN feedback rates and positions. Nevertheless, these simulations demonstrate the importance of uniformly high resolution in the extraplanar region for following both the hot and cool components of outflows and the interaction between phases.

Another potential issue in characterizing multiphase outflow properties from cosmological zoom or isolated global simulations is the adaptive resolution (either AMR or semi-Lagrangian method) that is usually employed (an exception is the CGOLS suite by \citealt{2018ApJ...860..135S}). Although semi-Lagrangian/adaptive resolution (generally at constant mass) provides better resolution at higher densities to improve treatment of star formation, the low-density hot gas, which carries the majority of outflowing energy and metals, can be quite under-resolved. The phase structure and overall energetics of the ISM and CGM depend sensitively on accurately resolving the mixing at interfaces between hot and cool gas \citep[e.g.][]{2020arXiv200308390F}, which would require extremely high \textit{mass} resolution given the low density of the hot gas.
Indeed, recent work employing fixed ``spatial'' rather than ``mass'' resolution in the CGM has revealed dramatic differences in multiphase gas properties  \citep[e.g.,][]{2019ApJ...882..156H,2019ApJ...873..129P,2019MNRAS.482L..85V}.
For winds, under-resolution of the hot gas raises potential concerns about numerical phase mixing that could lead to underestimated metal loading or overestimated mass loading in an artificially phase-mixed wind, depending on the halo potential  \citep[see discussion in][]{2018ApJ...853..173K}. 

In the large-box cosmological simulations that are necessary for predicting statistics of galactic populations and for connecting baryonic distributions on large scales to cosmological parameters, typical mass  resolutions are  $>10^5\Msun$ (see Figure~1 in \citealt{2019MNRAS.490.3234N} for a recent compilation). Even with significant improvements in computing  power, it will continue to be necessary into the future to apply subgrid methodology in treating star formation and winds, because  directly representing the physical processes involved would require several orders of magnitude higher resolution. Key wind parameterizations that are usually required by large-box cosmological simulations (as well as semi-analytic cosmological models) are (1) the dimensionless mass loading consisting of the mass (hydrogen and metals) carried out by  the wind per stellar mass formed, and (2) the energy loading, consisting of the fraction of the original supernova energy that is transferred to the outflowing gas.  In addition, it is necessary to set  (3) the wind velocity;  this is often scaled relative to the halo velocity,  but more generally a momentum loading (momentum ejection per stellar mass formed) or characteristic outflow velocity (or its distribution) can be given.
Currently, the standard practice \citep[e.g.,][]{2006MNRAS.373.1265O,2013MNRAS.436.3031V,2015MNRAS.450.1937C,2018MNRAS.473.4077P} is to tune the wind
parameters so that the resulting global galaxy properties match empirical constraints. 
Cosmological semi-analytic models (SAMs) also adopt empirical prescriptions for the impact of stellar driven winds on the ISM and CGM.  Although SAM feedback prescriptions have traditionally been tuned to match observations, it is also interesting to study how different these prescriptions are from those that emerge from detailed numerical simulations \citep{2020arXiv200616317P}.

In the present paper, as part of the first results from SMAUG,\footnote{\url{https://www.simonsfoundation.org/flatiron/center-for-computational-astrophysics/galaxy-formation/smaug/papersplash1}} we take a step towards the goal of a new subgrid approach by providing a detailed characterization of the mass, momentum, energy, and metal loading of multiphase outflows, based on pc-resolution  simulations of the star-forming ISM.
In contrast to earlier local simulations, the major advance of our work is to achieve self-consistency in the ISM evolution, star formation, and feedback as well as uniformly high resolution and long-term evolution. Using a new numerical framework called TIGRESS \citep[][ hereafter KO17]{2017ApJ...846..133K}, we resolve the self-gravitating collapse of star-forming cloud complexes, the energy-conserving stage of SNR evolution when hot gas is created, and the subsequent interactions between diffuse hot and denser warm gas. 
KO17 delineated the numerical methods involved, and demonstrated their application by running a Solar neighborhood model over 3 orbit times, covering $\sim10$ star-formation/feedback cycles. 
Over this time, the ISM achieves a quasi-steady state with self-regulated star formation, insensitive to the initial setups. KO17 presented a thorough resolution study and confirmed convergence of turbulence amplitudes, thermal phase balance, magnetic field strength, SFRs, and outflow rates. 
\citet[][hereafter KO18]{2018ApJ...853..173K} analyzed multiphase wind properties in  Solar-neighborhood TIGRESS simulations, focusing on the dichotomy of warm fountains and hot winds, loading factors as a function of heights and phases, and distributions of outflow velocities in the warm outflow. 

The current paper focuses on the systematic investigation of outflowing gas, separately characterizing \emph{multiple phases} of gas in  a suite of 7 TIGRESS simulations.  
We quantify  mass, momentum, energy, and metal loading factors, as well as characteristic outflow velocities and metal enrichment factors. 
We follow KO17's definition of thermal phases but merge the three lowest temperature phases to a single \emph{cool} ($T<2\times10^4\Kel$) component, while keeping the \emph{intermediate} ($2\times10^4\Kel<T<5\times10^5\Kel$) and \emph{hot} ($T>5\times10^5\Kel$) phases. 
Physical conditions in our set of 7 models span two orders of magnitude in gas surface density ($\Sigma_{\rm gas}\sim1-100\Surf$) and four orders of magnitude in SFR surface density ($\Sigma_{\rm SFR}\sim 10^{-4}-1\sfrunit$), covering typical local conditions of nearby star forming galaxies (e.g., \citealt{2017ApJ...846..159B}; see also \citealt{2020arXiv200616314M} for our parameter space coverage in comparison with Illustris-TNG galaxies).

To further characterize outflows, in a companion paper we shall present joint probability distribution functions (PDFs) of gas outflow velocities and sound speeds. We shall also provide a guide to combining the loading factors presented here and PDFs and applying them in large-scale models of galaxy formation. 

The structure of this paper is as follow. \autoref{sec:methodnmodel} summarizes the TIGRESS numerical framework and introduces the model suite  parameters employed in this paper. \autoref{sec:evolution} explicates  the cycles of star formation, feedback, and outflow/inflow that emerge in all the simulations. \autoref{sec:windprop} presents multiphase outflow properties including outflow fluxes, metal properties, and characteristic velocities as well as average loading factors. \autoref{sec:scaling} compares outflow characteristics with galactic properties including SFR surface density ($\Ssfr$), gas surface density ($\Sgas$), midplane density ($n_\mathrm{mid}$), midplane pressure ($\Pmid$), gas weight ($\mathcal{W}$), and gas depletion time ($\tdep$) to derive scaling relations. \autoref{sec:discussion} discusses our results in the context of other existing simulations and observations, provides physical interpretations, and summarizes strengths and weaknesses of our numerical model. In \autoref{sec:summary}, we summarize our results.

\section{Methods \& Models}\label{sec:methodnmodel}

This paper investigates properties of a suite of local ISM simulations in star-forming galactic disks to provide a comprehensive characterization of multiphase galactic outflows driven by stellar feedback. To evolve the ISM with star formation and stellar feedback self-consistently, we utilize the TIGRESS framework described in KO17. We refer the reader to KO17 for details of implementations and tests. In \autoref{sec:method}, we summarize key features and modifications of the TIGRESS framework from KO17. In \autoref{sec:model}, we introduce model parameters for our suite of simulations. 

\subsection{Methods}\label{sec:method}

The TIGRESS framework evolves the ISM by solving the ideal MHD equations, including gravity and cooling/heating, in a local, rotating frame with a galactic orbital frequency $\Omega(R_0)$ at a galactocentric distance $R_0$. Local Cartesian coordinates $x$ and $y$ respectively represent the local radial ($R-R_0$) and azimuthal ($R_0[\phi-\Omega t]$) directions, while $z$ represents the vertical distance from the midplane. 
Shearing-periodic and outflow boundary conditions are adopted in the horizontal and vertical directions, respectively.
We use the \textit{Athena} finite volume code for MHD \citep{2008ApJS..178..137S,2009NewA...14..139S} with additional physics modules.   
The shearing box approach \citep{2010ApJS..189..142S} allows us to model the ISM in the context of rotating disk galaxies with uniformly high resolution ($\sim\mathcal{O}(1)\pc$) everywhere.

To follow star formation by gravitational collapse, the TIGRESS framework includes self-gravity by solving Poisson's equation using FFTs \citep{2001ApJ...553..174G,2009ApJ...693.1346K} and forms sink particles to represent star cluster formation in cells undergoing unresolved gravitational collapse \citep{2013ApJS..204....8G}. The sink particles then further accrete if gas flows are converging into a virtual control volume ($3^3$ cells surrounding a particle) from all three directions. Gas accretion onto a given sink particle ceases as soon as the first SN explodes (the SN event is stochastically determined, with the first event typically 3-4~Myr after the birth; see below). When sink particles are first formed or actively accreting, we reset the gas density, momentum, and pressure within the control volume with the extrapolated values from the nearby cells and dump only the difference between original and extrapolated values of mass and momentum in the control volume into the star particle. In the original \citet{2013ApJS..204....8G} treatment adopted in KO17, all of the  mass flux into the control  volume is added to sink  particles and the control volume is treated as ghost zones.  Since control volume cells become active zones if a sink particle becomes a non-accreting passive particle or merges with other particles, this approach is not strictly mass conservative. As initially applied in KO17,  this non-conservation has a minimal effect in the total mass (net difference $\sim 10\%$ over 3$\torb$) of the R8 model, because the SFR is low and particle merging is not frequent, but it can be more significant for models with high SFRs. In the new approach, where the sink particle control volumes are treated as potential active cells, mass conservation is improved \citep[see also][]{2019MNRAS.489.5326L}; for example, the cumulative effect in mass is at 3\% over $\torb$ for model R4. By comparison, the total ISM mass reduction over $0.5<t/\torb<1.5$ is 23\%, with 17\% going into star formation and 8\% into winds.  Non-conservation is smaller for models with smaller SFRs, and $\sim 4-5\%$ for models R2 and LGR2. It should be kept in mind that instantaneous relationships among SFRs, outflow properties, and ISM properties are not affected by this slow secular variation, and the non-conservation does not affect any of the measures we report.  In particular, all measures of outflows are obtained directly from fluxes in the simulation. The non-conservation of mass (reflecting a small addition from ``re-activated'' control volume cells) simply makes the mean value of $\Sigma_{\rm gas}$ at most a few percent larger than it would otherwise be over the simulation duration.

Stellar feedback in the TIGRESS framework includes the effects of FUV radiation and SN explosions. 
We slightly update the treatment of the heating rate due to FUV radiation from young stars. FUV radiation absorbed by small grains (e.g. PAHs) produces photoelectrons that heat the gas \citep{1994ApJ...427..822B}. This is believed to be the dominant heating process in the neutral (atomic) ISM \citep{1995ApJ...443..152W} where FUV is not shielded (at low column densities) and where dust is not destroyed.
To first order, the heating rate is proportional to the mean FUV intensity. Allowing for background heating from the metagalactic UV \citep{2002ApJS..143..419S}, the heating rate is given by
\begin{equation}
    \Gamma = \Gamma_0\rbrackets{\frac{J_{\rm FUV}}{J_{\rm FUV,0}} + 0.0024},
\end{equation}
where we adopt as reference Solar neighborhood values a heating rate of $\Gamma_0=2\times10^{-26}\ergs$ \citep{2002ApJ...564L..97K} and a mean FUV intensity of $4\pi J_{\rm FUV,0}=2.7\times10^{-3}\ergs\cm^{-2}$ (or $G_0=1.7$; \citealt{1978ApJS...36..595D}). 
We note that $\Gamma_0$ is held fixed, implying that in the present treatment, we do not allow for variations in dust abundance or photoelectric efficiency \citep{1994ApJ...427..822B,2001ApJS..134..263W}.

In the TIGRESS framework, the total FUV luminosity is calculated by summing up FUV luminosity of individual star clusters 
\begin{equation}
    L_{\rm FUV}=\sum_{\rm sp}\Psi_{\rm FUV}(t_{\rm age,sp})M_{\rm sp}
\end{equation}
using a tabulated time-dependent mass-to-luminosity ratio $\Psi_{\rm FUV}$ (from STARBURST99 as in \citet{1999ApJS..123....3L}; see Figure 1 of KO17), star cluster age $t_{\rm age,sp}$ (mass-weighted average is taken when there is addition of mass from accretion and merging), and star cluster mass $M_{\rm sp}$. As our star clusters have masses $\gtrsim 10^3 \Msun$, we adopt a fully-sampled Kroupa initial mass function (IMF; \citealt{2001MNRAS.322..231K}) in setting $\Psi_{\rm FUV}$. In KO17, we calculated the mean FUV intensity as $4\pi J_{\rm FUV}=\Sigma_{\rm FUV}\equiv L_{\rm FUV}/(L_x L_y)$, assuming uniformly spread radiation over the horizontal area of $L_x L_y$.
This is valid for Solar neighborhood and outer disks (e.g., R8, R16, and LGR8), but may overestimate the interstellar radiation field in denser environments where attenuation is generally higher. To allow for attenuation in an average sense, here we set the mean intensity based on the plane-parallel solution of the equation of radiation transfer in a slab with a uniform source distribution,
\begin{equation}
    4\pi J_{\rm FUV} = \Sigma_{\rm FUV}\frac{(1-E_2(\tau_\perp/2))}{\tau_{\perp}},
\end{equation}
where $\tau_\perp=\kappa_{\rm FUV} \Sigma_{\rm gas}$ is the UV optical depth perpendicular to the slab and $E_2$ is the second exponential integral.\footnote{In follow-up work applying the adaptive ray-tracing method of \citet{2017ApJ...851...93K}, we are further testing this approximation.} We adopt $\kappa_{\rm FUV}= 10^3\cm^3{\,\rm g^{-1}}$. Note that our heating rate as a result is time-varying but uniform in space, modulo a turn-off at high temperatures $T>10^5\Kel$. 

The SN treatment is unchanged from KO17.  When a SN explodes, we first calculate the mean gas properties (total mass $M_{\rm SNR}$ and mean density $n_{\rm amb}$) for the cells whose cell-centered distances from the explosion center are smaller than $R_\mathrm{SNR}=3\Delta x$. We inject both thermal and kinetic energy with a ratio consistent with the Sedov-Taylor stage ($0.72:0.28$) if $M_{\rm SNR}/M_{\rm sf} <1$, where $M_{\rm sf} = 1540\Msun (n_{\rm amb}/\pcc)^{-0.33}$ is the shell formation mass at a given ambient medium density $n_{\rm amb}$ \citep{2015ApJ...802...99K}. If the shell formation mass is unresolved (i.e., $M_{\rm SNR}/M_{\rm sf}>1$ within the feedback region), we instead inject the terminal momentum of SNR $p_{\rm SNR}=2.8\times10^5\Msun\kms (n_{\rm amb}/\pcc)^{-0.17}$ \citep{2015ApJ...802...99K}. We find that more than  90\% SNe are well resolved (i.e., $M_{\rm SNR}/M_{\rm sf} <0.1$) in the simulations presented here.

With each SN explosion, we eject massless test particles with 50\% probability to represent a runaway originating from a binary OB star. The ejection velocity follows an exponential distribution with $\exp(-v_{\rm run}/50\kms)$ for $v_{\rm run}\in(20,200)\kms$ \citep{2011MNRAS.414.3501E}, and the direction is chosen isotropically. Each runaway moves under the total gravitational potential and explodes as a SN after a  pre-assigned explosion time. The total SN rate from a star cluster, including its runaways, is consistent with the SN rate from STARBURST99 \citep{1999ApJS..123....3L}. KO18 showed that the outflow properties are not sensitive to the inclusion of runaways.\footnote{\citet{2020MNRAS.494.3328A} explored the effect of runaways in isolated galaxy simulations and found large enhancement of mass outflow rates and corresponding loading factors ($\times~ 5-10$) when runaways were included. However, this result may reflect a numerical rather than a physical effect.  In particular, it is possible that their no-runaways simulation failed to drive strong outflows because the majority of SNe  had numerically-unresolved evolution which failed to create hot gas that breaks out from the disk. The adopted AMR scheme (the RAMSES code) has a refinement strategy of splitting cells when the cell mass exceeds a designated maximum mass, in this case $\sim 4\times 10^3\Msun$. The majority of SNe within star-forming, dense gas at the maximum level of refinement would then occur in cells with mass exceeding the shell-formation mass $\sim 10^3\Msun$, with feedback implemented via momentum injection.  However, runaway particles that have moved far from their birth places may be in lower-density environments, with higher mass refinement,  at the time of SN explosions. To some extent, the inclusion of runaways is a partial solution to numerical difficulties in resolving SNe and driving hot superbubble breakout in moderate resolution simulations like \citet{2020MNRAS.494.3328A}. In our simulations, however, resolution is much  higher and the majority of SNe in clusters ($>90\%$) resolve the Sedov-Taylor stage of evolution, so that inclusion of runaways has insignificant impact on outflows.}
We turn off runaways in the R2 model, in which a large number of runaways would otherwise be created, for the sake of simulation efficiency.

SN explosions involve energy, mass, and metal returns. For simplicity, we adopt single values for ``population-averaged'' SN explosion energy $E_{\rm SN}=10^{51}\erg$, ejecta mass $M_{\rm ej}=10\Msun$, and metallicity $Z_{\rm SN}=0.2$ \citep{1999ApJS..123....3L}. As we are focusing on the galactic winds driven by SNe, we do not consider other means of mass and metal returns from either young or old stars.
We note that we follow metal density with a separate passive scalar.

We use a tabulated cooling function from a combination of \citet{2002ApJ...564L..97K} at $T<10^{4.2}\Kel$ and \citet{1993ApJS...88..253S} at $T>10^{4.2}\Kel$ for solar metallicity. Although we include a metal tracer field, which is initialized with solar metallicity $Z_{\rm ISM,0}=0.02$, our cooling function does not depend on gas metallicity.
In the model with the highest $\Ssfr$ (R2, see below), we find that the ISM metallicity reaches $\sim 0.04$ at the end of simulation, which is well within a range of observational estimates of the ISM metallicity in nearby star forming galaxies \citep[e.g.,][]{2018MNRAS.476.3883L}. The hot outflow is typically more enriched by a factor of 2 than the ISM (although the cooling rate  of the diffuse hot gas is quite low in any case).
To address the effect of realistic, metallicity-dependent cooling, development of a second-generation TIGRESS framework, including more complete treatments of cooling, radiation, and chemistry, is now underway.

\subsection{Models}\label{sec:model}

For this paper, we use TIGRESS runs with 7 different parameter sets, covering conditions generally representative of inner and outer regions of Milky Way-like galaxies, including the Solar neighborhood model described in KO17 and KO18. We list key parameters of these models in \autoref{tbl:model}. The  gas surface density $\Sigma_{\rm gas,0}$ in Column (2) is the initial value in the simulation and decreases over time because gas turns into sink particles due to star formation and escapes vertically as a wind. SNe return mass to the gas in the form of ejecta, but on average this is only 10\% of that locked into stars. The galactic environment parameters such as angular speed of galactic rotation $\Omega$, stellar surface density $\Sigma_*$, stellar scale height $z_*$, dark matter halo density $\rho_{\rm dm}$, and galactocentric radius $R_0$ are fixed in time for each simulation; these parameters are important for setting the  gravitational potential (see KO17 for the analytic expression), and for setting the differential shear rate (important to dynamo activity; e.g., \citealt{2018A&A...611A..15K}) and Coriolis force. 
The ``LG'' models have external (stellar and dark-matter) vertical gravity reduced by about a factor of 8 near the midplane ($z\ll z_*$) and 4-5 far above the disk ($z\gg z_*$) compared to the corresponding R2, R4, R8 models. All simulations use $(N_x, N_y, N_z) = (256,256,1792)$ zones and uniform cubic grid cells with side length of $\Delta x$ (Column (8)). Our parameter choice covers typical ranges seen in nearby star-forming galaxies \citep[e.g.,][]{2020arXiv200208964S} as well as cosmological simulations \citep{2020arXiv200616314M}.

Additional parameters are used to set initial conditions of the gas in the simulation, including the temperature profile and  turbulent vertical velocity dispersion $\sigma_{z,0}$ and plasma beta $\beta_0\equiv P/P_\mathrm{mag}$ (see KO17). The initial vertical profiles of density, pressure, and azimuthal magnetic field are set to be in a hydrostatic equilibrium with given total vertical velocity dispersion, including thermal, turbulent, and magnetic terms. However, the initial thermal and turbulent support is lost quickly due to radiative cooling and turbulence dissipation. The gas soon falls toward the midplane and this density increase triggers a burst of star formation. As we shall discuss in \autoref{sec:evolution}, the first burst is not fully self-consistent because it is subject to the initialization. Over time, the evolution becomes self-regulated; our analysis therefore will focus on the time subsequent to the first burst. To offset the rapid initial cooling and turbulence dissipation, we introduce randomly placed star particles in the initial conditions with age and mass distributions corresponding to the $\Ssfr$ at later times (estimated from lower-resolution simulations). We adopt $\beta_0=10$ for all models, except R2 with $\beta_0=2$; and $\sigma_{z,0}=30$, 15, 10, and 10~km/s for R2/LGR2, R4/LGR4, R8/LGR8, and R16. 

From several independent simulation runs with different $\sigma_{z,0}$ and $\beta_0$ (typical ranges are $\sigma_{z,0}=10-30\kms$ and $\beta_0=1-10$), we have confirmed that the evolution is statistically converged irrespective of initial conditions unless the initial parameters are extreme; the initial magnetic field strength can impact the overall outcomes if it is too strong or too weak compared to the saturated value since the evolution of the regular magnetic field is much slower than all other time scales \citep[see][]{2015ApJ...815...67K}.

Magnetic fields in outflows do not contribute to momentum and energy fluxes significantly. However, inclusion of magnetic fields has indirect effects on outflows and associated galactic properties by increasing the vertical pressure support near the midplane and reducing SFRs. We find that contribution from magnetic stresses to the vertical pressure support can be as high as 50\% (depending on initial field strengths since saturation is not achieved within 1-2 $\torb$), but typically about 30\% (Ostriker \& Kim in prep.).

The numeral in each model name indicates the galactocentric radius of the simulation box; e.g. the box in model R8 is centered at $R_0= 8 \kpc$.
The spatial resolution in pc is progressively smaller from model R16 to model R2 (also implying smaller simulation box) as we move from outer (lower density) to inner (higher density) galactic regions. At higher densities, both thermal and dynamical length scales are smaller. 
For each model, we tested varying simulation box sizes, and the values ultimately adopted were optimized such that resolution is sufficiently high while still providing a large enough horizontal area such that superbubbles do not fill the entire horizontal domain. 
For our standard simulations, the horizontal box size, $L_x$ and $L_y$, decreases from 2048 pc for model R16, to 1024 pc for model R8, to 512 pc for models R2 and R4.  
In \autoref{sec:app-conv}, we briefly  discuss the role of box size and show the resolution dependence of our results to demonstrate convergence. 

Finally, we note that although a value of $R_0$ is adopted for each model, this is only used in setting the local background rotational velocity and the shape of dark matter halo gravity; the simulations are all local, and in principle could equally well describe similar conditions within a dwarf as a massive spiral (at a given metallicity).

\begin{deluxetable}{lCCCCCCC}
\tablecaption{Model Parameters\label{tbl:model}}
\tablehead{
\colhead{Model} &
\colhead{$\Sigma_{\rm gas,0}$} &
\colhead{$\Sigma_*$} &
\colhead{$\rho_{\rm dm}$} &
\colhead{$\Omega$} &
\colhead{$z_*$} &
\colhead{$R_0$} &
\colhead{$\Delta x$} 
}
\colnumbers
\startdata
R2    & 150 &  450 & 8.0\cdot 10^{-2}&100 & 245  &  2 &   2 \\
R4    &  50 &  208 & 2.4\cdot 10^{-2}& 54 & 245  &  4 &   2 \\
R8    &  12 &   42 & 6.4\cdot 10^{-3}& 28 & 245  &  8 &   4 \\
R16   & 2.5 & 1.71 & 1.4\cdot 10^{-3}& 12 & 245  & 16 &   8 \\
LGR2  & 150 &  110 & 1.5\cdot 10^{-2}& 50 & 500  &  2 &   2 \\
LGR4  &  60 &   50 & 5.0\cdot 10^{-3}& 30 & 500  &  4 &   2 \\
LGR8  &  12 &   10 & 1.6\cdot 10^{-3}& 15 & 500  &  8 &   4 \\
\enddata
\tablecomments{
Column (2): initial gas surface density in $M_\odot\pc^{-2}$.
Column (3): stellar surface density in $M_\odot\pc^{-2}$.
Column (4): dark matter volume density at the midplane in $M_\odot\pc^{-3}$.
Column (5): angular velocity of galactic rotation at the domain center in ${\rm km\, s^{-1}\, kpc^{-1}}$.
Column (6): scale height of stellar disk in pc.
Column (7): galactocentric radius in kpc.
Column (8): spatial resolution of simulation in pc.
}
\end{deluxetable}

\section{Overall Evolution}\label{sec:evolution}

\begin{deluxetable*}{lCCCCCCCCCC}
\tablecaption{Time Scales and Relevant Measured Quantities\label{tbl:time}}
\tablehead{
\colhead{Model} &
\colhead{$\torb$} &
\colhead{$\tvern$} &
\colhead{$\tvera$} &
\colhead{$\tdep$} &
\colhead{$H$} &
\colhead{$\sigma_{z,{\rm eff}}$} &
\colhead{$n_{\rm gas}$} &
\colhead{$\rho_{\rm tot}$} &
\colhead{$\Sigma_{\rm gas}$} &
\colhead{$\Sigma_{\rm SFR}$} \\
\colhead{} &
\colhead{(Myr)} &
\colhead{(Myr)} &
\colhead{(Myr)} &
\colhead{(Gyr)} &
\colhead{(pc)} &
\colhead{$({\rm km\, s^{-1}})$} &
\colhead{$({\rm cm^{-3}})$} &
\colhead{$(M_\odot\pc^{-3})$} &
\colhead{$(M_\odot\pc^{-2})$} &
\colhead{$(M_\odot\kpc^{-2}\yr^{-1})$} 
}
\colnumbers
\startdata
R2     &       61 &       32 &       23 &  6.6\cdot 10^{-2} &  3.3\cdot 10^{2} &       64 &      7.7 &      1.3 &       74 &      1.1\\
R4     &  1.1\cdot 10^{2} &       51 &       37 &     0.23 &  3.4\cdot 10^{2} &       41 &      1.4 &     0.50 &       30 &     0.13\\
R8     &  2.2\cdot 10^{2} &  1.2\cdot 10^{2} &       76 &      2.1 &  3.5\cdot 10^{2} &       18 &     0.86 &     0.12 &       11 &  5.1\cdot 10^{-3}\\
R16    &  5.2\cdot 10^{2} &  4.5\cdot 10^{2} &  3.1\cdot 10^{2} &       31 &  8.1\cdot 10^{2} &       11 &  6.1\cdot 10^{-2} &  7.1\cdot 10^{-3} &      2.5 &  7.9\cdot 10^{-5}\\
LGR2   &  1.2\cdot 10^{2} &       52 &       48 &     0.15 &  3.6\cdot 10^{2} &       43 &      5.1 &     0.31 &       75 &     0.49\\
LGR4   &  2.0\cdot 10^{2} &       87 &       80 &     0.42 &  4.2\cdot 10^{2} &       30 &      1.5 &     0.11 &       38 &  9.0\cdot 10^{-2}\\
LGR8   &  4.1\cdot 10^{2} &  2.2\cdot 10^{2} &  1.7\cdot 10^{2} &      3.3 &  6.0\cdot 10^{2} &       17 &     0.37 &  2.5\cdot 10^{-2} &       10 &  3.2\cdot 10^{-3}\\

\enddata
\tablecomments{
Column (2): orbit time (\autoref{eq:torb}). 
Column (3): vertical oscillation time defined by \autoref{eq:tversim} using numerically measured gas scale height (Column 6) and vertical velocity dispersion (Column 7).
Column (4): vertical oscillation time defined by \autoref{eq:tver} using total mass density at the midplane (Column 9).
Column (5): gas depletion time defined by the ratio of gas surface density (Column 10) and SFR surface density (Column 11).
Column (6): gas scale height (\autoref{eq:H}).
Column (7): effective vertical velocity dispersion (\autoref{eq:sz}).
Column (8): midplane number density of gas.
Column (9): total midplane mass density of gas, stars, and dark matter.
Column (10): gas surface density.
Column (11): SFR surface density.
Numerically measured quantities are averaged over $0.5~\torb<t<1.5~\torb$.}
\end{deluxetable*}

\begin{figure*}
    \centering
    \includegraphics[width=\textwidth]{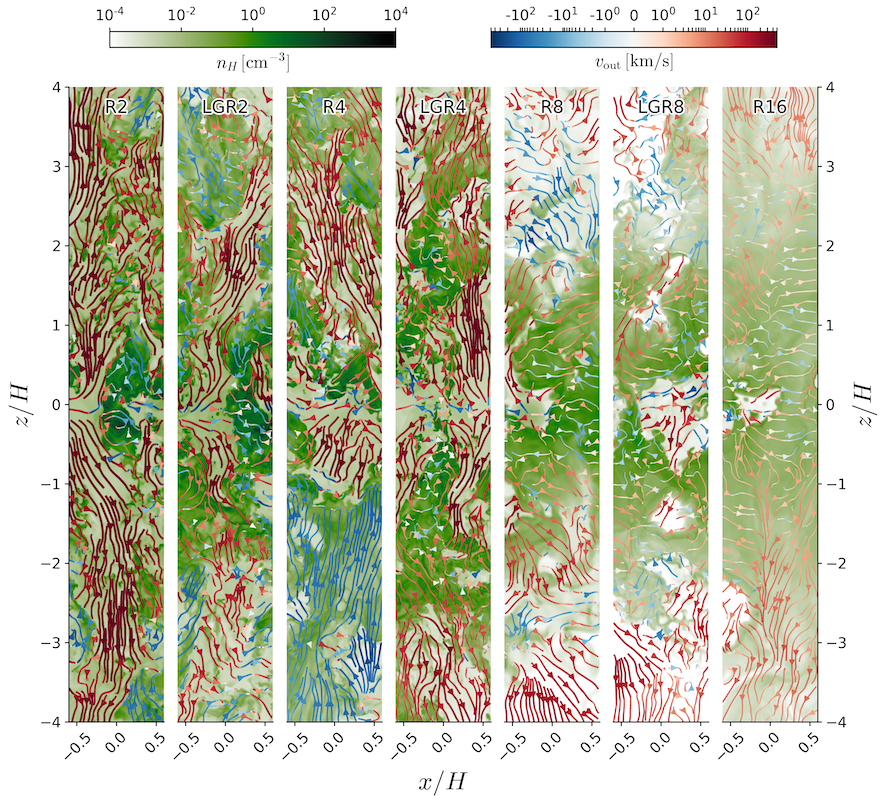}
    \caption{Snapshots from all of the simulations  at $t/\torb=0.7$. 
    The animation of this figure is available in the online journal as well as
at  \url{https://changgoo.github.io/tigress-wind-figureset/movies.html}. Number
density slices at $y=0$ are shown together with velocity streamlines (also
color coded by outward vertical velocity $v_{\rm out}=v_z\,{\rm sgn}(z)$).
Models are arranged in order of decreasing time-averaged $\Ssfr$ from left to
right. Spatial scales in each panel are normalized by the time-averaged scale
height $H$ of each model (see Column (6) in \autoref{tbl:time}). Note that the
full extent of the simulation domain is typically larger than what is shown in
this figure; $L_x/H$ and $L_y/H$ are 1-2 while $L_z/(2H)$ is 6-10.
An animation is available in the online journal as well as
at  \url{https://changgoo.github.io/tigress-wind-figureset/movies.html}.
The video begins at  $t/\torb=0$ and ends at $t/\torb=1.48$. The realtime duration of the video is 20 seconds.
}
    \label{fig:slices}
\end{figure*}

\begin{figure*}
    \centering
    \includegraphics[width=\textwidth]{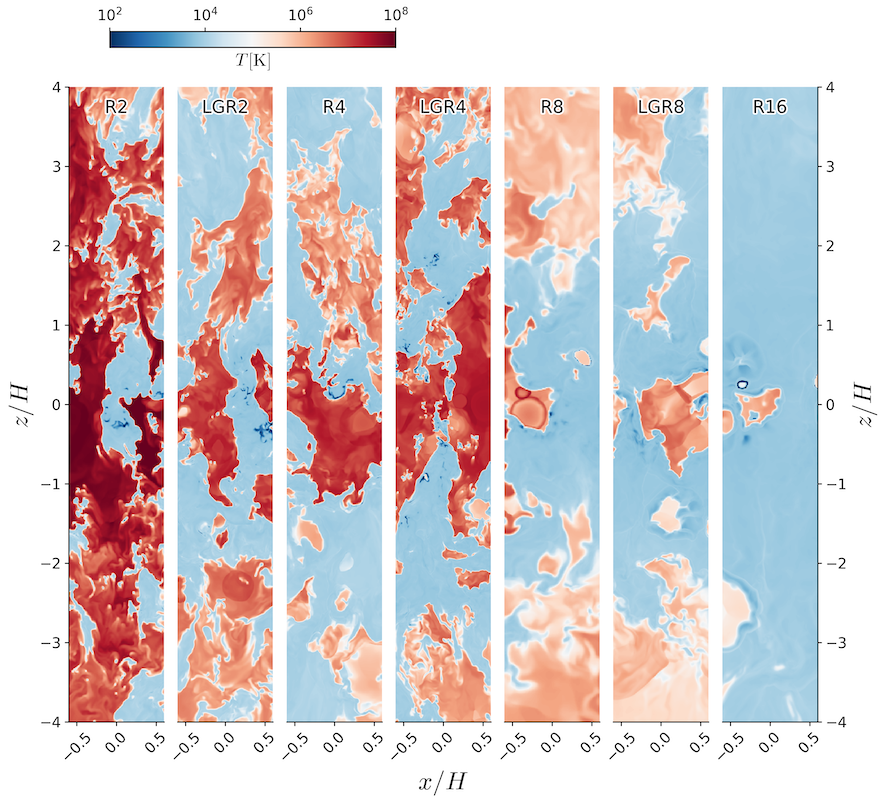}
    \caption{Same as \autoref{fig:slices}, but for temperature slices. 
An animation is available in the online journal as well as
at  \url{https://changgoo.github.io/tigress-wind-figureset/movies.html}.
The video begins at  $t/\torb=0$ and ends at $t/\torb=1.48$. The realtime duration of the video is 20 seconds.}
    \label{fig:slices_T}
\end{figure*}

\begin{figure*}
    \centering
    \includegraphics[width=\textwidth]{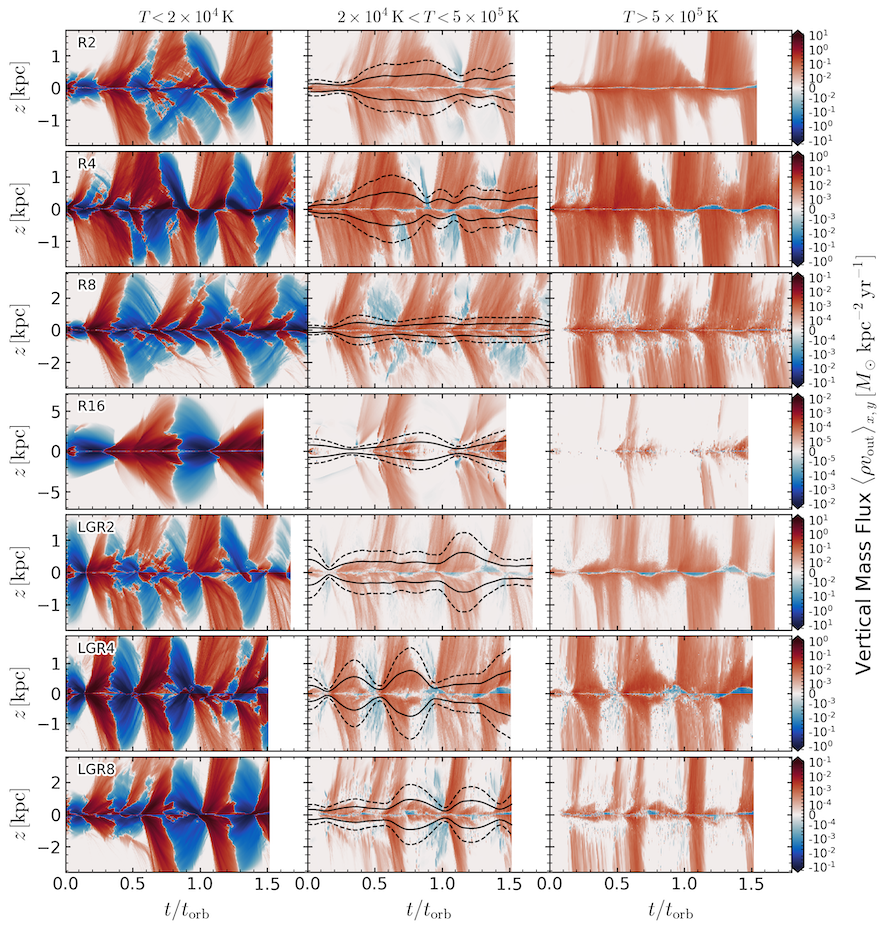}
    \caption{Space-time diagrams of the horizontally averaged mass flux  $\langle \rho v_{\rm out}  \rangle_{x,y}$. Outward fluxes ($\langle \rho v_{\rm out}\rangle_{x,y} >0$) are in red while inward fluxes ($ \langle \rho v_{\rm out} \rangle_{x,y} <0$) are in  blue. Each row represents different models, and each column represents different phases: {\bf left} for cool ($T<2\times10^4\Kel$), {\bf middle} for intermediate ($2\times10^4\Kel<T<5\times10^5\Kel$), and {\bf right} for hot ($T>5\times10^5\Kel$). Time is normalized by orbital time (see Column (2) in \autoref{tbl:time}; typically $\torb\sim2\tvern$ except R16). In the middle column, we plot one and two gas scale heights (see \autoref{eq:H} for definition of $H$) as solid and dashed black lines, which will be used to measure instantaneous fluxes through surfaces at $H$ and $2H$. While cool extraplanar gas is a fountain with alternating outflow and inflow (left column), hot extraplanar gas flows out consistently as a wind (right column).}
    \label{fig:massflux}
\end{figure*}

In previous papers (KO17 and KO18), we have presented the overall evolution of the Solar neighborhood model (R8 in this paper). The evolution exhibits multiple feedback cycles, reaching a quasi-steady state in which the SFR is self-regulated by stellar feedback. We will present a comprehensive analysis of the suite of simulations presented here in context of the theory of \emph{pressure-regulated, feedback-modulated star formation} in a separate paper (Ostriker \& Kim in prep.; see also \citealt{2010ApJ...721..975O,2011ApJ...731...41O,2011ApJ...743...25K}). In this paper, our focus is mainly on outflows from the main gas layer, above the scale height of gas. However, here we also briefly cover star formation self-regulation since the bursts and lulls  of star formation are responsible for the cyclic behavior of outflow/inflow. 

As soon as the simulation begins, the initial turbulent energy begins to dissipate and the denser gas cools to form the cold medium. Material falls vertically and dense, cold cloud complexes form and collect near the midplane. Star clusters are born in gravitationally collapsing parts of the cloud complexes; these heat the ISM by emitting UV radiation and drive turbulence through SN explosions, restoring the lost vertical support. As the disk puffs up, the overall SFR drops. The now-reduced stellar feedback cannot offset cooling and turbulent
dissipation, so that gas falls back to the midplane and the next star formation event follows. The cycle repeats, with the system entering a self-regulated, quasi-steady state.

In this state, the time-averaged total vertical pressure support (sustained by star formation feedback) balances the vertical weight of gas (as shown in simulations of \citealt{2013ApJ...776....1K,2015ApJ...815...67K}; see also \citealt{2005ApJ...630..167T,2010ApJ...721..975O,2011ApJ...731...41O,2011ApJ...743...25K, 2012ApJ...754....2S,2013MNRAS.433.1970F,2017MNRAS.465.1682H}). However, instantaneously there is always a mismatch between ``supply'' (pressure from feedback) and ``demand'' (weight from gravity). Especially, the injection of energy and momentum from SN feedback is highly concentrated in space and time, leading to an overshoot \citep[e.g.,][]{2016MNRAS.462.3053B,2019MNRAS.486.4724O}.
The resulting gas outflows carry the excess momentum and energy into the extraplanar region (above the gas scale height), and a portion is eventually vented to the CGM. As we shall show (see also KO18), in our simulation suite the hot gas created by SN shocks is mainly responsible for energy and momentum delivery to the extraplanar region and beyond, while the cooler gas delivers significant mass beyond the disk scale height. The self-regulation and outflow cycle is evident in all models, with some qualitative differences.

To help visualize feedback-driven outflows  in the simulation suite, \autoref{fig:slices} and \autoref{fig:slices_T} respectively  show density and temperature slices at $y=0$ for snapshots at $t/\torb=0.7$. Also, in \autoref{fig:slices} we show velocity streamlines color coded by outward vertical velocity,
\begin{equation}\label{eq:vout_def}
    \vout\equiv v_z\,{\rm sgn}(z).
\end{equation} The hot, fast outflows preferentially vent through low density chimneys, carved out of the denser warm ISM by superbubble breakout events. At the same time, a highly dynamic fountain of clumpy, cooler gas coexists with hot gas in the extraplanar region, and is both inflowing and outflowing. Turbulent flows of cool gas close off chimneys, limiting hot outflows and leading to significant interaction between hot winds and cool fountains.

\autoref{fig:massflux} shows the horizontally-averaged vertical mass flux, $\langle \rho v_{\rm out}\rangle_{x,y}$, from all models as a time series. The space-time diagram of mass flux profiles demonstrates the cycles of outflow/inflow. Outward fluxes ($\langle \rho v_{\rm out}\rangle_{x,y} >0$) are in red while inward fluxes ($\langle \rho v_{\rm out}\rangle_{x,y} <0$) are in  blue.
We separate gas using three temperature bins, $T<2\times10^4\Kel$ for cool (left column)\footnote{The cool phase in this paper includes cold/unstable/warm gas (or two-phase gas) as defined in KO17 and KO18. Although we do not explicitly distinguish cold, unstable, and warm gas as in previous work, the fractions of cold and unstable components are negligible at $|z|>H$. Therefore, the cool phase in this paper is essentially equivalent to the warm gas of KO18.}, $2\times10^4\Kel<T<5\times10^5\Kel$ for intermediate (middle column), and $T>5\times10^5\Kel$ for hot gas (right column). For reference, we plot $H$ and $2H$ as solid and dashed lines, respectively, in the middle column, where $H$ is the instantaneous scale height of gas (the mass-weighted dispersion of vertical gas positions; see \autoref{eq:H}).

Focusing on the left column, where we show the ``cool'' component, cyclic behavior of alternating outflow and inflow is evident for all models. The evolution is more regular for R16, and gets more complex at higher surface densities. For R8, LGR4, and LGR8, the evolution is still quite cyclic, while R2, R4, and LGR2 show complex interaction between outflows and inflows and generally less cyclic evolution, especially for the gas near the midplane.

Qualitative differences in the cyclic behavior among models can be understood from competition between key time scales: the vertical oscillation time $\tver$ and the  star cluster evolution time scale $\tevol \sim 40\Myr$ \citep{1999ApJS..123....3L}. The former controls the self-regulation cycle because gas pushed outward by feedback from a burst of star formation returns after $\tver$ and participates in the next star formation event. The latter sets the duration of energy/momentum injection from a given star formation event, during which star formation is generally reduced.

In \autoref{tbl:time}, we list three time scales along with relevant quantities measured from simulations to obtain these time scales. Column (2) gives the orbital period of galactic rotation ($\torb$) with the usual definition:
\begin{equation}\label{eq:torb}
    \torb \equiv \frac{2\pi}{\Omega}
    = 120\Myr \rbrackets{\frac{\Omega}{50\kms\kpc^{-1}}}^{-1}.
\end{equation}
For $\tver$, we list both a measure from the simulation and an analytic estimate in Columns (3) and (4), respectively.
The vertical gravity is nearly linear $g_z\approx -4\pi G \rho_{\rm tot}z$ for the majority of gas since the gas scale height is smaller than or comparable to the stellar height ($z_*$) and dark matter scale length ($R_0$) assumed in our potential model (except for model R16). The collisionless vertical oscillation time can then be approximated only in terms of the total midplane density $\rho_{\rm tot}=\rho_{\rm gas} + \Sigma_*/(2z_*) + \rho_{\rm dm}$ as
\begin{equation}\label{eq:tver}
    \tvera \approx \frac{2\pi}{(4\pi G \rho_{\rm tot})^{1/2}}= 37\Myr \rbrackets{\frac{\rho_{\rm tot}}{0.5\Msun\pc^{-3}}}^{-1/2}.
\end{equation}
Note that the midplane density of the gas $\rho_{\rm gas}$ (or $n_{\rm gas}=\rho_{\rm gas}/(\mu m_H)$) is calculated by taking the mean of density in the two horizontal planes at $z=\pm\Delta z/2$. Since the prediction for the scale height under linear gravity is $H=\sigma_z/(4\pi G \rho_{\rm tot})^{1/2}$, an alternative definition of the vertical oscillation time measurable directly from gas properties is
\begin{equation}\label{eq:tversim}
    \tvern\equiv \frac{2\pi H}{\sigma_{\rm z,eff}} 
    = 46\Myr\rbrackets{\frac{H}{300\pc}}\rbrackets{\frac{\sigma_{\rm z,eff}}{40\kms}}^{-1}.
\end{equation}
Here, we calculate $H$ and $\sigma_{\rm z,eff}$ from the mass-weighted height and effective velocity dispersion measured in the simulation, where they are respectively defined by
\begin{equation}\label{eq:H}
    H\equiv \rbrackets{\frac{\int \rho z^2 dV}{\int \rho dV}}^{1/2},
\end{equation}
and
\begin{equation}\label{eq:sz}
    \sigma_{z,{\rm eff}}\equiv\rbrackets{\frac{\int \sbrackets{\rho v_z^2 + P + B^2/(8\pi) -B_z^2/(4\pi)} dV}{\int \rho dV}}^{1/2}.
\end{equation}
Note that the effective vertical velocity dispersion includes the contributions from turbulent, thermal, and magnetic stresses. 
The values are all time averages over $0.5\torb<t<1.5\torb$. $\tvern\sim0.5\torb$ for our models, except R16.

If $\tver$ is sufficiently longer than $\tevol$ (e.g., for R16), a major star formation event cannot occur until after previously blown-out gas falls back. If $\tver$ is smaller than or comparable to $\tevol$ (e.g., R2, R4, and LGR2), the situation is more complicated. Since each major star formation event continuously injects energy/momentum for $\tevol$ (SN rates do not decline sharply for $\tevol$), gas that is launched and returns after $\tver$ can be re-launched before participating in the next star formation event. The self-regulation cycle is delayed until feedback shuts off after $\sim \tevol$.

The distinct oscillatory behavior seen in \autoref{fig:massflux} is in part due to the limited horizontal domain of the TIGRESS simulations.  Because the natural horizontal correlation scale of star formation is not extremely small compared to the size of our simulation domain, averages at a given time will not statistically sample many independent regions at different stages of the evolutionary cycle. Synchronization within a local patch can also be enhanced if initial conditions tend to trigger a  collapse of the entire disk, as in models LGR2 and LGR4. For LGR4, where $\tver>\tevol$, the prominent oscillation cycle persists for a long time. However, for LGR2, even though the initial collapse induces very coherent first outflows, the feedback regulation cycles become highly irregular since $\tver\sim\tevol$ so that the inflowing gas keeps interacting with outflows from previous feedback events. On the other hand, while the early evolution of LGR8 is quite irregular, it eventually shows a fairly regular oscillation at later times since $\tver\gg\tevol$. Overall, the late time evolution ($t>0.5\torb$) and regularity of the cyclic behavior are self-consistently set by the fundamental time scales of the system.

The orbital time of galactic rotation $\torb$ is relevant to the growth of the large scale gravitational instability, due to the effects of epicyclic oscillations and shear \citep[e.g.,][]{1965MNRAS.130...97G,1987ApJ...312..626E,2001ApJ...559...70K,2002ApJ...581.1080K}. In general, the gravitational timescale $t_g  \sim \sigma/(G\Sigma)$  must be  shorter than the epicyclic or shear times ($\sim \torb$) for gravitational instabilities to grow. Typically, $t_g  \sim \torb$ in normal galaxies, i.e. the Toomre parameter is order-unity \citep{1964ApJ...139.1217T}. If gravitational instability were the only important dynamical process acting on large scales, the inevitable result would be  a strong starburst. However, in our simulations, $\torb$ does not control  star formation by  itself because $\tver,\tevol \lesssim \torb, t_g$, such that coherent structures at large scales are not able to continue growing for very long  periods.  Instead, they are destroyed by feedback before high star formation  efficiency is  achieved. We note, however, that in model R2, $\torb, t_g  \sim \tevol$, so that feedback is less able to limit large scale gravitational instability.
In reality, conditions with very  short orbital and gravitational timescales may also be subject to strong radial flows.   Following this in detail would require global modeling, but unfortunately this is not yet tractable with the same uniformly high spatial resolution as our simulations.   

Finally, we note that the gas depletion time is generally longer than $\tver$, $\tevol$, and $\torb$, so that secular evolution has a minimal effect on the average properties in the self-regulated state. 

In \autoref{fig:massflux}, it is generally possible to link cool outflows (left) with the outflows of intermediate (middle) and hot (right) phases. Simultaneous, distinct outflows in all phases are realized when there is breakout of superbubbles produced by spatially and temporally correlated SNe. In an ``outflowing'' epoch, the hot outflows easily reach the domain boundaries without significant loss of mass flux. 
However, the cool gas launched with  the hot gas after  a burst eventually falls back. That is, red turns to blue for the cool gas. Notably, even during an ``inflowing'' epoch of the cool gas, the high-entropy hot gas continues to rise. 
That is, the hot gas shows only outflows (red), with no returning mass flux. 
We note, however, that even if there is a reasonably high SN rate producing hot gas, hot outflows are sometimes blocked by returning inflows of cool gas, and cannot reach the boundary (see e.g., R2 (top row) at $t/\torb\sim1$). Thus, successful breakout is  not solely determined by the SFRs (or SN rates), but is subject to the complex interaction between superbubble expansion and inflowing gas from previous events (sometimes in neighboring regions).

To  summarize:  outflows in our simulation suite show both regular and complex behaviors, depending on the model parameters (\autoref{fig:massflux}). In the extraplanar region, outflows and inflows coexist in different phases, which we resolve  in our  simulations (\autoref{fig:slices} and \autoref{fig:slices_T}). In what follows, we explain how we characterize key properties of outflows and relate them to global properties, thereby deriving scaling relations.  

\section{Characterizations of Multiphase Outflows}\label{sec:windprop}

In this section, we present characterization of multiphase outflows using outward mass, momentum, energy, and metal fluxes, separating the different thermal phases. We first present results for time evolution of outward fluxes (\autoref{sec:flux}) and metal properties (\autoref{sec:metal}) through surfaces (both upper and lower sides of the disk) at different heights, including two fixed heights at 500~pc and 1~kpc,  and two time-dependent heights using the instantaneous gas scale height at $H$ and $2H$.\footnote{The main motivation of this work is to report emergent multiphase outflow properties from resolved, self-consistent simulations of the  star-forming ISM. In this undertaking, there is a tension between competing desiderata. On the one hand, it may be desired to measure outflowing gas properties at heights far from the disk midplane where interactions with the ``ISM'' gas have been left behind.  Larger distances are also closer to the resolution of big-box cosmological simulations.  
On the other  hand, there is a countervailing need to choose a height closer to the midplane where the local approximation is valid  (and climbing out of the global potential has not affected the outflow velocities).    
We shall show that interactions are minimized above $\sim 2H$, while the local assumption is reasonable (with the local potential dominating over the global disk + halo potential and the flow streamlines not affected by global geometry) up to $H$ or $2H$.
Locations between $H$ and $2H$ are thus a good compromise for making our measurements of outflow properties.}
We then show time averaged vertical profiles of loading factors (\autoref{sec:loading}), outflow velocities, and metal properties (\autoref{sec:velmetal}). Figures and Tables in this section are for model R4 or for values at $|z|=H$. Figures for other models and the data for Tables at different heights are available at \url{https://changgoo.github.io/tigress-wind-figureset/figureset.html} and \href{http://doi.org/10.5281/zenodo.3872049}{doi:10.5281/zenodo.3872049}.

\subsection{Outflow Fluxes}\label{sec:flux}
\begin{figure*}
    \centering
    \includegraphics[width=\textwidth]{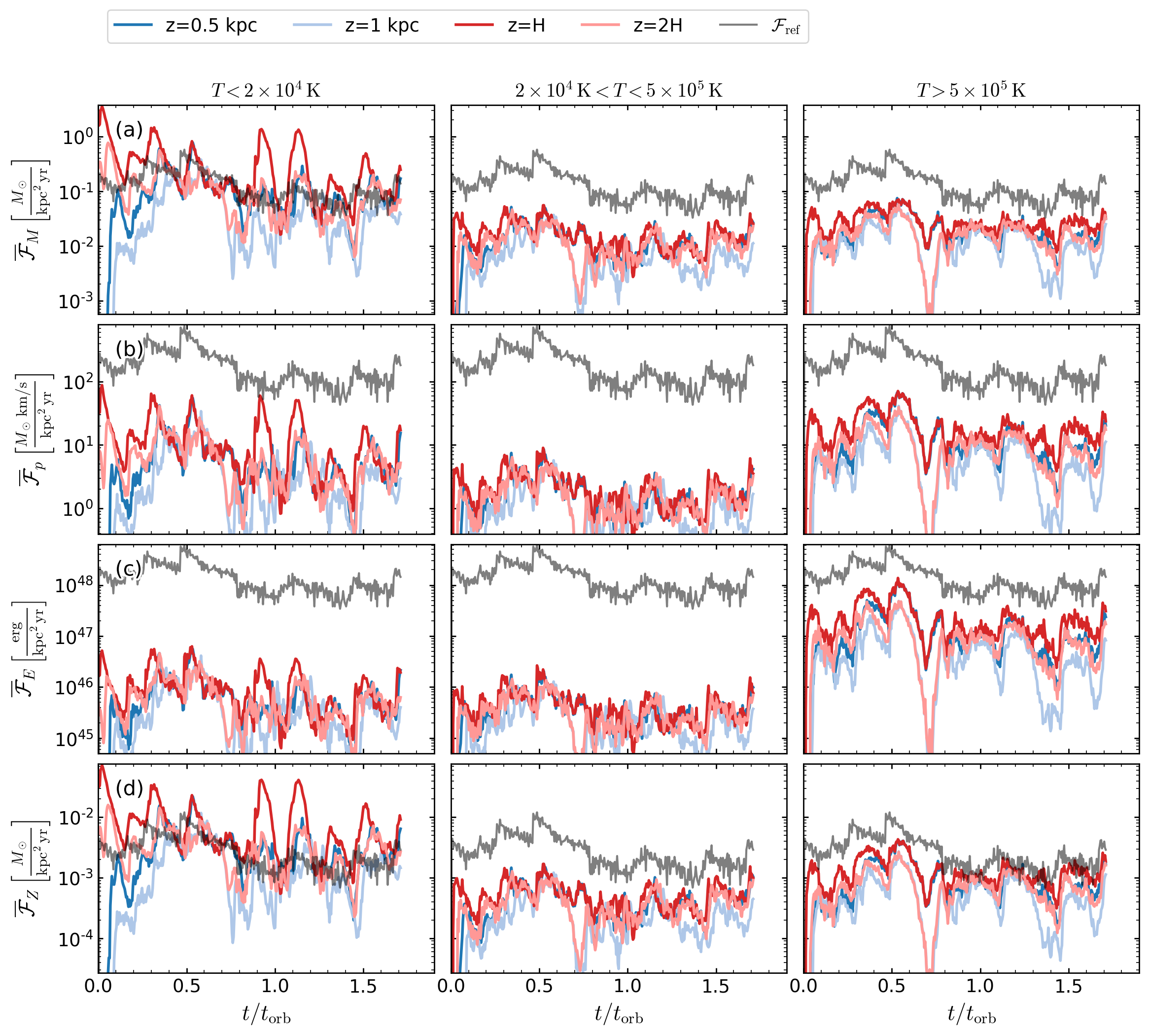}
    \caption{Time evolution of outward fluxes in model R4. Figures for other models are available 
    at \url{https://changgoo.github.io/tigress-wind-figureset/figureset.html}.
    The fluxes are measured through surfaces at $|z|=H$ (red), $2H$ (light red), 0.5~kpc (blue), and 1~kpc (light blue). Each column represents different phases: left for cool, middle for intermediate, and right for hot. Each row represents flux of  a different quantity (see \autoref{eq:outflowrate} for definition):
    {\bf (a)-row:} mass flux in 
    $\mfluxunit$ (\autoref{eq:massflux});
    {\bf (b)-row:} $z$-momentum flux in 
    $\pfluxunit$ (\autoref{eq:momflux});
    {\bf (c)-row:} energy flux in 
    $\efluxunit$ (\autoref{eq:energyflux}); and
    {\bf (d)-row:} total metal flux in 
    $\mfluxunit$ (\autoref{eq:metalflux}).
    The grey solid lines are corresponding reference fluxes ($\mathcal{F}_{\rm ref}$) defined by \autoref{eq:fluxref}, representing (a) SFR surface density and (b) momentum, (c) energy, and (d) metal injection rates by SNe per horizontal area. Mass is predominantly carried by the  cool component, while energy is predominantly carried by the hot component.
    }
    \label{fig:R4flux}
\end{figure*}

The instantaneous outflow fluxes of each thermal phase through a horizontal surface at height $z$ are calculated by summing up the vertical fluxes of cells with a positive outward vertical velocity $\vout>0$ and temperature in the specified temperature range of each phase. Formally, the outflow flux of the quantity `$q$' in phase `ph' is defined by summing up the phase-selected flux over the horizontal domain with an area of $L_x L_y$ as
\begin{equation}\label{eq:outflowrate}
\flux{q}{ph} = \sum_{i,j} \mathcal{F}_q(i,j,k;t)
\Theta(C) \frac{\Delta x \Delta y}{L_x L_y},
\end{equation}
where $(i,j,k)$ is grid zone index, $\Delta x=\Delta y$ is grid resolution (Column (8) in \autoref{tbl:model}), and $\Theta(C)$ is the top-hat-like filter that returns 1 if the conditional argument is true or 0 otherwise. Here, the conditional argument is $\vout>0$ for the outflowing gas and $T$ in three temperature bins $T<2\times10^4\Kel$, $2\times10^4\Kel <T<5\times10^5\Kel$, and $T>5\times10^5\Kel$ for the cool, intermediate, and hot phases, respectively. We consider four physical quantities $q=M$, $p$, $E$, and $M_Z$ (we simply use $Z$ in the subscript for succinctness; e.g., $\mathcal{F}_Z$ is not metallicity flux but metal mass flux) to denote mass, $z$-momentum, energy, and metal mass. The corresponding vertical outgoing fluxes are defined by
\begin{eqnarray}
\mathcal{F}_M& =& \rho \vout \label{eq:massflux}\\
\mathcal{F}_p & =& \rho \vout^2 + P\label{eq:momflux}\\
\mathcal{F}_E& =& \rho \vout\rbrackets{\frac{1}{2} v^2 + \frac{\gamma}{\gamma-1}c_s^2}, \label{eq:energyflux}\\
\mathcal{F}_{Z}& =& \rho Z \vout \label{eq:metalflux}
\end{eqnarray}
where $v^2=(\vel\cdot\vel)$, $c_s^2\equiv P/\rho$, and $\gamma=5/3$, while other symbols have their usual meaning. We note that \autoref{eq:momflux} and \autoref{eq:energyflux} include both the kinetic and thermal components of the vertical momentum and total energy fluxes, ignoring the magnetic terms that are negligible in outflows. We analyze the total momentum flux instead of just the kinetic term separately since the contribution from the thermal pressure in the hot gas is substantial in outflows. The thermal pressure in cool gas is largely set by the balance between photoheating and cooling (rather than being driven by SNe), but is negligible in outflows. The simulation has nonzero metallicity in the beginning ($Z_{\rm ISM,0}=0.02$), and the metal flux consists of two origins, metals from the ISM and newly injected by SNe ($Z_{\rm SN}=0.2$). Although we separately trace the total and SN-origin metals using independent passive scalar variables, separation between SN-origin and ISM-origin metals is non-trivial due to enrichment of the ISM and recycling. We discuss this in more detail in the next section (\autoref{sec:metal}). For now, we show the total metal flux.

\autoref{fig:R4flux} plots, from top to bottom, the horizontally-averaged mass, momentum, energy, and metal fluxes of model R4. We report combined fluxes through both upper and lower horizontal surfaces at a fixed height (dark and light blues for $|z|=0.5$ and $1\kpc$, respectively) or at a time dependent height (dark and light reds for $|z|=H$ and $2H$, respectively; see grey solid and dashed lines in the middle column of \autoref{fig:massflux} for the time variation of the scale heights). From the left to right column, we show cool, intermediate, and hot outflows separately.

From \autoref{fig:R4flux}, it is evident that the features of multiphase outflows at launch seen in the Solar neighborhood TIGRESS model (R8) (as reported by KO18) are generic across galactic conditions considered in our model suite. The cool component delivers most of  the  mass to the extraplanar region; in these simulations this cool gas subsequently returns to the midplane (the cool ``fountain'' is clear in the left column of \autoref{fig:massflux}), but in a shallower global potential this cool outflow could escape. The hot component carries most of energy, and escapes from the simulation domain as a wind.  
The intermediate component is subdominant in all fluxes. Due to the short cooling time of the intermediate component, a significant fraction of the outflow in this temperature range cools and mixes into the cooler gas in the course of its evolution \citep[e.g.,][for more quantitative analyses]{2019arXiv191107872V}.\footnote{The exception is model R16 (an outer disk model with the lowest surface density), in which the intermediate component contains mass, momentum, and energy comparable to those in the hot component at launching. Also, the intermediate-temperature gas behaves more or less similarly to the hot component due to the increased cooling time at the  low densities in this model.}
The fluxes, especially for the cool component, generally decrease with  distance from the midplane. Occasionally (e.g., at around $t\sim0.7\torb$; see also \autoref{fig:massflux}), cool inflows are strong enough to shut off outflows nearly completely, showing dramatic drops from lower heights ($|z|=H$ or 500 pc; darker lines) to upper heights ($|z|=2H$ or 1 kpc; lighter lines).

In every panel  of \autoref{fig:R4flux}, the gray solid line shows a corresponding reference flux calculated based on the instantaneous SN rate, defined by
\begin{equation}\label{eq:fluxref}
    \mathcal{F}_{q, {\rm ref}} \equiv q_{\rm ref}\frac{\dot{N}_{\rm SN}}{L_x L_y}.
\end{equation}
Here, $\dot{N}_{\rm SN}$ is the instantaneous SN rate calculated with an adaptive time window, within which the number of SNe is 100. The coefficient $q_\mathrm{ref}$ adopted in each reference flux is set based on simple physical considerations, as follows:
\begin{eqnarray}
    M_{\rm ref}= m_* &=& 95.5\Msun \label{eq:mref}\\
    p_{\rm ref}= E_{\rm SN}/(2v_{\rm cool}) &=& 1.25\times10^5\Msun\kms \label{eq:pref}\\
    E_{\rm ref} = E_{\rm SN} &=&10^{51}\erg \label{eq:Eref}\\
    M_{\rm Z, ref} = M_{\rm ej} Z_{\rm SN} &=& 2\Msun.\label{eq:Zref}
\end{eqnarray}
Here, we adopt a total mass of new stars formed for each SN of $m_*=95.5\Msun$ \citep{2001MNRAS.322..231K}, a SN explosion energy of $E_{\rm SN}=10^{51}\erg$, a mean mass in ejecta from each SN of $M_{\rm ej}=10\Msun$, and a mean SN ejecta metallicity of $Z_{\rm SN}=0.2$. We note that the combination of \autoref{eq:fluxref} and \autoref{eq:mref} is equivalent to a reference mass flux of  $\mathcal{F}_{M, {\rm ref}} = \Ssfr$, the mean SFR per unit area averaged over the star cluster life time.  Because SNe from a given star cluster persist over $\tevol\sim40\Myr$ for a fully-sampled IMF, the reference fluxes defined by \autoref{eq:fluxref} depend on the SFR over the last 40~Myr and are therefore smoother than they would be if an instantaneous (or time-delayed) value of $\Ssfr$ were employed, while still giving the  same long-term  average.

For the reference momentum per SN, we adopt a value $E_{\rm SN}/v_{\rm cool}$ with  $v_{\rm cool}=200 \kms$, which represents the spherical momentum at the end of the Sedov stage when a SN blast wave cools and a shell forms  \citep{2011piim.book.....D,2015ApJ...802...99K}, also applying the geometric factor $1/2$ to account for the vertical component of midplane-centered sources \citep{2011ApJ...731...41O}.
An alternative reference momentum choice that is sometimes adopted is the initial SN ejecta momentum $p_{\rm ej}\equiv (2M_{\rm ej}E_{\rm SN})^{1/2} = 3.2 \times10^4\Msun\kms$.  This (reduced by a  factor 2) would be more instructive choice if the SNR evolution remains in the free expansion stage until it reaches the height where a wind is launched. We generally find that this is not the case.
We note that $p_\mathrm{ref}$ is an order of magnitude greater than the vertical momentum from initial SN ejecta, $p_{\rm ej}/2$.

From a large number of recent investigations of individual SNR evolution in inhomogeneous environments  \citep[e.g.,][]{2015ApJ...802...99K,2015MNRAS.450..504M,2015A&A...576A..95I,2015MNRAS.451.2757W}, and of superbubble evolution driven by multiple SNe \citep[e.g.,][]{2017ApJ...834...25K,2018MNRAS.481.3325F,2019MNRAS.483.3647G,2019MNRAS.490.1961E}, there is an emerging consensus in the community that the momentum injection to the ISM per event is relatively insensitive to ambient conditions. In particular, the momentum depends only very weakly on density, as shown in earlier uniform-background simulations \citep[e.g.][]{ 1988ApJ...334..252C,1998ApJ...500..342B,1998ApJ...500...95T}. The terminal momentum per SN from clustered SNe is comparable to that from a single SN event as long as shocks from individual SNe remain supersonic, but can be factor of a few smaller if blast waves from individual SNe become subsonic before reaching the shell \citep[e.g.,][]{2017ApJ...834...25K,2019MNRAS.490.1961E}. Since most SNe are clustered, with the reference value of \autoref{eq:pref} we expect the kinetic momentum loading near the midplane to smaller than unity, and to be lower in models with higher SFR.

The dimensionless ratios between measured and reference fluxes are often termed \emph{loading factors} (e.g., \citealt{2015ARA&A..53...51S}; see \autoref{sec:loading}). Although actual and reference fluxes in \autoref{fig:R4flux} share similar evolutionary trends, the reference fluxes do not show the same large modulations as some of the measured fluxes. 
As we discussed in \autoref{sec:evolution}, complicated interaction between outflows and inflows makes one-to-one correspondence between strength of feedback (outflow driving) and emergent fluxes non-trivial (see also \autoref{sec:app-delay}).

\subsection{Metallicity and Enrichment of Outflows}\label{sec:metal}

To understand the role of SN feedback in metal evolution within and beyond galaxies, simply measuring the total metal flux is insufficient. Every SN explosion injects metal mass $Z_{\rm SN}M_{\rm ej}=2\Msun$, some of which goes directly to the extraplanar region as outflows, and some of which mixes with the ISM near the midplane, which has initialized with solar metallicity $Z_{\rm ISM,0}=0.02$. At a given epoch, outflowing gas can thus originate from one of three components: the ISM at the beginning of the simulation, $\mdotio$, SN ejecta from \emph{previous} SN events that have mixed into the ISM, $\mdotsio$, and SN ejecta from  \emph{current} SN events, $\mdotso$. Note that $\mdotsio$ in principle includes metals recycled from fountain flows, which we do not separate track in this study. The total mass and metal outflow rates can be respectively written as
\begin{eqnarray}
    \dot{M} &=& \mdotio+\mdotsio+\mdotso\label{eq:mdot_decomp}\\
    \dot{M}_{Z} &=& Z_{\rm ISM,0}\mdotio+Z_{\rm SN}\mdotsio+Z_{\rm SN}\mdotso.\label{eq:mzdot_decomp}
\end{eqnarray}
In TIGRESS, we employ passive scalars for total and SN-injected metals, with densities that evolve under the mass conservation equation with a given velocity field. The total metal scalar allows us to measure $\dot{M}_Z$, while the SN-injected scalar traces the sum of last two terms $Z_{\rm SN}[\mdotsio+\mdotso]$ in \autoref{eq:mzdot_decomp}, corresponding to the ``cumulative'' SN-origin metal flux.

While not directly calculated, the  ``instantaneous'' SN-origin metal flux, $Z_{\rm SN}\mdotso$, is of great interest to quantify how much of injected metals go promptly into outflows and how enriched the outflow is compared to the ISM. We use the following procedure to estimate this quantity.  
Theoretically, the instantaneous metallicity of ISM-origin outflows is
\begin{equation}\label{eq:ZISMdef}
    \overline{Z}_{\rm ISM} \equiv \frac{Z_{\rm ISM,0}\mdotio + Z_{\rm SN}\mdotsio}{\mdotio+\mdotsio},
\end{equation}
while the mean metallicity of outflows,
\begin{equation}
    \overline{Z}=\frac{\dot{M}_Z}{\dot{M}},\label{eq:Zbar}
\end{equation}
is directly measured in the simulation using mass and total metal scalar fluxes at specified $z$.  
Combining with \autoref{eq:mdot_decomp} and \autoref{eq:mzdot_decomp}, we obtain the instantaneous SN-origin mass outflow rate
\begin{equation}
\mdotso = \frac{\overline{Z}-\overline{Z}_{\rm ISM}}{Z_{\rm SN}-\overline{Z}_{\rm ISM}}\dot{M} \equiv f_M^{\rm SN}\dot{M} \label{eq:mfluxsn}
\end{equation}
and the instantaneous SN-origin metal outflow rate
\begin{equation}
\mzdotso = Z_{\rm SN}\mdotso= \frac{Z_{\rm SN}}{\overline{Z}}f_M^{\rm SN}\dot{M}_Z\equiv f_Z^{\rm SN}\dot{M}_Z.\label{eq:Zfluxsn}
\end{equation}
\autoref{eq:mfluxsn} and \autoref{eq:Zfluxsn}  define the  instantaneous SN-origin mass and metal fractions in outflows, $f_M^{\rm SN}$ and $f_Z^{\rm SN}$, respectively. In the rest of the paper, we will use the superscript ``SN'' to refer to the   \emph{instantaneous} SN-origin component, e.g., $\dot{M}^{\rm SN}_{\rm out}$ and $\dot{M}^{\rm SN}_{\rm Z,out}$ for $\mdotso$ and $\mzdotso$, respectively, and $\overline{\mathcal{F}}_M^{\rm SN}$ and $\overline{\mathcal{F}}_Z^{\rm SN}$ for the corresponding mass and metal fluxes. 

As a proxy for the instantaneous metallicity of ISM-origin outflows in \autoref{eq:ZISMdef}, we use the {instantaneous} metallicity of the ISM itself.
In practice, in the simulation we measure the ISM metallicity based on the cool phase gas within $|z|<50\pc$; this defines $\overline{Z}_{\rm ISM}$ (we find no strong variation with different thickness used in this definition if smaller than the scale height). For a given $\overline{Z}_{\rm ISM}$, we use the phase-separated mean outflow metallicity ($\overline{Z}_{\rm ph}$) and mass outflow rate ($\dot M_{\rm ph}$) to obtain $f_{M, {\rm ph}}^{\rm SN}$ and $f_{Z, {\rm ph}}^{\rm SN}$ phase-by-phase. Note that our definition for $\overline{Z}_{\rm ISM}$ is not a perfect tracer of the instantaneous metallicity in ISM-origin outflows, so that occasionally we get negative $f_M^{\rm SN}$; we simply set it to zero in such occasions. This occurs only for the cool outflow at $|z|=H$ (at most 20\% of the time). R16 is the only exception, where the genuine cool ISM is easily pushed out to large distance so that $\overline{Z}_{\rm cool}\le\overline{Z}_{\rm ISM}$ for most snapshots at all heights (up to 80\% of the time). For this  reason, we exclude R16 in analysis regarding SN-origin metals of cool outflows (e.g., \autoref{tbl:scaling} and \autoref{fig:scaling-loading}). For the hot gas, $f_M^{\rm SN}$ is always positive.

Given instantaneous outflow and ISM metallicities, we obtain the instantaneous outflow enrichment factor for each phase `ph'
\begin{equation}\label{eq:yZ}
\zeta_{\rm ph}\equiv \frac{\overline{Z}_{\rm ph}}{\overline{Z}_{\rm ISM}}.
\end{equation} 

\begin{figure*}
    \centering
    \includegraphics[width=\textwidth]{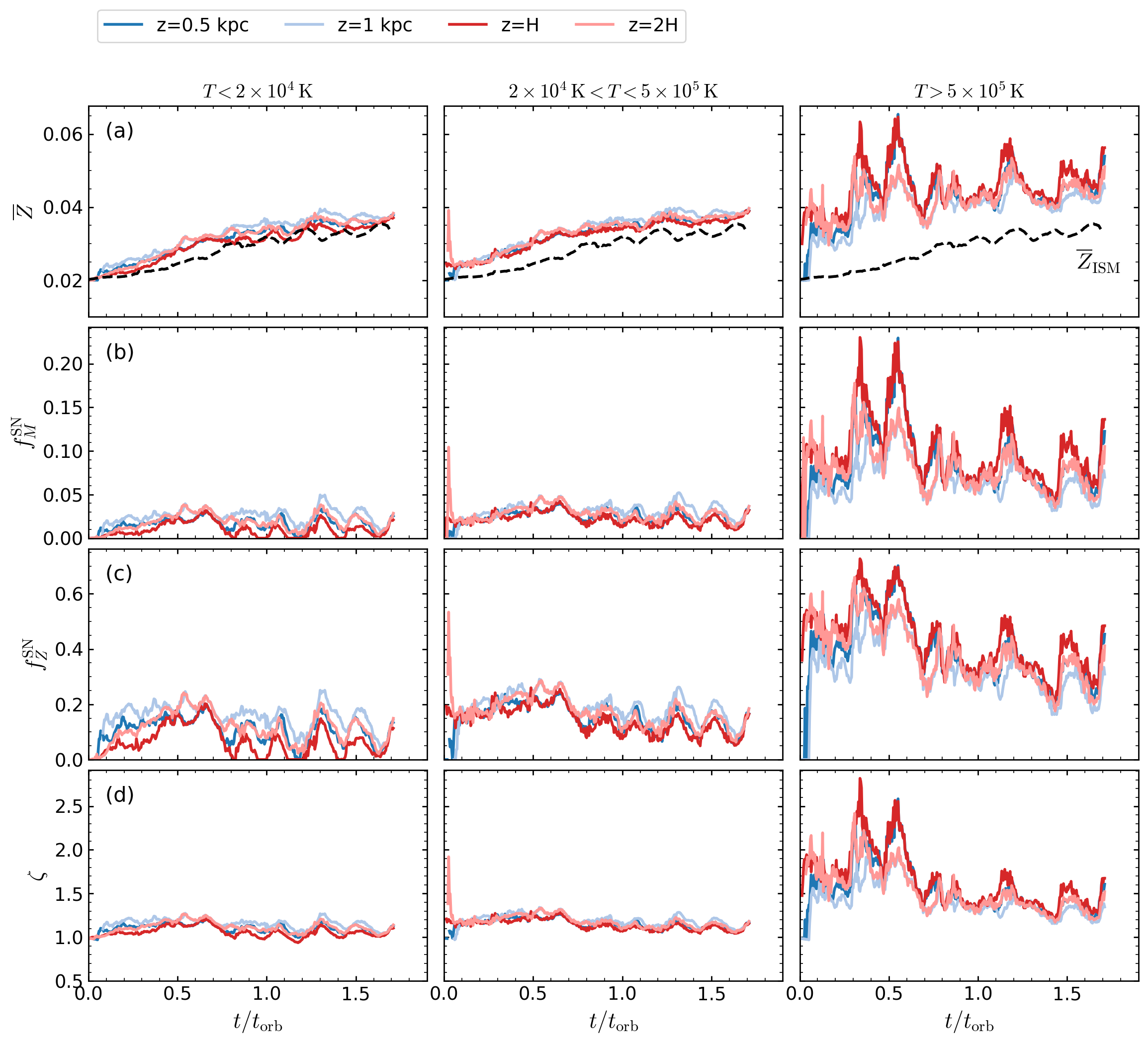}
    \caption{Time evolution of metal properties in  the outflow of model R4. Figures for other models are available at \url{https://changgoo.github.io/tigress-wind-figureset/figureset.html}. Each quantity is measured through a surface at $|z|=H$ (red), $2H$ (light red), 0.5 kpc (blue), and 1 kpc (light blue). Each column represents different phases: left for cool, middle for intermediate, and  right for hot. {\bf (a)-row:} mean metallicity $\overline{Z}$ of the outflows along with  the instantaneous ISM metallicity $\overline{Z}_\mathrm{ISM}$ (black dashed, from the cool gas within $|z|<50\pc$). {\bf (b)-row:} ratio of instantaneous SN-origin outflowing mass flux to total. {\bf (c)-row:} ratio of instantaneous SN-origin outflowing metal flux to total. {\bf (d)-row:} instantaneous outflow enrichment factor. See  \autoref{eq:mdot_decomp} - \autoref{eq:yZ} for definitions.   
    The metallicity of cool outflowing gas is essentially the same as that of the ISM near the midplane, whereas hot outflowing gas is significantly enriched by the ejecta from recent SNe.
    }
    \label{fig:R4metalflux}
\end{figure*}

\autoref{fig:R4metalflux} shows, from the top to bottom, (a) the mean metallicity of outflow along with the instantaneous ISM metallicity (solid dashed), (b/c) the fractions of instantaneous SN-origin mass and metal in the outflow, and (d) the instantaneous outflow enrichment factor for R4. The ISM is gradually enriched, and cool outflows consist mostly of the pre-enriched ISM ($\zeta_{\rm cool}\sim1$). The hot outflow is more metal-enriched than cooler components and the ISM by a factor of 1.5-2. The contribution of recent-SN-origin metals to the outflowing metal flux is $\sim30$-60\% in the hot outflow and $\sim 10\%$ in the cooler components.

\subsection{Loading Factors}\label{sec:loading}

\begin{figure}
    \centering
    \includegraphics[width=0.5\textwidth]{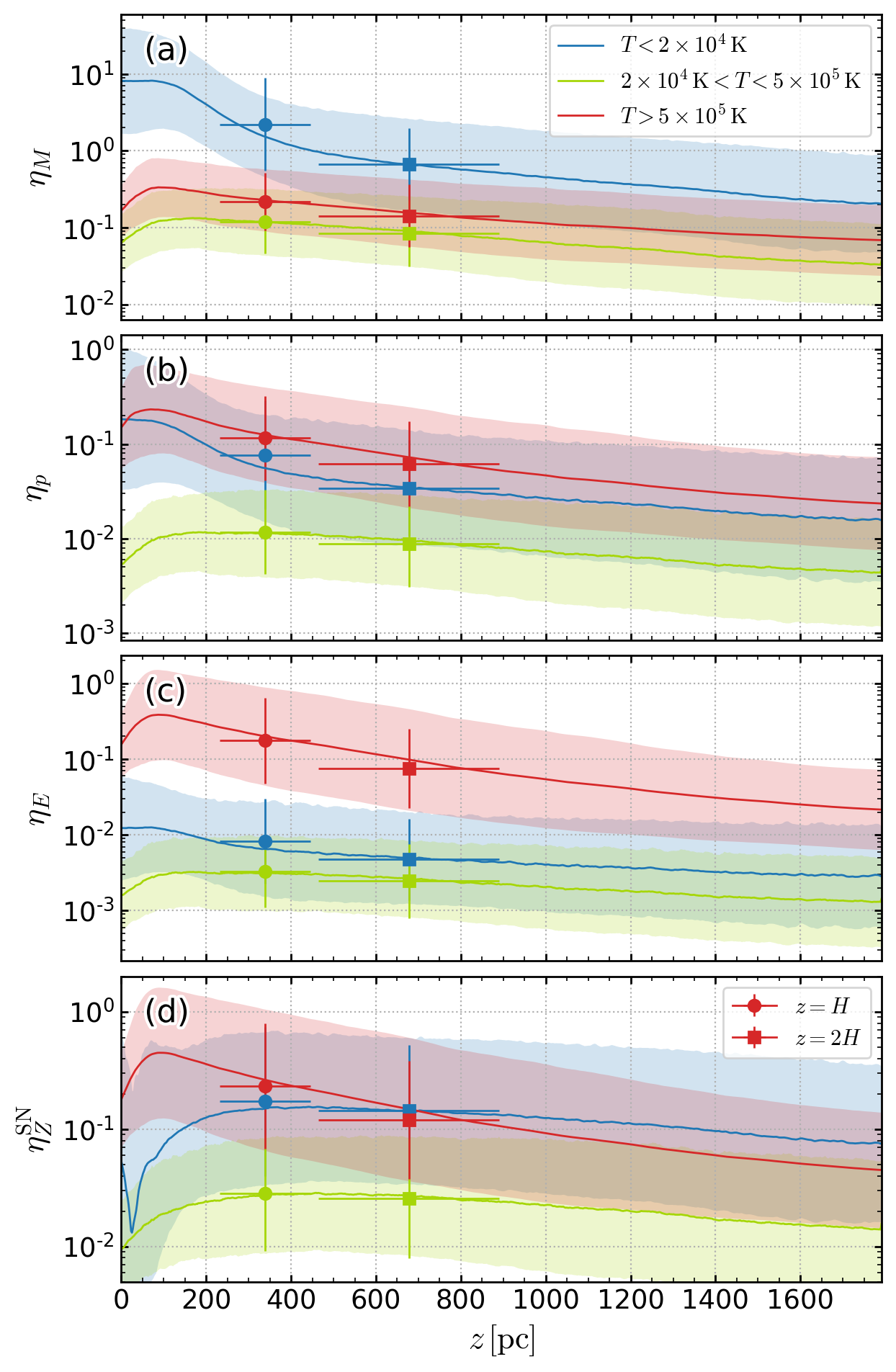}
    \caption{Vertical profiles of loading factors in model R4. Figures for other models are available at \url{https://changgoo.github.io/tigress-wind-figureset/figureset.html}. Rows show {\bf (a)} mass, {\bf (b)} $z$-momentum, {\bf (c)} energy, and {\bf (d)} SN-origin metal loading factors. These loading factors represent fluxes carried by  outflowing gas relative to time-averaged reference fluxes defined by \autoref{eq:fluxref}.
    The solid line denotes the mean value $\abrackets{q}_t$ at a given height averaged over $0.5<t/\torb<1.5$. The shaded area represents temporal variations at each height using the standard deviation in temporal fluctuations $\delta q$, i.e., $\abrackets{q}_t\exp(\pm\delta q/\abrackets{q}_t)$.
    The symbols with errorbars denote the mean and fractional temporal variation range of instantaneous measurements at $|z|=H$ and $2H$. The  mass in outflows is primarily loaded in the cool gas, but this declines with height as cool gas velocities are insufficient to escape and the flow turns around as a fountain. The energy is primarily loaded in a hot wind; while velocities are high enough to escape, the energy loading declines with $z$ due to  interactions with the warm gas.}
    \label{fig:R4loading}
\end{figure}

We now calculate \emph{loading factors}; the ratios of outgoing mass, momentum, energy, and metal to mass locked into stars and momentum, energy, and metal injected by SNe. With the definition of the reference outflow fluxes in Equation (\ref{eq:fluxref}), we get outflow loading factors simply by
\begin{equation}\label{eq:loading}
\eta_{q}\equiv\frac{\overline{\mathcal{F}}_q}{\mathcal{F}_{q, {\rm ref}}},
\end{equation}
where $q=M$, $p$, $E$, and $Z$. 
The definitions of $\eta_M$ and $\eta_E$ are identical to the conventional definition \citep[e.g.,][]{1985Natur.317...44C}.

In principle, \autoref{eq:loading} can give \emph{instantaneous} loading factors as a function of time, but care needs to be taken with this.
The quantity of interest is the outflow rate normalized by the injection (or star formation) rate that is responsible for the outflow. The injection (or star formation) generally occurs near the disk midplane, while the outflow is measured at a certain height above the midplane. 
There is inevitably a time delay between injection and outflow rates. Therefore, instantaneous loading factors measured by Equation (\ref{eq:loading}) can be misleading regarding the physical impact of stellar feedback. This issue is more serious when (1) star formation is more bursty, and the SFR rather than SNR is used for the reference flux\footnote{For example, \citet{2019arXiv190710623M} recently reported an large instantaneous mass loading factor ($\sim 100$), which they suggested was due to clustered star formation under self-gravity. However, from Figures 7 and 10 in \citet{2019arXiv190710623M}, the peak in the  instantaneous mass loading factor from Model S100\_WSG at $t/t_{\rm dyn}\sim3$ occurs at both a maximum of the outflow rate and a minimum of the SFR. If one simply reads off the peaks of both outflow and SFRs and takes the ratio, the mass loading factor is 0.1, comparable to their non-self-gravitating (non-clustering) model. The drop in star formation (after an initial big burst) is the major reason for the high instantaneous mass loading factor.} 
and (2) the distance between locations where feedback injection (or star formation) and outflow rates are measured is larger. One would need to either carefully model the reference fluxes including determination of an appropriate time delay in computing instantaneous loading factors, or else report time-averaged loading factors with averaging timescale longer than the time delay and timescale of the feedback cycle \citep[e.g.,][]{2015MNRAS.454.2691M}. 

Taking advantage of the long duration of our simulation suite (covering a few feedback/outflow cycles), we report the ratio of time-averaged flux to time-averaged reference flux as time-averaged loading factors.
For model R4, \autoref{fig:R4loading} shows the vertical profiles of the time-averaged loading factors, as well as temporal variation ranges. We note that in contrast to \autoref{fig:R4flux}, rather than the total metal loading factor we now show the instantaneous 
SN-origin metal loading factor $\eta_Z^{\rm SN}$ (using \autoref{eq:Zfluxsn}) in the bottom panel of \autoref{fig:R4loading}.  
We also note that the reference fluxes are not height dependent, so that \autoref{fig:R4loading} essentially shows rescaled flux profiles. We plot as symbols with errorbars the mean and standard deviation of loading factors measured at the instantaneous $H$ (circle) and $2H$ (square), which are generally in good agreement with the values from the time averaged profiles at the mean $H$ and $2H$. In \autoref{sec:app-delay}, we make use of the time-delayed reference fluxes to find the mean time delay and calculate instantaneous loading factors. The time averaged profiles of the instantaneous loading factors are almost identical with \autoref{fig:R4loading}, providing reassurance  of the robustness of mean loading factors we report here.

In \autoref{fig:R4loading}, we see a steep drop of all loading factors of the cool phase from the midplane to $H$. The hot (and intermediate) phase loading factors peak at $\sim50\pc$ (most SNe explode below this height).  Above $|z|\sim 50\pc$, outflow fluxes gradually drop. The decreasing trend is moderated above $H$, but is still significant in cooler phases for mass and hotter phases for energy. The mass loading factor of the cool phase decreases with $|z|$ as lower velocity components drop out (see \autoref{fig:R4vout} and KO18). The energy loading factor of the hot phase decreases from a maximum slightly above the midplane as some of the energy (mostly thermal) in the hot gas transferred to cooler phases, from which it is quickly radiated away \citep{2019arXiv191107872V}. The intermediate phase is subdominant for all loading factors at all heights (except R16).

The momentum loading factor of the sum of the cool and hot components near the midplane is $\eta_p\sim 0.5$, implying that SNRs have built up momentum exceeding the initial ejecta momentum (which would yield  $\eta_p\sim 0.1$). We note that $\eta_p$ is not as large as unity since the terminal momentum per SN from clustered SNe is generally smaller than $p_{\rm ref}$ from a single SN, especially when SN events are nearly continuous and blast waves become subsonic in the hot ISM before reaching the cool shell \citep{2017ApJ...834...25K,2019MNRAS.490.1961E}. Also, a portion  of the injected SNe momentum flux is converted to magnetic stresses, and  near the midplane these are comparable to the vertical kinetic momentum flux. The momentum flux also decreases as function of $z$, especially for the cool component, since it must contribute support against the weight of the ISM (the thermal plus turbulent pressure is approximately twice $\mathcal{F}_p$, allowing for $\vout < 0$). At $|z|=H$ and above, the leftover kinetic vertical momentum flux is only 10\% of the reference momentum flux.
This is generally true except in R16, in which fluxes are all dominated by the cool component, and SN events  are more or less discrete.

In \autoref{tbl:fluxloading}, we provide the mean values over $0.5<t/\torb<1.5$ of the measured fluxes and loading factors of all models and phases at $|z|=H$.

\begin{deluxetable*}{lcCCCCCCCCCC}
\tabletypesize{\footnotesize}
\tablecaption{Time averaged fluxes and loading factors at $|z|=H$\label{tbl:fluxloading}}
\tablehead{
\colhead{Model} &
\colhead{phase} &
\dcolhead{\mathcal{F}_M} &
\dcolhead{\mathcal{F}_p} &
\dcolhead{\mathcal{F}_E} &
\dcolhead{\mathcal{F}_{Z}} &
\dcolhead{\mathcal{F}_{Z}^{\rm SN}} &
\dcolhead{\eta_M} &
\dcolhead{\eta_p} &
\dcolhead{\eta_E} &
\dcolhead{\eta_{Z}} &
\dcolhead{\eta_{Z}^{\rm SN}} 
}
\colnumbers
\startdata
R2     &   cool &     0.75 &       51 &  7.3\cdot 10^{46} &  2.9\cdot 10^{-2} &  3.2\cdot 10^{-3} &     0.69 &  3.5\cdot 10^{-2} &  6.4\cdot 10^{-3} &      1.3 &     0.14\\
       &    int &  6.3\cdot 10^{-2} &       10 &  2.8\cdot 10^{46} &  2.6\cdot 10^{-3} &  5.6\cdot 10^{-4} &  5.8\cdot 10^{-2} &  7.1\cdot 10^{-3} &  2.5\cdot 10^{-3} &     0.11 &  2.5\cdot 10^{-2}\\
       &    hot &     0.13 &  1.4\cdot 10^{2} &  2.8\cdot 10^{48} &  9.6\cdot 10^{-3} &  6.2\cdot 10^{-3} &     0.12 &     0.10 &     0.24 &     0.42 &     0.27\\
\hline
R4     &   cool &     0.27 &       12 &  1.1\cdot 10^{46} &  8.3\cdot 10^{-3} &  4.4\cdot 10^{-4} &      2.2 &  7.7\cdot 10^{-2} &  8.2\cdot 10^{-3} &      3.3 &     0.17\\
       &    int &  1.4\cdot 10^{-2} &      1.9 &  4.2\cdot 10^{45} &  4.8\cdot 10^{-4} &  7.2\cdot 10^{-5} &     0.12 &  1.2\cdot 10^{-2} &  3.3\cdot 10^{-3} &     0.19 &  2.8\cdot 10^{-2}\\
       &    hot &  2.7\cdot 10^{-2} &       19 &  2.2\cdot 10^{47} &  1.3\cdot 10^{-3} &  6.0\cdot 10^{-4} &     0.22 &     0.12 &     0.17 &     0.51 &     0.23\\
\hline
R8     &   cool &  3.3\cdot 10^{-2} &     0.79 &  4.4\cdot 10^{44} &  7.2\cdot 10^{-4} &  2.1\cdot 10^{-5} &      6.4 &     0.12 &  8.2\cdot 10^{-3} &      6.7 &     0.20\\
       &    int &  1.3\cdot 10^{-3} &     0.12 &  2.3\cdot 10^{44} &  3.0\cdot 10^{-5} &  2.9\cdot 10^{-6} &     0.25 &  1.8\cdot 10^{-2} &  4.3\cdot 10^{-3} &     0.28 &  2.7\cdot 10^{-2}\\
       &    hot &  1.3\cdot 10^{-3} &     0.67 &  5.5\cdot 10^{45} &  4.1\cdot 10^{-5} &  1.5\cdot 10^{-5} &     0.26 &     0.10 &     0.10 &     0.39 &     0.14\\
\hline
R16    &   cool &  5.5\cdot 10^{-3} &  8.5\cdot 10^{-2} &  2.3\cdot 10^{43} &  1.1\cdot 10^{-4} &  3.3\cdot 10^{-9} &       56 &     0.67 &  2.2\cdot 10^{-2} &       54 &  1.6\cdot 10^{-3}\\
       &    int &  3.6\cdot 10^{-5} &  2.9\cdot 10^{-3} &  3.8\cdot 10^{42} &  7.8\cdot 10^{-7} &  5.3\cdot 10^{-8} &     0.37 &  2.3\cdot 10^{-2} &  3.8\cdot 10^{-3} &     0.39 &  2.6\cdot 10^{-2}\\
       &    hot &  1.4\cdot 10^{-5} &  9.3\cdot 10^{-3} &  6.1\cdot 10^{43} &  4.5\cdot 10^{-7} &  1.8\cdot 10^{-7} &     0.15 &  7.3\cdot 10^{-2} &  6.0\cdot 10^{-2} &     0.22 &  8.7\cdot 10^{-2}\\
\hline
LGR2   &   cool &     0.55 &       27 &  2.8\cdot 10^{46} &  1.8\cdot 10^{-2} &  1.5\cdot 10^{-3} &      1.2 &  4.2\cdot 10^{-2} &  5.7\cdot 10^{-3} &      1.9 &     0.15\\
       &    int &  2.6\cdot 10^{-2} &      3.6 &  8.9\cdot 10^{45} &  9.7\cdot 10^{-4} &  1.9\cdot 10^{-4} &  5.4\cdot 10^{-2} &  5.7\cdot 10^{-3} &  1.8\cdot 10^{-3} &  9.7\cdot 10^{-2} &  1.9\cdot 10^{-2}\\
       &    hot &  5.4\cdot 10^{-2} &       48 &  6.7\cdot 10^{47} &  3.2\cdot 10^{-3} &  1.8\cdot 10^{-3} &     0.11 &  7.6\cdot 10^{-2} &     0.14 &     0.33 &     0.18\\
\hline
LGR4   &   cool &     0.46 &       14 &  8.4\cdot 10^{45} &  1.2\cdot 10^{-2} &  2.1\cdot 10^{-4} &      5.1 &     0.12 &  9.0\cdot 10^{-3} &      6.3 &     0.11\\
       &    int &  1.0\cdot 10^{-2} &      1.2 &  2.5\cdot 10^{45} &  3.0\cdot 10^{-4} &  3.7\cdot 10^{-5} &     0.11 &  1.0\cdot 10^{-2} &  2.7\cdot 10^{-3} &     0.16 &  2.0\cdot 10^{-2}\\
       &    hot &  1.5\cdot 10^{-2} &       10 &  1.0\cdot 10^{47} &  6.5\cdot 10^{-4} &  2.8\cdot 10^{-4} &     0.17 &  8.4\cdot 10^{-2} &     0.11 &     0.34 &     0.15\\
\hline
LGR8   &   cool &  4.0\cdot 10^{-2} &     0.85 &  3.6\cdot 10^{44} &  8.6\cdot 10^{-4} &  7.8\cdot 10^{-6} &       13 &     0.20 &  1.1\cdot 10^{-2} &       13 &     0.12\\
       &    int &  7.3\cdot 10^{-4} &  7.2\cdot 10^{-2} &  1.3\cdot 10^{44} &  1.7\cdot 10^{-5} &  1.4\cdot 10^{-6} &     0.23 &  1.7\cdot 10^{-2} &  4.0\cdot 10^{-3} &     0.26 &  2.2\cdot 10^{-2}\\
       &    hot &  8.8\cdot 10^{-4} &     0.43 &  3.3\cdot 10^{45} &  2.7\cdot 10^{-5} &  8.5\cdot 10^{-6} &     0.28 &     0.10 &  9.9\cdot 10^{-2} &     0.41 &     0.13\\
\hline

\enddata
\tablecomments{
Columns (3)-(7) are time averaged outflow fluxes defined by \autoref{eq:outflowrate} for (3) mass flux in $\mfluxunit$ (\autoref{eq:massflux}),
(4) momentum flux in $\pfluxunit$ (\autoref{eq:momflux}),
(5) energy flux in $\efluxunit$ (\autoref{eq:energyflux}),
(6) total metal flux in $\mfluxunit$ (\autoref{eq:metalflux}), and (7) SN-origin metal flux in $\mfluxunit$ (\autoref{eq:Zfluxsn}).
Columns (8)-(12) are time averaged loading factors (dimensionless) defined by \autoref{eq:loading} and \autoref{eq:fluxref} with 
(8) \autoref{eq:mref} for mass,
(9) \autoref{eq:pref} for momentum,
(10) \autoref{eq:Eref} for energy,
and \autoref{eq:Zref} for 
(11) total metal and (12) SN-origin metal.
Time averages are taken over $0.5\,\torb<t<1.5\,\torb$. The data for additional tables at different heights ($|z|=2H$, 500~pc, and 1~kpc) as well as standard deviations are available at \href{http://doi.org/10.5281/zenodo.3872049}{doi:10.5281/zenodo.3872049}.
}
\end{deluxetable*}

\subsection{Characteristic Velocities and Metal Properties}\label{sec:velmetal}

\begin{figure}
    \centering
    \includegraphics[width=0.5\textwidth]{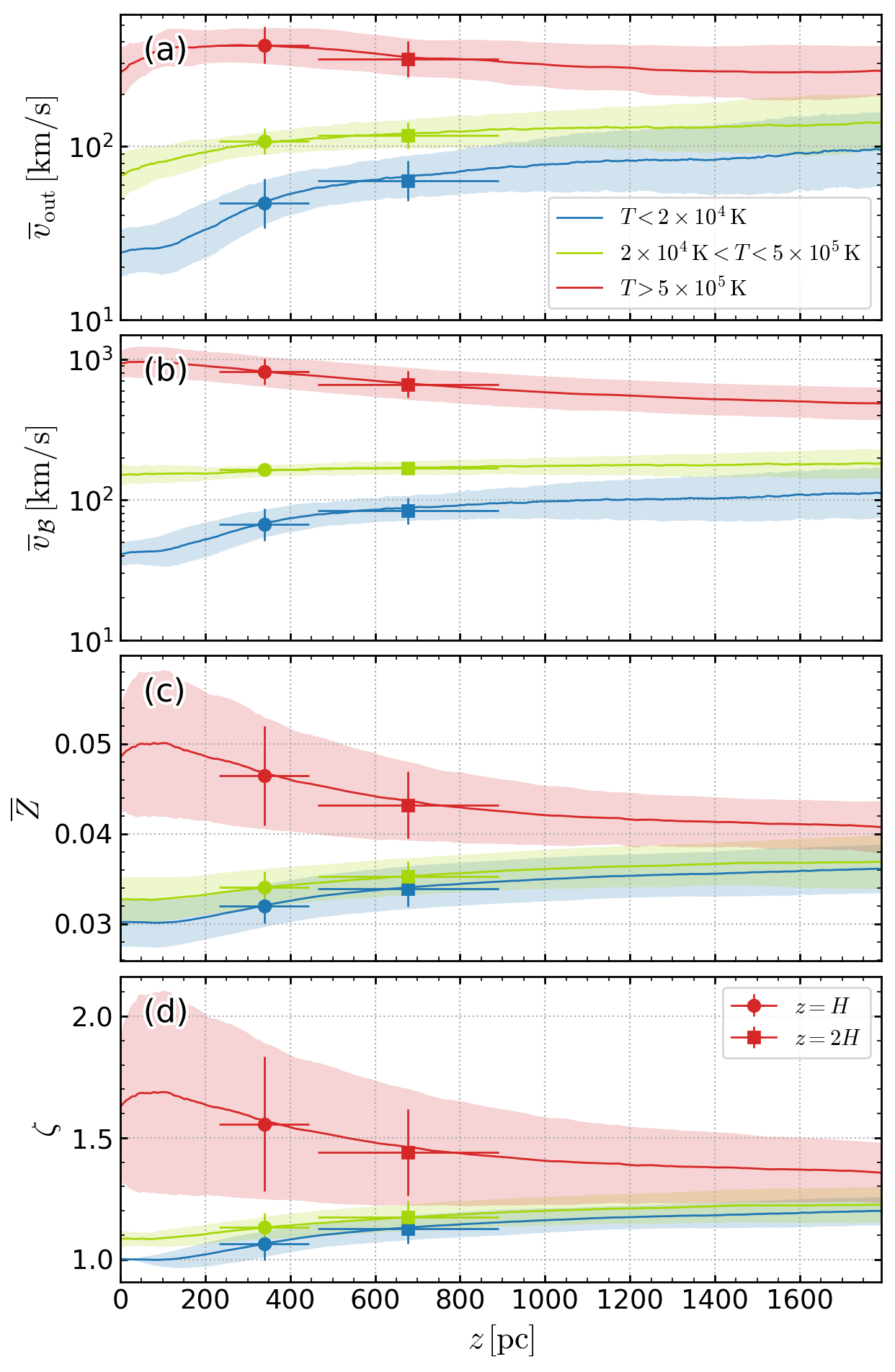}
    \caption{Vertical profiles of characteristic velocities and metal properties in model R4. Figures for other models are available at \url{https://changgoo.github.io/tigress-wind-figureset/figureset.html}. Rows show {\bf (a)} outflow velocity (\autoref{eq:vout}), {\bf (b)} Bernoulli velocity (\autoref{eq:vB}), {\bf (c)} outflow metallicity (\autoref{eq:Zbar}), and {\bf (d)} metal enrichment factor (\autoref{eq:yZ}). 
    Lines and symbols have the same meaning as in \autoref{fig:R4loading}.  The hot wind has both larger characteristic velocities and larger metal enrichment than the cool and intermediate-temperature outflows.}
    \label{fig:R4vout}
\end{figure}

\autoref{fig:R4vout} plots time-averaged vertical profiles of additional quantities of interest, including (a) outflow velocity, (b) Bernoulli velocity, (c) metallicity, and (d) metal enrichment factor for model R4. The characteristic outflow velocity is defined as
\begin{equation}\label{eq:vout}
    \overline{v}_{\rm out, ph}(z;t)\equiv
    \frac{\flux{p}{ph}^{\rm kin}(z;t)}{\flux{M}{ph}(z;t)},
\end{equation}
where $\flux{p}{ph}^{\rm kin}$ is the kinetic component of momentum flux defined by only the first term of \autoref{eq:momflux}.
The Bernoulli velocity is defined by
\begin{equation}\label{eq:vB}
    \overline{v}_{\mathcal{B}, {\rm ph}}(z;t)\equiv \rbrackets{\frac{2\flux{E}{ph}(z;t)}{\flux{M}{ph}(z;t)}}^{1/2},
\end{equation}
including both the kinetic and thermal term in \autoref{eq:energyflux}. For an adiabatic steady flow, the Bernoulli velocity at $z$ must exceed the escape speed $v_\mathrm{esc} \equiv [2(\Phi(z_\mathrm{max})-\Phi(z))]^{1/2}$ in order for the flow to reach $z_\mathrm{max}$; this criterion also applies for a completely cold ballistic flow, where $v_\mathcal{B}\rightarrow v$.   

The outflow velocity of cool outflows is as low as $\sim 30\kms$ near the midplane and increases as the outflow moves farther away, reaching as high as $\sim100\kms$ near the simulation boundaries. The increasing trend of the outflow velocity of the cool phase with $|z|$ seen in \autoref{fig:R4vout}(a) is often interpreted as an acceleration, but the mean outflow velocity can also increase as low velocity gas drops out. The former is more important near the midplane $|z|<H$ where actual acceleration of the cool phase by superbubble expansion is occurring, but the latter dominates the trend at higher altitudes \citep[e.g., KO18,][]{2019arXiv191107872V}. 
Indeed, panel (b) of \autoref{fig:R4loading} shows a trend of steadily decreasing momentum flux with $|z|$, due to the dropout of low-velocity gas.  
In the extraplanar region $|z|>H$, some  acceleration of cool outflows occurs due to the hot-cool interaction, which helps to maintain high velocity tails of cool outflows \citep{2019arXiv191107872V}, but this is not the dominant reason for the increasing trend of outflow velocity. 
We also note that cooling of intermediate-temperature gas is preferentially at low velocity and adds to the cool gas inflow; cooling of intermediate-temperature gas has minimal impact on the momentum transfer to the cool phase  \citep[see][]{2019arXiv191107872V}. 

For model R4, the escape velocity from the box relative to $|z|=H=340\pc$ (where we tabulate $\ovB$) is 
$v_\mathrm{esc}=154\kms$. Given the low mean outflow velocity of cool outflows, it is evident that the majority of cool (and intermediate) phase outflows cannot travel far from the disk midplane and escape the simulation domain. This is also clearly demonstrated by the steep decrease of mass loading factor as a function of $z$ in \autoref{fig:R4loading}(a).

The outflow velocity of hot outflows, in contrast, is higher than the escape velocity of the system, as clearly illustrated in \autoref{fig:massflux}. The Bernoulli velocity is much larger than the outflow velocity as the thermal term dominates, implying  the  possibility of further acceleration of hot outflows. In our simulations, the outflow velocity for the hot gas flattens out above $|z|>H$ or $2H$, reaching $\overline{v}_{\rm out,hot}\sim250\kms$ for model R4. This flattening is mainly due to the limited volume of the local box simulations. 
When a volume much larger than the source  region is available, hot outflows expand and increase outflow velocity at the expense of thermal energy \citep[e.g.,][]{1985Natur.317...44C}. In order for a hot flow to fully accelerate, the simulation box must be large compared to the source region, so that the streamlines can open and transition through a sonic point before reaching the boundaries \citep[e.g.][]{2017MNRAS.470L..39F,2018ApJ...860..135S,2020arXiv200210468S}, which generally does not occur when there is distributed star formation in a local box \citep[e.g.,][]{2016MNRAS.459.2311M}.

Since the asymptotic velocity is $v=(v_\mathcal{B}^2 - v_\mathrm{esc}^2)^{1/2}$ for an adiabatic wind, the Bernoulli velocity can be used as a proxy for the terminal velocity that the hot gas would reach in the case that $v_\mathcal{B}\gg v_\mathrm{esc}$.  
In \autoref{fig:R4vout}(b), the Bernoulli velocity of hot outflow decreases with  $|z|$, from $\overline{v}_{\mathcal{B}, \rm hot}(H)\sim820\kms$ to $\overline{v}_{\mathcal{B}, \rm hot}(2H)\sim660\kms$ to $\overline{v}_{\mathcal{B}, \rm hot}(L_z/2)\sim  490\kms$, while  the escape velocity decreases from $v_{\rm esc}(H)=154\kms$ to $v_{\rm esc}(2H)=137\kms$.
Since only the combination $v_\mathcal{B}^2-v_\mathrm{esc}^2$ is expected to be conserved in an adiabatic flow, the decrease of $v_\mathcal{B}$ with distance owes in part to the decrease of $v_\mathrm{esc}$ with distance, although this effect is small when $v_\mathcal{B} \gg v_\mathrm{esc}$.
Within the main body of the disk ($|z|<H$ or $2H$), a decrease in $v_\mathcal{B}$ is also expected since the strong interaction between hot and cool components transfers energy from hot to cooler gas. After the hot  gas emerges into the extraplanar region, the interaction between phases is reduced, but there is still  substantial loss of energy flux from the hot component due to interaction with cool fountain flows populated by previous events \citep{2019arXiv191107872V}. 
Even with these losses, the Bernoulli velocity of the hot outflow in all models is large enough ($>600\kms$ at $|z|=2H$) that the hot gas could be  expected to travel far out to the CGM. 

\autoref{fig:R4vout}(c) plots the mean metallicity of outflows. As shown in \autoref{fig:R4metalflux}(a), the metallicity of outflows (and the ISM) gradually increases over time. SNe inject metals mainly near the midplane.
As hot, metal-enriched bubbles expand and mix into surrounding cooler gas, the metallicity of hotter/cooler component decreases/increases as outflows travel farther. The mean metallicity in each phase significantly changes as a function of $z$ up to $|z|=2H$, again indicating active interaction and mixing between phases within $|z|<2H$. 
For $|z|\in (H,2H)$, the hot and cool outflows are respectively $\sim 50\%$ and $10$--$20\%$ more metal enriched than the ISM near the midplane (\autoref{fig:R4vout}(d)). 

In \autoref{tbl:velmetal}, we provide the mean values of the mass weighted outflow velocity, Bernoulli velocity, mean metallicity, and enrichment factors of all models and phases at $|z|=H$, averaged over $0.5<t/\torb<1.5$. 

\begin{deluxetable*}{lcCCCCCC}
\tabletypesize{\footnotesize}
\tablecaption{Time averaged velocities and metal properties at $|z|=H$\label{tbl:velmetal}}
\tablehead{
\colhead{Model} &
\colhead{phase} &
\dcolhead{\overline{v}_{\rm out}} &
\dcolhead{\overline{v}_{\mathcal{B}}} &
\dcolhead{\overline{Z}} &
\dcolhead{\zeta} &
\dcolhead{f_M^{\rm SN}} &
\dcolhead{f_Z^{\rm SN}} 
} 
\colnumbers
\startdata
R2     &   cool &       69 &  1.0\cdot 10^{2} &  3.9\cdot 10^{-2} &      1.1 &  2.6\cdot 10^{-2} &     0.14\\
       &    int &  1.4\cdot 10^{2} &  2.1\cdot 10^{2} &  4.2\cdot 10^{-2} &      1.2 &  4.4\cdot 10^{-2} &     0.21\\
       &    hot &  5.8\cdot 10^{2} &  1.4\cdot 10^{3} &  7.2\cdot 10^{-2} &      2.1 &     0.23 &     0.63\\
\hline
R4     &   cool &       47 &       67 &  3.2\cdot 10^{-2} &      1.1 &  1.1\cdot 10^{-2} &  6.9\cdot 10^{-2}\\
       &    int &  1.1\cdot 10^{2} &  1.6\cdot 10^{2} &  3.4\cdot 10^{-2} &      1.1 &  2.3\cdot 10^{-2} &     0.13\\
       &    hot &  3.8\cdot 10^{2} &  8.2\cdot 10^{2} &  4.6\cdot 10^{-2} &      1.6 &  9.6\cdot 10^{-2} &     0.40\\
\hline
R8     &   cool &       20 &       37 &  2.2\cdot 10^{-2} &      1.0 &  3.5\cdot 10^{-3} &  3.2\cdot 10^{-2}\\
       &    int &       69 &  1.3\cdot 10^{2} &  2.4\cdot 10^{-2} &      1.1 &  1.2\cdot 10^{-2} &     0.10\\
       &    hot &  2.4\cdot 10^{2} &  6.0\cdot 10^{2} &  3.1\cdot 10^{-2} &      1.4 &  5.4\cdot 10^{-2} &     0.34\\
\hline
R16    &   cool &      7.9 &       20 &  2.0\cdot 10^{-2} &      1.0 &  9.3\cdot 10^{-6} &  1.0\cdot 10^{-4}\\
       &    int &       36 &       96 &  2.2\cdot 10^{-2} &      1.1 &  6.3\cdot 10^{-3} &  7.1\cdot 10^{-2}\\
       &    hot &  1.3\cdot 10^{2} &  5.4\cdot 10^{2} &  3.2\cdot 10^{-2} &      1.6 &  5.0\cdot 10^{-2} &     0.36\\
\hline
LGR2   &   cool &       44 &       69 &  3.5\cdot 10^{-2} &      1.1 &  1.5\cdot 10^{-2} &  8.5\cdot 10^{-2}\\
       &    int &  1.1\cdot 10^{2} &  1.8\cdot 10^{2} &  3.8\cdot 10^{-2} &      1.2 &  3.6\cdot 10^{-2} &     0.19\\
       &    hot &  4.2\cdot 10^{2} &  1.0\cdot 10^{3} &  5.7\cdot 10^{-2} &      1.8 &     0.15 &     0.51\\
\hline
LGR4   &   cool &       29 &       45 &  2.8\cdot 10^{-2} &      1.0 &  4.5\cdot 10^{-3} &  3.2\cdot 10^{-2}\\
       &    int &       92 &  1.5\cdot 10^{2} &  3.0\cdot 10^{-2} &      1.1 &  1.8\cdot 10^{-2} &     0.12\\
       &    hot &  3.1\cdot 10^{2} &  7.4\cdot 10^{2} &  4.1\cdot 10^{-2} &      1.5 &  8.0\cdot 10^{-2} &     0.38\\
\hline
LGR8   &   cool &       13 &       26 &  2.2\cdot 10^{-2} &      1.0 &  1.4\cdot 10^{-3} &  1.3\cdot 10^{-2}\\
       &    int &       50 &  1.2\cdot 10^{2} &  2.4\cdot 10^{-2} &      1.1 &  1.4\cdot 10^{-2} &     0.11\\
       &    hot &  1.6\cdot 10^{2} &  4.6\cdot 10^{2} &  3.0\cdot 10^{-2} &      1.4 &  3.9\cdot 10^{-2} &     0.29\\
\hline

\enddata
\tablecomments{
Column (3): characteristic outflow velocity in $\kms$ (\autoref{eq:vout_def}).
Column (4): Bernoulli velocity in $\kms$ (\autoref{eq:vB}).
Column (5): outflow metallicity (\autoref{eq:Zbar}).
Column (6): metal enrichment factor (\autoref{eq:yZ}).
Column (7): mass fraction of SN-origin materials in outflows (\autoref{eq:mfluxsn}).
Column (8): metal mass fraction of SN-origin materials in outflows (\autoref{eq:Zfluxsn}).
Time averages are taken over $0.5\,\torb<t<1.5\,\torb$. 
The data for additional tables at different heights ($|z|=2H$, 500~pc, and 1~kpc) as well as standard deviations are available at \href{http://doi.org/10.5281/zenodo.3872049}{doi:10.5281/zenodo.3872049}.
}
\end{deluxetable*}

\section{Scaling relations}\label{sec:scaling}

In this section, we systematically investigate the dependence of outflow characteristics (loading factors, metal properties, and outflow velocities) on a variety of galactic properties in our simulations, including SFR surface density ($\Ssfr$), gas surface density ($\Sgas$), midplane gas number density ($n_{\rm mid}$), midplane total pressure ($\Pmid$), gas weight ($\mathcal{W}$), and gas depletion time ($\tdep$).
At any  time, $\Ssfr$ is calculated from the total mass of star cluster particles with age younger than $\tbin$ such that
\begin{equation}\label{eq:SFR}
\Sigma_{\rm SFR,\tbin} \equiv \frac{M_{\rm sp} (t_{\rm age}<\tbin)}{\tbin L_x L_y}.
\end{equation}
As a default, we use $\tbin=\tevol=40\Myr$, corresponding to the SFR definition that best traces the SN rates used in the reference flux calculations, but we also explored different $\tbin=10\Myr$ and $100\Myr$. $\Sgas = M_{\rm gas}/(L_xL_y)$ is directly calculated from the total gas mass divided by horizontal area.
Midplane averages are computed by taking averages in two horizontal slices at $z=\pm\Delta x/2$, with $n_{\rm mid}$ and $P_{\rm mid}$ defined using volume averaged number density and total pressure (including turbulent, thermal, and magnetic terms) just for cool gas. $\mathcal{W}$ is obtained by directly integrating  $\rho d\Phi/dz$ for cool gas from the top or bottom of the simulation domain to the midplane and averaging the  two values. The depletion time $\tdep=\Sgas/\Sigma_{\rm SFR,\tbin}$ for $\tbin=40\Myr$.
Here, we will present dependencies on galactic properties as scaling relations for cool and hot phase loading factors, characteristic velocities, and metal enrichment measured at $|z|=H$.
We also have fit scaling relations at different heights, and these results are available  at \href{http://doi.org/10.5281/zenodo.3872049}{doi:10.5281/zenodo.3872049} (see \href{https://github.com/changgoo/tigress-wind-figureset/blob/paper1/tables/Example_scripts.ipynb}{Jupyter notebook}).
We generally find smaller intrinsic scatter and better correlation at $|z|=H$ and $2H$ than at fixed heights $|z|=500\pc$ and $1\kpc$.
Because the intermediate phase is subdominant, we do not include these results in this section, but the data is available at \href{http://doi.org/10.5281/zenodo.3872049}{doi:10.5281/zenodo.3872049}.

To quantify scaling relations between two variables, we report linear regression results in log-log space. We first construct time series of quantities of interest with $0.01\torb$ interval over $0.5<t/\torb<1.5$ for each model. We then perform bootstrap resampling 500 times with a sample size of 10 (we find typical auto-correlation time scales of time series $t_{\rm corr}/\torb\in (0.05, 0.1)$) to obtain mean ($\tilde{q}$) and its error ($\delta \tilde{q}$). We feed in log of the mean ($\log \tilde{q}$) and error $\delta \tilde{q}/(\tilde{q}\ln 10 )$ for linear regression using a python version of the {\tt linmix} package.\footnote{\url{https://github.com/jmeyers314/linmix}} This is a widely tested Bayesian estimator for linear regression \citep{2007ApJ...665.1489K} to derive posterior distributions of intercept $\alpha$ and slope $\beta$ as well as intrinsic scatter $\sigma_{\rm int}$ and Pearson correlation coefficient $\rho$.

\begin{deluxetable*}{ccCCCCCCCCCC}
\tabletypesize{\footnotesize}
\tablecaption{Fitting results for cool and hot phases at $|z|=H$\label{tbl:scaling}}
\tablehead{
\colhead{} &
\colhead{} &
\multicolumn{5}{c}{{\rm cool}} &
\multicolumn{5}{c}{{\rm hot}} \\
\cline{3-7} \cline{8-12}
\colhead{X} &
\colhead{Y} &
\dcolhead{\alpha} &
\dcolhead{\beta} &
\dcolhead{{\rm Cov}(\alpha,\beta)} &
\dcolhead{\sigma_{\rm int}} &
\dcolhead{\rho} &
\dcolhead{\alpha} &
\dcolhead{\beta} &
\dcolhead{{\rm Cov}(\alpha,\beta)} &
\dcolhead{\sigma_{\rm int}} &
\dcolhead{\rho} 
} 
\colnumbers
\startdata
$\Sigma_{\rm SFR,40}$ &$\eta_M$ &-0.07^{+0.16}_{-0.15}  &  -0.44^{+0.08}_{-0.08}  &  0.019  &  0.20^{+0.19}_{-0.11}  &  -0.98^{+0.08}_{-0.02} &-0.86^{+0.14}_{-0.11}  &  -0.07^{+0.08}_{-0.06}  &  0.019  &  0.17^{+0.16}_{-0.09}  &  -0.64^{+0.71}_{-0.31}\\
                &$\eta_p$ &-1.43^{+0.14}_{-0.14}  &  -0.29^{+0.07}_{-0.07}  &  0.033  &  0.18^{+0.17}_{-0.09}  &  -0.96^{+0.14}_{-0.03} &-1.01^{+0.10}_{-0.10}  &  0.02^{+0.06}_{-0.06}  &  0.009  &  0.12^{+0.13}_{-0.07}  &  0.25^{+0.59}_{-0.86}\\
                &$\eta_E$ &-2.23^{+0.13}_{-0.13}  &  -0.12^{+0.06}_{-0.06}  &  0.013  &  0.15^{+0.15}_{-0.08}  &  -0.86^{+0.40}_{-0.12} &-0.70^{+0.12}_{-0.14}  &  0.14^{+0.08}_{-0.08}  &  0.012  &  0.16^{+0.16}_{-0.09}  &  0.87^{+0.11}_{-0.44}\\
                &$\eta_{Z}^{\rm SN}$ &-0.85^{+0.17}_{-0.17}  &  -0.02^{+0.12}_{-0.11}  &  0.112  &  0.20^{+0.26}_{-0.11}  &  -0.16^{+0.77}_{-0.64} &-0.61^{+0.11}_{-0.12}  &  0.11^{+0.07}_{-0.07}  &  0.010  &  0.14^{+0.14}_{-0.07}  &  0.87^{+0.12}_{-0.46}\\
                &$\overline{v}_{\rm out}$ &1.78^{+0.07}_{-0.07}  &  0.23^{+0.04}_{-0.04}  &  0.004  &  0.11^{+0.09}_{-0.05}  &  0.97^{+0.02}_{-0.08} &2.72^{+0.05}_{-0.06}  &  0.16^{+0.03}_{-0.03}  &  0.002  &  0.07^{+0.07}_{-0.03}  &  0.98^{+0.02}_{-0.08}\\
                &$\overline{v}_{\mathcal{B}}$ &1.92^{+0.07}_{-0.07}  &  0.17^{+0.03}_{-0.03}  &  0.004  &  0.10^{+0.08}_{-0.04}  &  0.95^{+0.03}_{-0.11} &3.04^{+0.08}_{-0.08}  &  0.11^{+0.04}_{-0.04}  &  0.005  &  0.12^{+0.09}_{-0.05}  &  0.86^{+0.10}_{-0.29}\\
                &$\zeta$ &0.04^{+0.01}_{-0.01}  &  0.01^{+0.00}_{-0.00}  &  0.000  &  0.01^{+0.01}_{-0.01}  &  0.87^{+0.10}_{-0.30} &0.25^{+0.05}_{-0.05}  &  0.03^{+0.02}_{-0.03}  &  0.002  &  0.08^{+0.06}_{-0.03}  &  0.64^{+0.26}_{-0.53}\\
\hline
$\Sigma_{\rm gas}$ &$\eta_M$ &2.17^{+0.36}_{-0.34}  &  -1.16^{+0.24}_{-0.25}  &  -0.168  &  0.27^{+0.23}_{-0.13}  &  -0.96^{+0.13}_{-0.03} &-0.47^{+0.26}_{-0.31}  &  -0.20^{+0.21}_{-0.18}  &  -0.086  &  0.16^{+0.16}_{-0.09}  &  -0.67^{+0.69}_{-0.28}\\
                &$\eta_p$ &0.04^{+0.28}_{-0.28}  &  -0.76^{+0.20}_{-0.19}  &  -0.266  &  0.19^{+0.18}_{-0.10}  &  -0.95^{+0.16}_{-0.04} &-1.06^{+0.24}_{-0.26}  &  0.02^{+0.17}_{-0.16}  &  -0.062  &  0.12^{+0.13}_{-0.07}  &  0.11^{+0.68}_{-0.81}\\
                &$\eta_E$ &-1.62^{+0.26}_{-0.25}  &  -0.32^{+0.18}_{-0.18}  &  -0.075  &  0.16^{+0.16}_{-0.08}  &  -0.85^{+0.41}_{-0.12} &-1.36^{+0.33}_{-0.33}  &  0.34^{+0.22}_{-0.22}  &  -0.101  &  0.18^{+0.16}_{-0.09}  &  0.81^{+0.16}_{-0.49}\\
                &$\eta_{Z}^{\rm SN}$ &-0.71^{+0.48}_{-0.50}  &  -0.07^{+0.32}_{-0.31}  &  -0.441  &  0.19^{+0.25}_{-0.11}  &  -0.21^{+0.78}_{-0.60} &-1.15^{+0.29}_{-0.29}  &  0.27^{+0.19}_{-0.19}  &  -0.085  &  0.16^{+0.14}_{-0.08}  &  0.81^{+0.16}_{-0.52}\\
                &$\overline{v}_{\rm out}$ &0.65^{+0.21}_{-0.21}  &  0.59^{+0.15}_{-0.15}  &  -0.070  &  0.18^{+0.13}_{-0.06}  &  0.93^{+0.05}_{-0.17} &1.92^{+0.15}_{-0.15}  &  0.41^{+0.10}_{-0.10}  &  -0.026  &  0.11^{+0.09}_{-0.04}  &  0.94^{+0.05}_{-0.16}\\
                &$\overline{v}_{\mathcal{B}}$ &1.09^{+0.17}_{-0.17}  &  0.43^{+0.12}_{-0.12}  &  -0.037  &  0.15^{+0.10}_{-0.05}  &  0.91^{+0.07}_{-0.19} &2.52^{+0.18}_{-0.17}  &  0.27^{+0.12}_{-0.12}  &  -0.035  &  0.14^{+0.11}_{-0.05}  &  0.81^{+0.14}_{-0.34}\\
                &$\zeta$ &-0.02^{+0.02}_{-0.02}  &  0.03^{+0.01}_{-0.01}  &  -0.000  &  0.02^{+0.01}_{-0.01}  &  0.83^{+0.13}_{-0.34} &0.10^{+0.10}_{-0.09}  &  0.08^{+0.07}_{-0.07}  &  -0.013  &  0.08^{+0.06}_{-0.03}  &  0.60^{+0.28}_{-0.54}\\
\hline
  $n_{\rm mid}$ &$\eta_M$ &0.95^{+0.11}_{-0.10}  &  -0.75^{+0.11}_{-0.11}  &  -0.010  &  0.16^{+0.17}_{-0.09}  &  -0.99^{+0.06}_{-0.01} &-0.69^{+0.11}_{-0.12}  &  -0.12^{+0.14}_{-0.11}  &  -0.025  &  0.17^{+0.18}_{-0.09}  &  -0.64^{+0.70}_{-0.30}\\
                &$\eta_p$ &-0.76^{+0.10}_{-0.10}  &  -0.49^{+0.11}_{-0.11}  &  -0.008  &  0.14^{+0.15}_{-0.08}  &  -0.98^{+0.10}_{-0.02} &-1.04^{+0.10}_{-0.11}  &  0.02^{+0.11}_{-0.11}  &  -0.010  &  0.13^{+0.14}_{-0.07}  &  0.17^{+0.65}_{-0.81}\\
                &$\eta_E$ &-1.96^{+0.10}_{-0.10}  &  -0.20^{+0.11}_{-0.11}  &  -0.008  &  0.15^{+0.14}_{-0.08}  &  -0.87^{+0.38}_{-0.11} &-1.01^{+0.12}_{-0.12}  &  0.23^{+0.13}_{-0.13}  &  -0.015  &  0.16^{+0.15}_{-0.08}  &  0.87^{+0.11}_{-0.42}\\
                &$\eta_{Z}^{\rm SN}$ &-0.80^{+0.17}_{-0.19}  &  -0.03^{+0.21}_{-0.19}  &  -0.178  &  0.18^{+0.24}_{-0.10}  &  -0.14^{+0.77}_{-0.67} &-0.88^{+0.11}_{-0.11}  &  0.19^{+0.11}_{-0.12}  &  -0.011  &  0.13^{+0.14}_{-0.07}  &  0.87^{+0.11}_{-0.44}\\
                &$\overline{v}_{\rm out}$ &1.27^{+0.06}_{-0.06}  &  0.38^{+0.07}_{-0.07}  &  -0.004  &  0.12^{+0.10}_{-0.05}  &  0.97^{+0.03}_{-0.10} &2.35^{+0.04}_{-0.04}  &  0.28^{+0.05}_{-0.05}  &  -0.002  &  0.06^{+0.07}_{-0.03}  &  0.98^{+0.01}_{-0.07}\\
                &$\overline{v}_{\mathcal{B}}$ &1.55^{+0.05}_{-0.05}  &  0.28^{+0.06}_{-0.05}  &  -0.002  &  0.10^{+0.08}_{-0.04}  &  0.96^{+0.03}_{-0.11} &2.80^{+0.06}_{-0.06}  &  0.18^{+0.06}_{-0.07}  &  -0.003  &  0.12^{+0.09}_{-0.05}  &  0.88^{+0.09}_{-0.28}\\
                &$\zeta$ &0.01^{+0.01}_{-0.01}  &  0.02^{+0.01}_{-0.01}  &  -0.000  &  0.01^{+0.01}_{-0.01}  &  0.89^{+0.09}_{-0.26} &0.18^{+0.04}_{-0.03}  &  0.05^{+0.04}_{-0.04}  &  -0.001  &  0.07^{+0.05}_{-0.03}  &  0.66^{+0.25}_{-0.51}\\
\hline
$P_{\rm mid}/k_B$ &$\eta_M$ &3.16^{+0.40}_{-0.41}  &  -0.51^{+0.08}_{-0.08}  &  -0.059  &  0.18^{+0.17}_{-0.09}  &  -0.98^{+0.07}_{-0.01} &-0.35^{+0.41}_{-0.45}  &  -0.08^{+0.09}_{-0.07}  &  -0.059  &  0.16^{+0.16}_{-0.09}  &  -0.66^{+0.70}_{-0.29}\\
                &$\eta_p$ &0.69^{+0.39}_{-0.39}  &  -0.34^{+0.08}_{-0.08}  &  -0.051  &  0.16^{+0.16}_{-0.09}  &  -0.97^{+0.12}_{-0.03} &-1.11^{+0.36}_{-0.38}  &  0.02^{+0.07}_{-0.07}  &  -0.038  &  0.12^{+0.12}_{-0.07}  &  0.21^{+0.61}_{-0.82}\\
                &$\eta_E$ &-1.39^{+0.37}_{-0.38}  &  -0.13^{+0.07}_{-0.07}  &  -0.046  &  0.15^{+0.14}_{-0.08}  &  -0.85^{+0.41}_{-0.12} &-1.68^{+0.47}_{-0.44}  &  0.16^{+0.08}_{-0.09}  &  -0.061  &  0.15^{+0.15}_{-0.08}  &  0.87^{+0.11}_{-0.41}\\
                &$\eta_{Z}^{\rm SN}$ &-0.75^{+0.69}_{-0.71}  &  -0.02^{+0.13}_{-0.13}  &  -1.878  &  0.19^{+0.25}_{-0.11}  &  -0.11^{+0.74}_{-0.67} &-1.47^{+0.41}_{-0.38}  &  0.14^{+0.07}_{-0.08}  &  -0.046  &  0.13^{+0.13}_{-0.07}  &  0.88^{+0.11}_{-0.42}\\
                &$\overline{v}_{\rm out}$ &0.14^{+0.15}_{-0.17}  &  0.26^{+0.03}_{-0.03}  &  -0.008  &  0.07^{+0.06}_{-0.04}  &  0.99^{+0.01}_{-0.04} &1.54^{+0.14}_{-0.14}  &  0.19^{+0.03}_{-0.03}  &  -0.006  &  0.05^{+0.05}_{-0.03}  &  0.99^{+0.01}_{-0.05}\\
                &$\overline{v}_{\mathcal{B}}$ &0.71^{+0.14}_{-0.15}  &  0.19^{+0.03}_{-0.03}  &  -0.008  &  0.08^{+0.06}_{-0.03}  &  0.98^{+0.02}_{-0.07} &2.26^{+0.22}_{-0.20}  &  0.12^{+0.04}_{-0.04}  &  -0.015  &  0.11^{+0.08}_{-0.04}  &  0.90^{+0.08}_{-0.25}\\
                &$\zeta$ &-0.04^{+0.02}_{-0.02}  &  0.01^{+0.00}_{-0.00}  &  -0.000  &  0.01^{+0.01}_{-0.01}  &  0.90^{+0.08}_{-0.24} &0.01^{+0.15}_{-0.14}  &  0.04^{+0.03}_{-0.03}  &  -0.009  &  0.07^{+0.05}_{-0.03}  &  0.67^{+0.24}_{-0.49}\\
\hline
$\mathcal{W}/k_B$ &$\eta_M$ &3.23^{+0.39}_{-0.42}  &  -0.54^{+0.08}_{-0.08}  &  -0.064  &  0.17^{+0.16}_{-0.09}  &  -0.99^{+0.06}_{-0.01} &-0.31^{+0.40}_{-0.48}  &  -0.09^{+0.09}_{-0.08}  &  -0.064  &  0.17^{+0.16}_{-0.09}  &  -0.66^{+0.69}_{-0.29}\\
                &$\eta_p$ &0.73^{+0.38}_{-0.38}  &  -0.35^{+0.07}_{-0.08}  &  -0.064  &  0.16^{+0.16}_{-0.08}  &  -0.97^{+0.12}_{-0.03} &-1.12^{+0.39}_{-0.40}  &  0.02^{+0.08}_{-0.07}  &  -0.048  &  0.12^{+0.13}_{-0.07}  &  0.24^{+0.60}_{-0.88}\\
                &$\eta_E$ &-1.35^{+0.40}_{-0.39}  &  -0.14^{+0.08}_{-0.08}  &  -0.050  &  0.16^{+0.15}_{-0.08}  &  -0.84^{+0.40}_{-0.13} &-1.73^{+0.49}_{-0.47}  &  0.17^{+0.09}_{-0.09}  &  -0.208  &  0.15^{+0.15}_{-0.08}  &  0.88^{+0.10}_{-0.41}\\
                &$\eta_{Z}^{\rm SN}$ &-0.74^{+0.70}_{-0.73}  &  -0.02^{+0.14}_{-0.13}  &  -0.435  &  0.19^{+0.24}_{-0.11}  &  -0.11^{+0.76}_{-0.66} &-1.46^{+0.42}_{-0.41}  &  0.14^{+0.08}_{-0.08}  &  -0.048  &  0.13^{+0.13}_{-0.07}  &  0.88^{+0.10}_{-0.42}\\
                &$\overline{v}_{\rm out}$ &0.10^{+0.17}_{-0.17}  &  0.27^{+0.03}_{-0.03}  &  -0.013  &  0.08^{+0.06}_{-0.03}  &  0.99^{+0.01}_{-0.04} &1.52^{+0.14}_{-0.15}  &  0.19^{+0.03}_{-0.03}  &  -0.007  &  0.06^{+0.06}_{-0.03}  &  0.99^{+0.01}_{-0.05}\\
                &$\overline{v}_{\mathcal{B}}$ &0.68^{+0.15}_{-0.15}  &  0.20^{+0.03}_{-0.03}  &  -0.010  &  0.08^{+0.06}_{-0.03}  &  0.97^{+0.02}_{-0.07} &2.25^{+0.23}_{-0.21}  &  0.13^{+0.04}_{-0.05}  &  -0.017  &  0.11^{+0.08}_{-0.04}  &  0.89^{+0.08}_{-0.26}\\
                &$\zeta$ &-0.04^{+0.02}_{-0.02}  &  0.01^{+0.00}_{-0.00}  &  -0.000  &  0.01^{+0.01}_{-0.01}  &  0.90^{+0.08}_{-0.24} &0.01^{+0.15}_{-0.13}  &  0.04^{+0.03}_{-0.03}  &  -0.008  &  0.07^{+0.05}_{-0.03}  &  0.66^{+0.24}_{-0.50}\\
\hline
$t_{\rm dep,40}$ &$\eta_M$ &-1.44^{+0.32}_{-0.29}  &  0.70^{+0.09}_{-0.10}  &  -0.085  &  0.16^{+0.17}_{-0.09}  &  0.99^{+0.01}_{-0.06} &-1.01^{+0.34}_{-0.31}  &  0.09^{+0.11}_{-0.12}  &  -0.069  &  0.19^{+0.17}_{-0.09}  &  0.51^{+0.38}_{-0.68}\\
                &$\eta_p$ &-2.32^{+0.33}_{-0.32}  &  0.45^{+0.10}_{-0.10}  &  -0.098  &  0.17^{+0.18}_{-0.09}  &  0.96^{+0.03}_{-0.14} &-0.95^{+0.26}_{-0.25}  &  -0.03^{+0.09}_{-0.09}  &  -0.040  &  0.12^{+0.13}_{-0.06}  &  -0.30^{+0.81}_{-0.55}\\
                &$\eta_E$ &-2.61^{+0.29}_{-0.29}  &  0.19^{+0.09}_{-0.09}  &  -0.048  &  0.15^{+0.14}_{-0.08}  &  0.86^{+0.11}_{-0.39} &-0.27^{+0.30}_{-0.32}  &  -0.22^{+0.11}_{-0.10}  &  -0.059  &  0.14^{+0.15}_{-0.07}  &  -0.89^{+0.37}_{-0.10}\\
                &$\eta_{Z}^{\rm SN}$ &-0.87^{+0.48}_{-0.47}  &  0.02^{+0.17}_{-0.18}  &  -0.550  &  0.19^{+0.24}_{-0.11}  &  0.11^{+0.68}_{-0.76} &-0.25^{+0.26}_{-0.27}  &  -0.18^{+0.10}_{-0.09}  &  -0.037  &  0.12^{+0.13}_{-0.07}  &  -0.90^{+0.37}_{-0.09}\\
                &$\overline{v}_{\rm out}$ &2.46^{+0.11}_{-0.11}  &  -0.34^{+0.03}_{-0.04}  &  -0.006  &  0.06^{+0.05}_{-0.03}  &  -0.99^{+0.03}_{-0.01} &3.22^{+0.09}_{-0.10}  &  -0.25^{+0.03}_{-0.03}  &  -0.005  &  0.04^{+0.05}_{-0.02}  &  -0.99^{+0.04}_{-0.01}\\
                &$\overline{v}_{\mathcal{B}}$ &2.44^{+0.11}_{-0.11}  &  -0.26^{+0.03}_{-0.03}  &  -0.007  &  0.07^{+0.05}_{-0.03}  &  -0.98^{+0.06}_{-0.01} &3.37^{+0.16}_{-0.18}  &  -0.17^{+0.06}_{-0.05}  &  -0.017  &  0.11^{+0.08}_{-0.04}  &  -0.89^{+0.25}_{-0.09}\\
                &$\zeta$ &0.07^{+0.02}_{-0.02}  &  -0.02^{+0.01}_{-0.01}  &  -0.000  &  0.01^{+0.01}_{-0.01}  &  -0.91^{+0.23}_{-0.07} &0.36^{+0.11}_{-0.12}  &  -0.05^{+0.04}_{-0.04}  &  -0.008  &  0.07^{+0.06}_{-0.03}  &  -0.66^{+0.50}_{-0.24}\\
\hline

\enddata
\tablecomments{The data at different heights along with python scripts for fitting are available at \href{http://doi.org/10.5281/zenodo.3872049}{doi:10.5281/zenodo.3872049}. Linear regression results for $\log X$ and $\log Y$. We exclude R16 for fitting of $\loading{Z}{cool}^{\rm SN}$. The values given for the intercept $\alpha$, slope $\beta$, intrinsic scatter $\sigma_{\rm int}$, and Pearson correlation coefficient $\rho$ are the median and interval containing 68\% of the estimates over the posterior distributions. Covariance of $\alpha$ and $\beta$ is given in Columns (5) and (10).}
\end{deluxetable*}

\subsection{Loading Factors with SFRs}\label{sec:loading-sfr}

\begin{figure*}
    \centering
    \includegraphics[width=\textwidth]{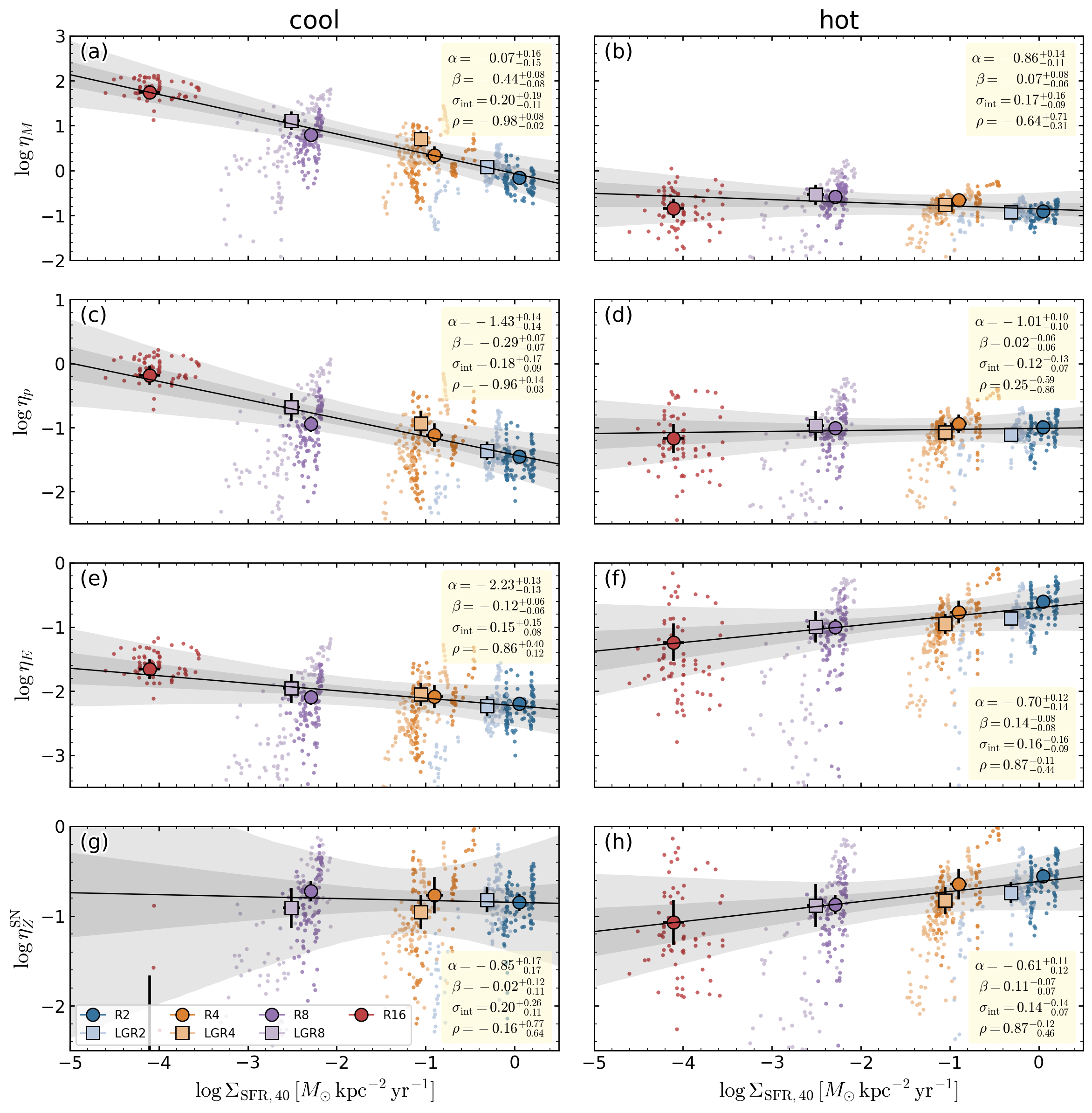}
    \caption{Scalings of loading factors with SFR surface density at $|z|=H$. Figures at different heights are available at \url{https://changgoo.github.io/tigress-wind-figureset/figureset.html}. The mass, kinetic $z$-momentum, energy, and SN-origin metal loading factors are shown from top to bottom. We exclude R16 for fitting of $\loading{Z}{cool}^{\rm SN}$. Left and right columns are for cool and hot outflows, respectively.
    Symbols with errorbars denote mean and standard deviation over $0.5<t/\torb<1.5$, while corresponding colored points represent full time evolution sampled with an interval of $0.01\torb$. The results of linear regression using {\tt linmix} are shown as black solid lines (median) and grey bands (68\% and 95\% confidence intervals).  For each panel the key gives the values of the median and interval containing 68\% of the estimates for the intercept ($\alpha$), slope ($\beta$), intrinsic scatter ($\sigma_{\rm int}$), and Pearson correlation coefficient ($\rho$).  While all of the loading factors for the hot wind are nearly independent of $\Ssfr$, the mass loading of the cool outflow decreases at larger $\Ssfr$.}
    \label{fig:scaling-loading}
\end{figure*}

\autoref{fig:scaling-loading} shows scaling relations of mass, momentum, energy, and SN-origin metal loading factors measured at $H$ as a function of $\Ssfr$ for cool (left) and hot (right) outflows. We present the mean and error of measured quantities from bootstrapping as symbols and errorbars, which we use for the fitting, along with small points denoting time evolution of each model over $0.5<t/\torb<1.5$ with sampling interval of $0.01\torb$. 
In each panel, the solid line and shaded regions denote the median and 68\% and 95\% confidence intervals of model posterior distributions; the median and 16th/84th percentile values of the intercept $\alpha$, slope $\beta$, intrinsic scatter $\sigma_{\rm int}$, and Pearson correlation coefficient $\rho$ are shown in the box of each panel (also shown in \autoref{tbl:scaling}).

The hot mass loading factors are nearly flat with $\loading{M}{hot}\sim 0.1-0.2$ over a wide range in $\Ssfr$.
This level of the hot gas loading is consistent with what has been reported in other simulations \citep[e.g.,][]{2017ApJ...841..101L,2020ApJ...890L..30L} and with the expectation from superbubble breakout after shell formation \citep{2017ApJ...834...25K}. The hot gas energy loading factors $\loading{E}{hot}$  show a weakly increasing trend from 0.06-0.25 with $\Ssfr$, and is larger than $\loading{E}{cool}$ by more than an order of magnitude. In general, the models with higher SFR have greater temporal and spatial correlation of star formation (and SNe), providing a potential explanation for the enhancement of energy loading factor. However, the effect is
less dramatic than suggested by previous idealized numerical simulations \citep{2018MNRAS.481.3325F}. This is partly because we are reporting time-averaged loading factors (averaging over both high and low states) and partly because our self-consistent simulations always have fountain flow gas at high altitudes, with which hot gas must interact. In addition, the larger horizontal velocity dispersions at higher SFR tend to close off chimneys. Thus, even though there is a burst of star formation that creates a superbubble, the energy  loading  is reduced below what it would be if the superbubble were to vent into an almost-vacuum region.

The cool mass loading factors $\loading{M}{cool}$ decrease steeply with $\Ssfr$, with values ranging from 100 to 1. However, it is noteworthy that much of the cool gas is at low velocities ($\overline{v}_{\rm out, cool}\sim 10-100\kms$; see \autoref{tbl:velmetal}), as evidenced by the low energy loading factor (see also KO18). 
Therefore, the high mass loading factor of the cool phase at $|z|=H$ shown here does not immediately imply heavily mass-loaded winds at large distances. Indeed, the mass loading factor in model R4 drops by a factor of 3 from $|z|=H$ to $|z|=2H$, and keep decreasing as a function of $|z|$ (see \autoref{fig:R4loading} and also KO18). 
In our simulation suite, most of the mass in cool outflows cannot reach the vertical boundary of the simulation box, and falls back toward the midplane (see \autoref{fig:massflux}). It is still possible to anticipate a higher mass loading factor at large distances (e.g., 0.1-1 virial radius) in dwarf galaxies that have a shallower global gravitational potential, as reported in cosmological zoom-in \citep{2015MNRAS.450..504M} or isolated galaxy simulations \citep{2019MNRAS.483.3363H}. We refrain from extrapolating our results to that regime since those outflows may consist of both directly launched cool outflows, and swept-up CGM driven out by energy delivered by hot outflows. 

The momentum loading factors in the cool gas at  $|z|=H$ decrease from $\loading{p}{cool} \sim 0.7$ to $0.04$ with increasing $\Ssfr$.
When combined with the nearly constant $\loading{p}{hot}\sim0.1$ and decreasing trend of $\eta_p$ as a function of $z$ in general (see \autoref{fig:R4loading}), this implies that most of the vertical momentum injection from SN feedback goes into the bulk of the ISM in the disk, rather than escaping from galaxies. Further analysis of the momentum injection to the  ISM from SN feedback, quantifying its contribution to supporting the gravitational weight of the disk and regulating SFRs, will be given in a separate paper (Ostriker \& Kim in prep.; see also \citealt{2011ApJ...743...25K,2013ApJ...776....1K,2015ApJ...815...67K}).
Consistent with our previous result in KO18 (see also \citealt{2020ApJ...890L..30L}), we find that the energy loading factor of the cool gas is significantly lower than in the hot gas, $\loading{E}{cool}\sim 0.02-0.005$, decreasing with increasing $\Ssfr$.   

A key conclusion from our simulation suite is that energy is carried by hot outflows while mass is carried by cool outflows.  \textit{Including two distinct wind components is therefore crucial in any physically-motivated wind model}. 

\subsection{Dependence of Loading Factors on Galactic Properties}\label{sec:loading-galprop}

\begin{figure*}
    \centering
    \includegraphics[width=\textwidth]{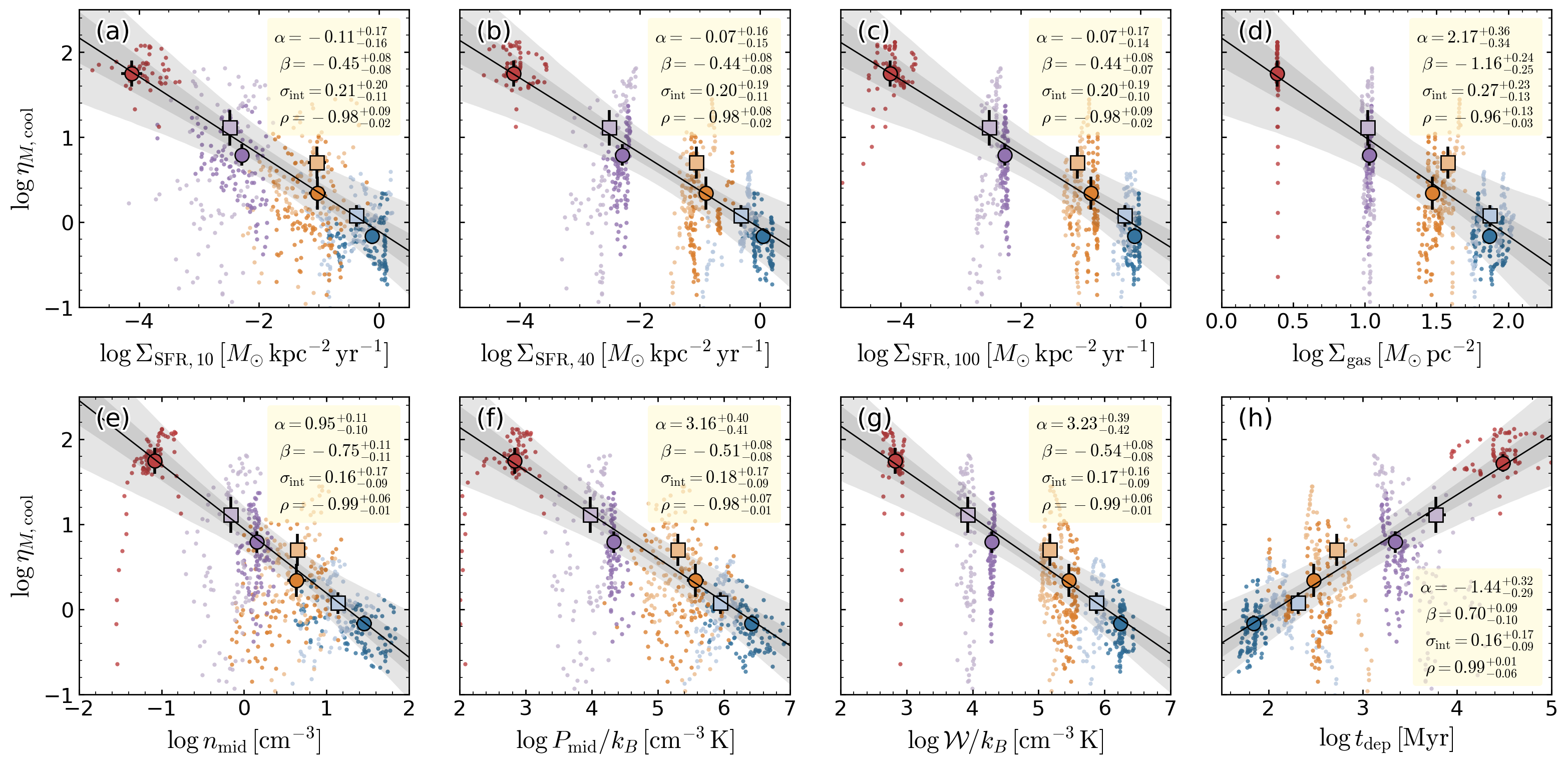}
    \caption{Scaling relations of cool mass loading factor $\loading{M}{cool}$ with galactic properties. 
    The mass flux is measured at $|z|=H$. Figures at different heights are available at \url{https://changgoo.github.io/tigress-wind-figureset/figureset.html}. The abscissas are 
    {\bf (a)-(c)} SFR surface density with $\tbin=10\Myr$, 40~Myr, and 100~Myr,
    {\bf (d)} gas surface density, 
    {\bf (e)} midplane gas volume density, 
    {\bf (f)} midplane total pressure, 
    {\bf (g)} total gas weight, and 
    {\bf (h)} gas depletion time.
    The simulation results and fitting results are presented as in \autoref{fig:scaling-loading}.}
    \label{fig:scaling-Mloading}
\end{figure*}

\begin{figure*}
    \centering
    \includegraphics[width=\textwidth]{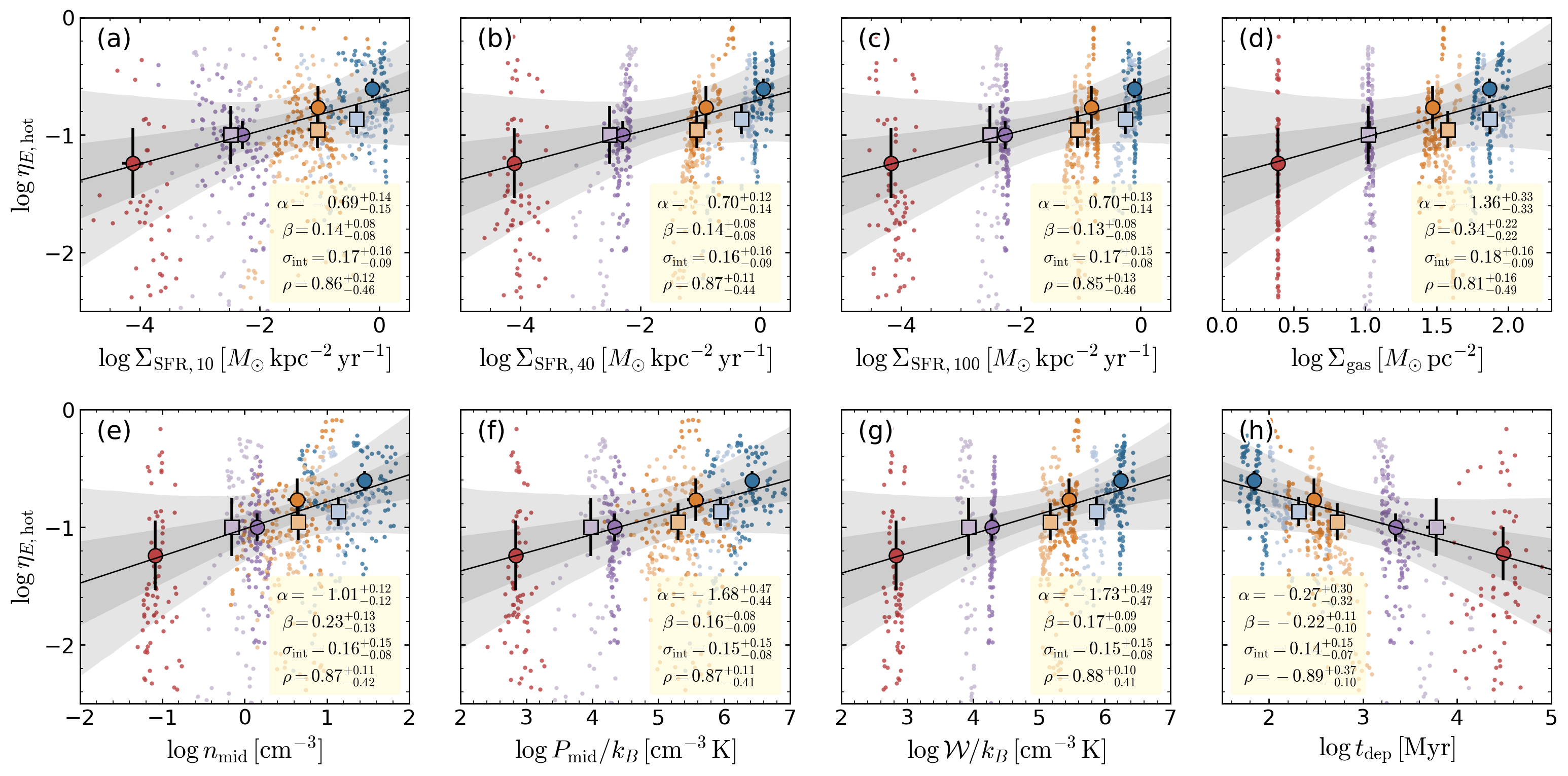}
    \caption{Scaling relations of hot energy loading factor $\loading{E}{hot}$ with galactic properties. 
    The energy flux is measured at $|z|=H$.
    Figures at different heights are available at \url{https://changgoo.github.io/tigress-wind-figureset/figureset.html}.
    The abscissas are 
    {\bf (a)-(c)} SFR surface density with $\tbin=10\Myr$, 40~Myr, and 100~Myr,
    {\bf (d)} gas surface density, 
    {\bf (e)} midplane gas volume density, 
    {\bf (f)} midplane total pressure, 
    {\bf (g)} total gas weight, and 
    {\bf (h)} gas depletion time.
    The simulation results and fitting results are presented as in \autoref{fig:scaling-loading}.}
    \label{fig:scaling-Eloading}
\end{figure*}

\autoref{fig:scaling-Mloading} and \autoref{fig:scaling-Eloading} show 
$\loading{M}{cool}$ and $\loading{E}{hot}$ as a function of different galactic conditions, including $\Ssfr$ with different $\tbin$, $\Sgas$, $n_{\rm mid}$, $\Pmid$, $\mathcal{W}$, and $\tdep$ with $\tbin=40\Myr$.
These parameters are chosen both because they represent important physical properties of the ISM, and because we expect outflows to correlate with them. These parameters are also quantities that can be estimated in large volume cosmological simulations or semi-analytic models, and therefore would be available as inputs to a subgrid model for wind launching. 
The level of $\Ssfr$ (which is not imposed, but obtained self-consistently in each simulation) sets the overall strength of feedback, while $\Sgas$ and $n_{\rm mid}$ characterize the conditions that affect superbubble propagation and breakout as well as transfer of momentum and energy to the bulk ISM. The values of $\Pmid$ and $\mathcal{W}$ are related to each other and to $\Ssfr$ through self-regulation.
These also reflect the vertical gravitational field, and therefore encode oscillation timescales that control fountain flows that limit gas escape.  The value of $\tdep = \Sgas/\Ssfr$ represents a local  evolutionary timescale.

In the first three panels, (a)-(c), of \autoref{fig:scaling-Mloading} and \autoref{fig:scaling-Eloading}, we compare scaling relations for three different choices of the averaging timescale for $\Ssfr$. These values, $\tbin=10$, 40, and 100~Myr, are rough proxies for different observational tracers. As expected, $\Ssfr$  traced by younger star clusters (e.g., $\Sigma_{\rm SFR,10}$) exhibits larger amplitude fluctuations. 
Nevertheless, the scaling relations from time-averaged points with all $\tbin$ choices are consistent with each other.

Panels (d)-(h), of \autoref{fig:scaling-Mloading} and \autoref{fig:scaling-Eloading} show scaling  relations of $\loading{M}{cool}$ and $\loading{E}{hot}$ with respect to $\Sgas$, $n_\mathrm{mid}$, $P_\mathrm{mid}$, $\mathcal{W}$, and $t_\mathrm{dep}$.  
Overall, we do not find particularly better correlation with one parameter over another (except $\Sgas$ has poorer  correlation). This is mainly because the quantities are not physically independent, but mutually connected through self-regulation. Fundamentally, the SFR surface density is self-regulated to provide the vertical pressure support through feedback that is required by the gas weight $\Sigma_{\rm SFR}\propto P_{\rm mid} \approx \mathcal{W}$  \citep{2010ApJ...721..975O,2011ApJ...731...41O,2011ApJ...743...25K} with near-linear relationships demonstrated in both simulations \citep[Ostriker \& Kim in prep.; see also][]{2012ApJ...754....2S,2013ApJ...776....1K,2015ApJ...815...67K} and observations \citep{2017ApJ...835..201H,2020arXiv200208964S}.  
All three of these quantities therefore are fundamental measures of the feedback strength, while including the local vertical gravity and gas density implicitly/explicitly. The midplane pressure and the weight are the same on average, but their instantaneous response to feedback is different; the midplane pressure responds more immediately and directly to SN rates ($\propto\Sigma_{\rm SFR,40}$) and FUV luminosity ($\propto\Sigma_{\rm SFR,10}$), while the weight varies only indirectly through the change of gas scale height (or velocity dispersion). The temporal variations in $\Pmid$ and $\mathcal{W}$ are thus similar to those in $\Ssfr$ with shorter and longer averages, respectively, so that the scatter in the points in panels (f) and (g) is more or less similar to panels (a)/(b) and (c), respectively.

The scaling with $\Sgas$ (panel (d)) is related to the scaling with gas weight.
If the external gravity dominates the weight, $\mathcal{W}\approx \Sigma_{\rm gas}\sigma_z(2 G \rho_{\rm sd})^{1/2}$, where $\rho_{\rm sd}\equiv \Sigma_*/(2z_*)+ \rho_{\rm dm}$ is the midplane density of stars and dark matter; however, a large range of $\Ssfr \propto \mathcal{W}$ is possible at a given $\Sgas$.
As a consequence, correlation with $\Sgas$ is indeed slightly worse than other parameters considered including $\Ssfr$, $\Pmid$, and $\mathcal{W}$, judging from the larger intrinsic scatter and the smaller Pearson correlation coefficient derived by linear regression. A wider parameter space survey and more experiments with extreme combinations between gas and gravity parameters would help to uncover which properties are the most fundamental in setting the loading factors.

The scaling with $n_{\rm mid}$ (panel (e)) is a measure of cooling in the ISM ($\dot\mathcal{E}_{\rm cool}\sim n^2\Lambda$) and is also related to the scaling with midplane pressure, since $\rho_{\rm mid}\sigma_{\rm z, eff}^2 = P_{\rm mid}$. Over more than three orders of magnitude variation in $P_{\rm mid}$ covered by our simulation suite, the effective vertical velocity dispersion $\sigma_{\rm z,eff}$ increases by no more than a factor 3 from the lowest to the highest $\Sigma_{\rm SFR}$ and $P_{\rm mid}$ (see \autoref{tbl:time}; see also \citealt{2009ApJ...704..137J}). Therefore, panels (e) and (f) are similar.

Finally, the scatter in the relation  for $\tdep$ (panel (h)) in each model simply arises from the scatter in $\Sigma_{\rm SFR}$ since variations in $\Sgas$ are (by design) narrow for each simulation. The gas depletion time is useful since it is not specific to geometry and can be defined either locally or globally (the area factor cancels out in the definition of $\tdep$). Although values of $\tdep$ in some of our simulations may be somewhat low (see Ostriker \& Kim in prep. for discussions on potential causes and missing physics), the scaling may still hold true.

\subsection{Outflow Velocity, Bernoulli Velocity, and Metal Enrichment}\label{sec:vscaling}

\begin{figure*}
    \centering
    \includegraphics[width=\textwidth]{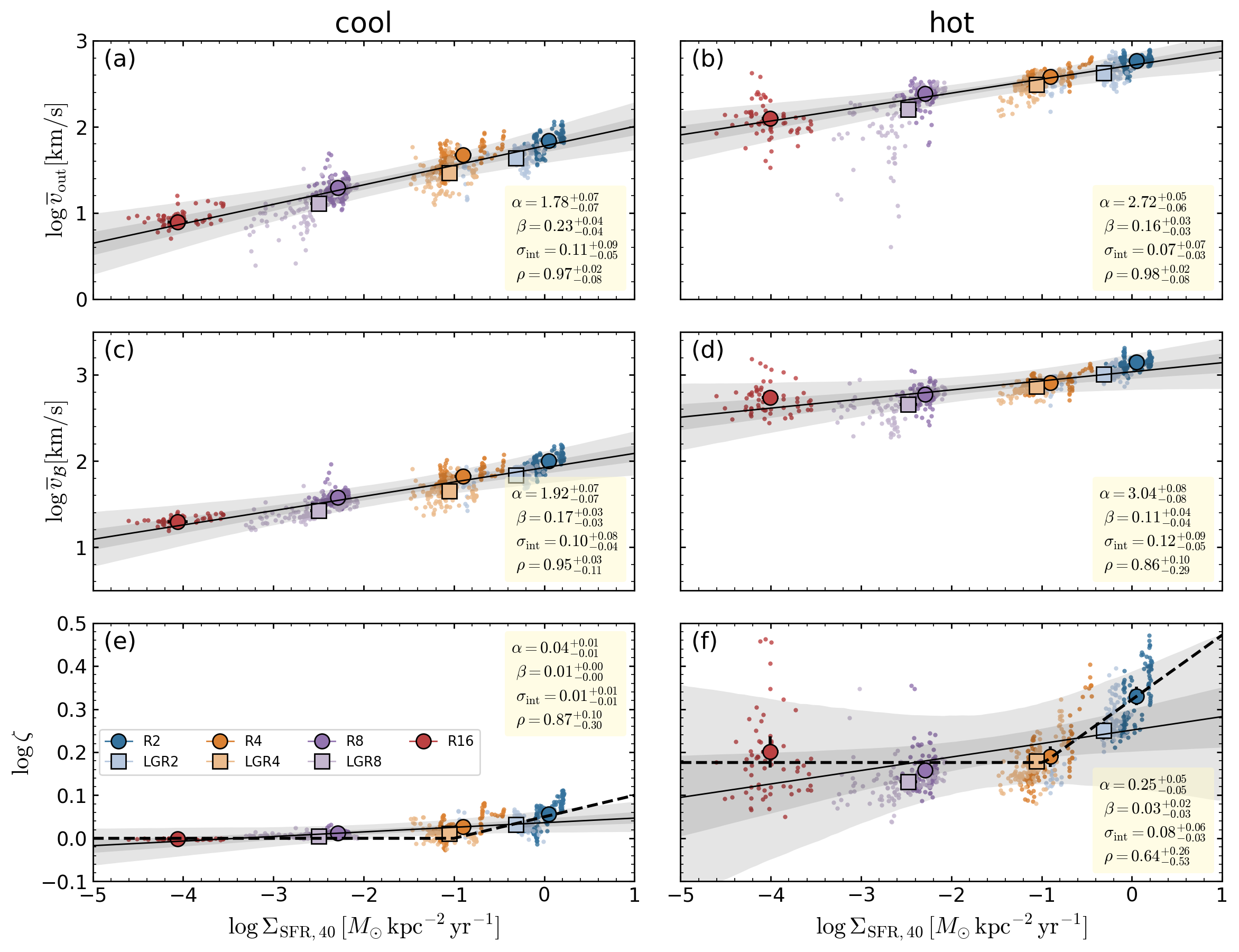}
    \caption{Scaling relations of characteristic velocities $v_\mathrm{out}$, $v_\mathcal{B}$, and metal enrichment factor $\zeta$, with SFR surface density $\Ssfr$ at $|z|=H$. Figures at different heights are available at \url{https://changgoo.github.io/tigress-wind-figureset/figureset.html}. Top row ((a) and (b)) is for outflow velocity (\autoref{eq:vout}), middle row ((c) and (d)) is for Bernoulli velocity (\autoref{eq:vB}), and bottom row ((e) and (f)) is for metal enrichment factor (\autoref{eq:yZ}). All quantities are measured at $|z|=H$ for cool (left column) and hot (right column) outflows separately.  The dashed lines in (e) and (f) denote simple models describing flattening behaviors of $\zeta$ at low $\Sigma_{\rm SFR}$ as in \autoref{eq:yZmodel_cool} and \autoref{eq:yZmodel}. The simulation results and fitting results are presented as in \autoref{fig:scaling-loading}.  Characteristic velocities of the hot component are an order of magnitude higher than those of the cool component, although the cool component velocities increase with $\Ssfr$ slightly more steeply. }
    \label{fig:scaling-v}
\end{figure*}

In \autoref{fig:scaling-v}, we present scaling relations for additional wind characteristics, including outflow velocity (\autoref{eq:vout}), Bernoulli velocity (\autoref{eq:vB}), and metal enrichment factor (\autoref{eq:yZ}) at $|z|=H$, as a function of $\Sigma_{\rm SFR}$ with $\tbin=40\Myr$. 

Both outflow velocity and Bernoulli velocity scale weakly with $\Ssfr$. The power-law exponent for the $\ovout$ vs. $\Ssfr$ relation in cool outflows is shallow $\sim0.2-0.25$ (depending on where $\ovout$ is measured). The power-law exponent for the  $\ovB$ vs. $\Ssfr$ relation in hot outflows is even shallower $\sim 0.07-0.1$ (depending on where $\ovB$ is measured). These weak scalings seem to be related to the characteristic shell velocity and specific energy of superbubble driven by SNe at the time of break out, which are largely insensitive to galactic properties (see \autoref{sec:dis-interpretation}).

The hot outflow clearly shows a metal enrichment factor larger than unity, while the cool outflow is only marginally enriched compared to the bulk of the ISM, and only for high SFR models. The metal enrichment factor for hot outflows seems to flatten out at low SFRs, so that simple linear regression is not a good description of the behavior. Given the limited number of models, we do not attempt to find a quantitative model from more sophisticated fitting. Instead, we provide simple models shown as the dashed lines in \autoref{fig:scaling-v}(e) and (f) given by
\begin{equation}
    \zeta_{{\rm cool}}=\left\{
    \begin{array}{lc}
        1.12(\Sigma_{\rm SFR}/\sfrunit)^{0.05} & \textrm{if}\quad \Sigma_{\rm SFR}>0.1\sfrunit\\
        1.0 & \textrm{otherwise}  
    \end{array}\right.\label{eq:yZmodel_cool}
\end{equation}
and
\begin{equation}
    \zeta_{{\rm hot}}=\left\{
    \begin{array}{lc}
        2.1(\Sigma_{\rm SFR}/\sfrunit)^{0.15} & \textrm{if}\quad \Sigma_{\rm SFR}>0.1\sfrunit\\
        1.5 & \textrm{otherwise}  
    \end{array}\right..\label{eq:yZmodel}
\end{equation}

\section{Discussion}\label{sec:discussion}

\subsection{Comparison with other simulations}\label{sec:dis-sims}

\begin{figure*}
    \centering
    \includegraphics[width=\textwidth]{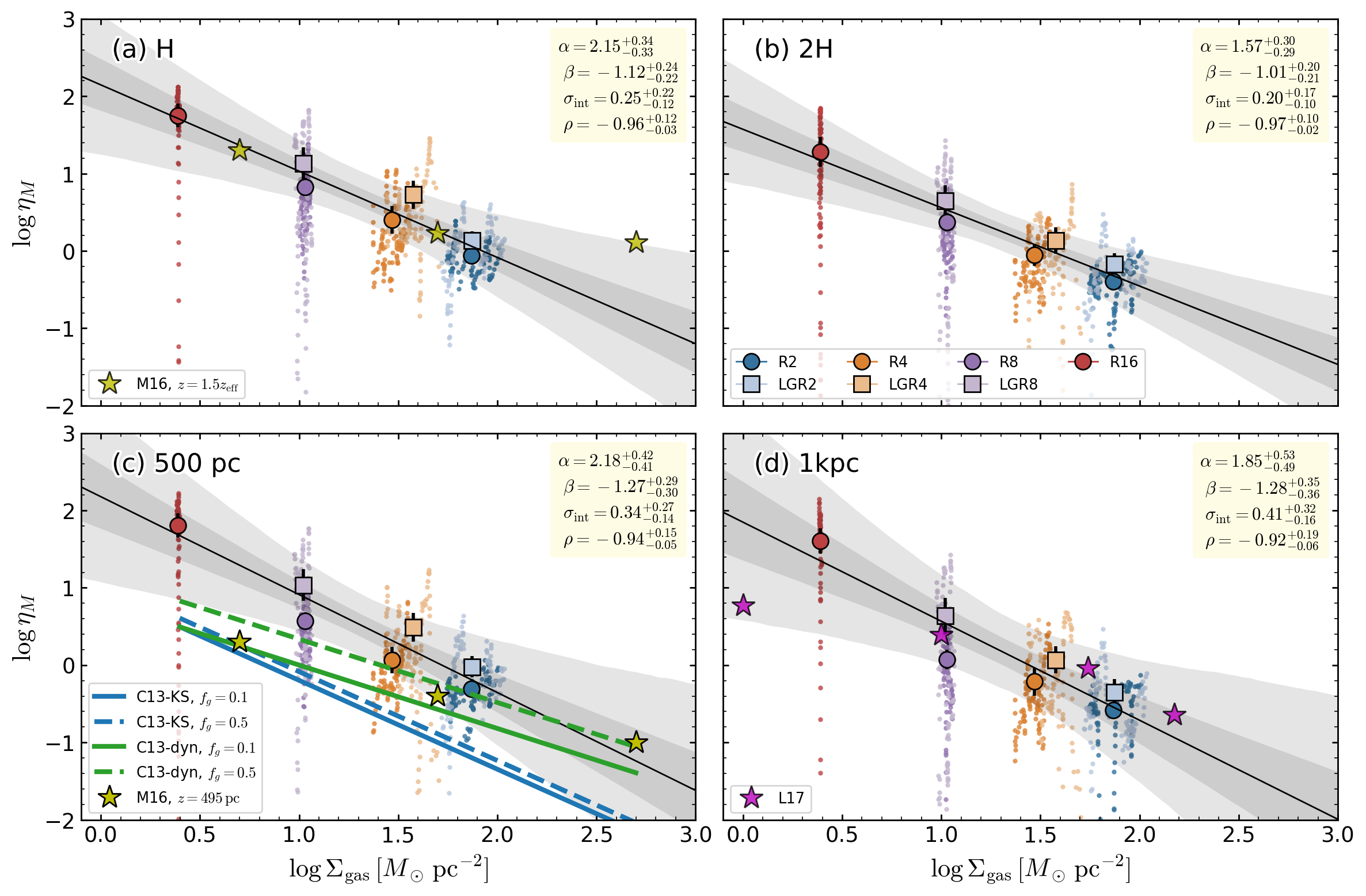}
    \caption{Mass loading factor of total outflowing gas as a function of 
    $\Sgas$, in comparison to other work.
    For our simulations, we show mass fluxes measured at $|z|=$ (a) $H$, (b) $2H$, (c) 500~pc, and (d) 1~kpc. The simulation results and fitting results for our models are presented as in \autoref{fig:scaling-loading}. Scaling relations reported in C13 are shown as blue (C13-KS; \autoref{eq:etaM-C13KS}) and green (C13-dyn; \autoref{eq:etaM-C13dyn}) with $f_g=0.1$ (solid) and 0.5 (dashed). Note that the extent of lines represents the parameter coverage of C13. Magenta stars denote fiducial local models from L17 and yellow stars denote local model FX of M16.
    }
    \label{fig:mloading-surf-comp}
\end{figure*}

\begin{figure*}
    \centering
    \includegraphics[width=\textwidth]{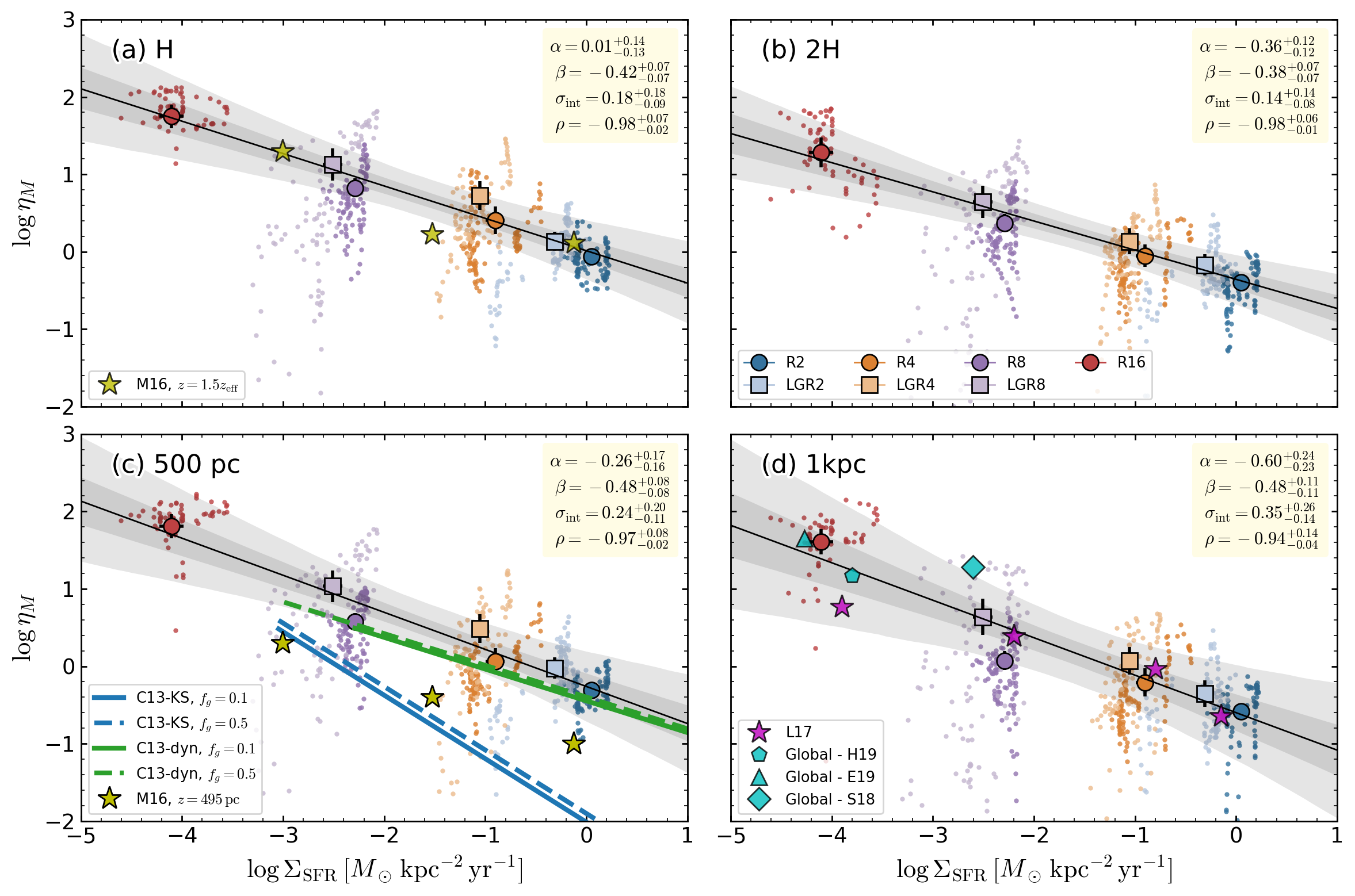}
    \caption{Mass loading factor of total outflowing gas as a function of 
     $\Ssfr$, in comparison to other work.
    For our simulations, mass fluxes are measured at $|z|=$ (a) $H$, (b) $2H$, (c) 500~pc, and (d) 1~kpc. The simulation results and fitting results for our models are presented as in \autoref{fig:scaling-loading}.
    Scaling relations reported in C13 are shown as blue (C13-KS; \autoref{eq:etaM-sfr-C13KS}) and green (C13-dyn; \autoref{eq:etaM-sfr-C13dyn}) with $f_g=0.1$ (solid) and 0.5 (dashed). Note that the extent of lines represent the parameter coverage of C13. Magenta stars denote fiducial local models from L17, yellow stars denote local model FX of M16, and cyan symbols show results from global dwarf galaxy models  of  H19, E19, and S18.
    }
    \label{fig:mloading-sfr-comp}
\end{figure*}

\begin{figure*}
    \centering
    \includegraphics[width=\textwidth]{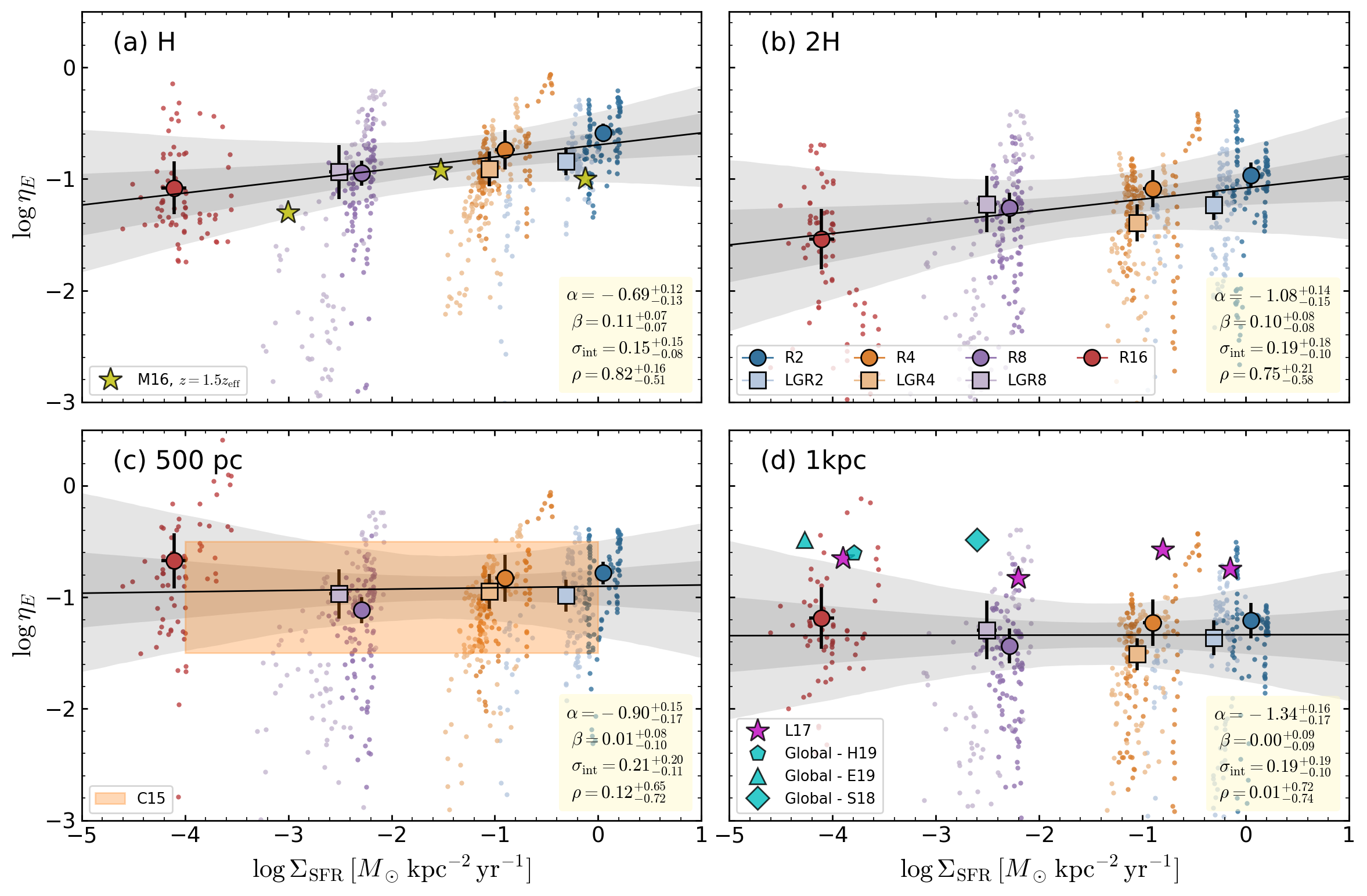}
    \caption{Energy loading factor of total outflowing gas as a function of 
     $\Ssfr$, in comparison to other work.
    For our simulations, energy fluxes are measured at $|z|=$ (a) $H$, (b) $2H$, (c) 500~pc, and (d) 1~kpc. The simulation results and fitting results for our models are presented as in \autoref{fig:scaling-loading}.
    The orange region covers the result of C15, $\eta_{E}=0.05-0.5$. 
    Magenta stars denote fiducial local models from L17, yellow stars denote local model FX of M16, and cyan symbols show results from global dwarf galaxy models  of  H19, E19, and S18.
    }
    \label{fig:Eloading-sfr-comp}
\end{figure*}

\begin{figure*}
    \centering
    \includegraphics[width=\textwidth]{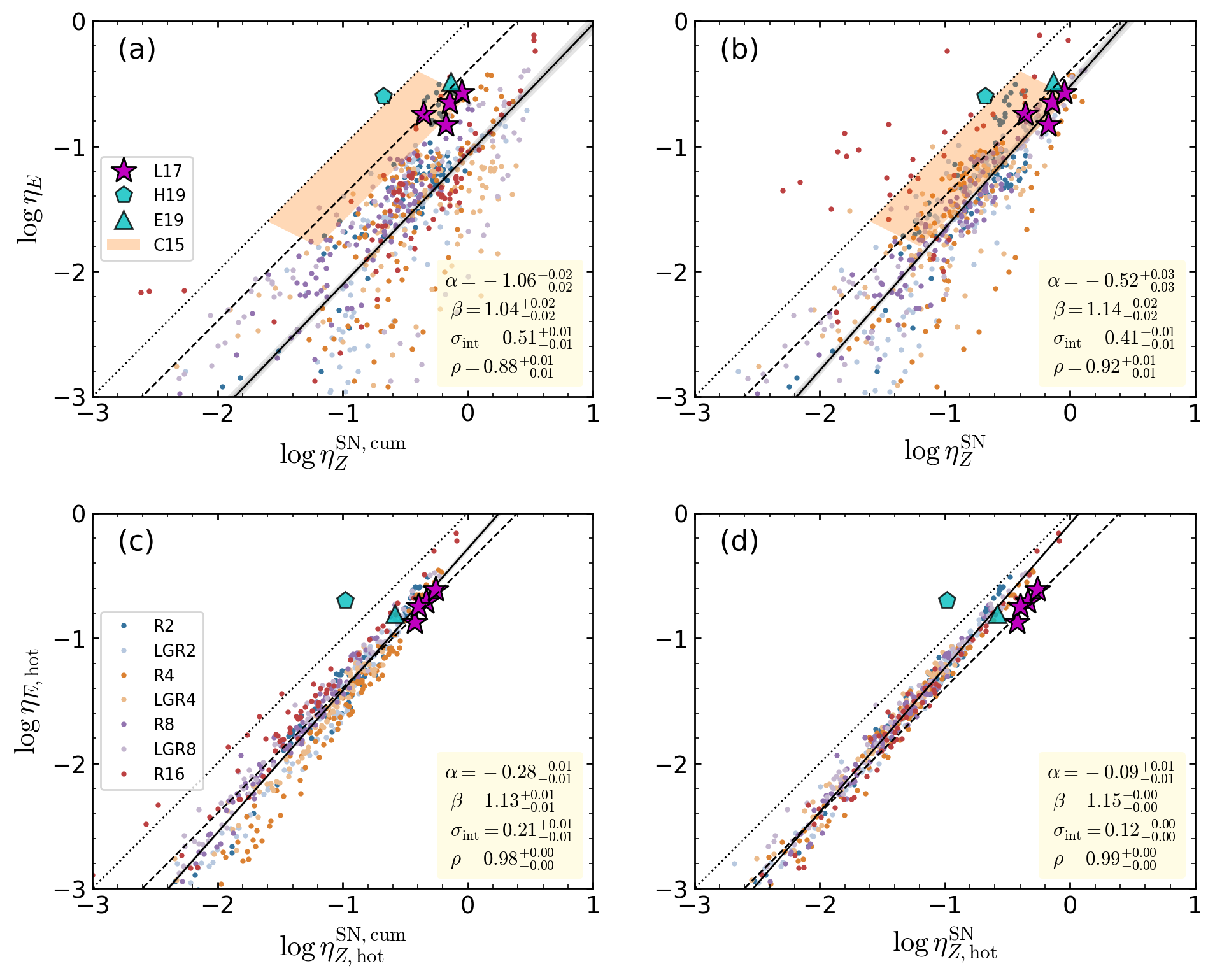}
    \caption{Correlation between energy loading factor and SN-origin metal loading factor at $|z|=1\kpc$, with comparison to other work. Figures at different heights are available at \url{https://changgoo.github.io/tigress-wind-figureset/figureset.html}. Left ((a) and (c)) and right ((b) and (d)) columns are for cumulative and instantaneous measures of SN-origin metal loading factor obtained by \autoref{eq:Zfluxsn} with initial and instantaneous ISM metallicity, respectively. Top ((a) and (b)) and bottom ((c) and (d)) rows are for total and hot outflows, respectively. The reference lines for $\eta_E=\eta_{Z}$ (dotted) and $\eta_E=0.4\eta_{Z}$ (dashed; C15) are also shown. The orange region covers the result of C15. The fitting results for our models are presented as in \autoref{fig:scaling-loading}.
    }
    \label{fig:ZvsE}
\end{figure*}

There have been a wide range of local simulations in vertically stratified disks including SN feedback  \citep[e.g.,][]{1999ApJ...514L..99K,2000MNRAS.315..479D,2006ApJ...653.1266J,2009ApJ...704..137J,2008A&A...486L..35G,2012ApJ...750..104H,2013MNRAS.432.1396G,2013MNRAS.430L..40G,2015MNRAS.454..238W,2017MNRAS.466.3293P,2018MNRAS.480.3511G,2014A&A...570A..81H,2017A&A...604A..70I,2018A&A...620A..21C}. 
Notably for present purposes,
quantitative analyses of SN-driven outflow properties have been provided in some papers \citep[e.g.,][]{2000MNRAS.315..479D,2013MNRAS.429.1922C,2015MNRAS.446.2125C,2016MNRAS.459.2311M,2016MNRAS.456.3432G,2017MNRAS.466.1903G,2017ApJ...841..101L,2018MNRAS.481.3325F,2020MNRAS.491.2088K};
including  a few where cosmic rays were part of the physics model \citep[e.g.,][]{2016ApJ...827L..29S,2016ApJ...816L..19G,2018MNRAS.479.3042G}. Still, among these studies, only a few have treated SN rates and positions self-consistently with explicit modeling of star formation from self-gravitating collapse, and these have been limited to short-term evolution and considered only a particular galactic condition \citep[e.g.,][]{2017MNRAS.466.1903G,2020MNRAS.491.2088K}.
The present work  is the first, to our knowledge, that considers a wide parameter space of local models with different galactic conditions, simulates star formation and feedback self-consistently at high resolution, and follows long-term evolution.  
As we shall show below, there are interesting similarities and differences between our results on wind scaling and those from  other local models.

Comparison with global simulations is of great interest, since these are not subject to some of the limitations of local models. Very recently, cosmological zoom-in simulations have begun to model the disk ISM and star formation without \emph{ad hoc} subgrid models for wind driving \citep[e.g.,][]{2014MNRAS.445..581H,2015MNRAS.454...83W,2018MNRAS.480..800H,2019MNRAS.489.4233M}. Quantitative analyses of outflows have been presented, aiming at understanding cosmic baryonic cycles \citep[e.g.,][]{2015MNRAS.454.2691M,2017MNRAS.468.4170M,2017MNRAS.470.4698A,2016ApJ...824...57C,2019MNRAS.485.2511T}. However, direct comparison of our results on outflow scaling relations with those  from zoom-in simulations is beyond the scope of this work because:
(1) most present scaling of loading factors to global properties, e.g., stellar/halo mass and circular velocity; (2) outflow properties are measured far from their galactic disk origin and interactions with CGM may have strongly altered the initial outflow properties; (3) outflow analyses mostly do not differentiate by thermal phase \citep[but, see][Pandya et al. in prep.]{2019MNRAS.485.2511T}; and (4) although the treatment of the ISM becomes more explicit, individual SN feedback is still unresolved, necessitating the adoption either of artificially delayed cooling \citep{2016ApJ...824...57C,2019MNRAS.485.2511T} or momentum feedback for most SNe \citep{2015MNRAS.454.2691M,2017MNRAS.468.4170M,2017MNRAS.470.4698A}. In particular, while the ``momentum'' feedback approach at mass resolution $\sim 10^3 -10^5 \Msun$ is able to control star formation, multiphase wind driving from SNe is not resolved.

Idealized global galaxy simulations are a good way to bridge the gap between local and cosmological zoom-in simulations.
Currently, only global simulations of dwarf galaxies employ sufficiently high resolution to resolve both star formation and feedback equivalently to this work \citep[e.g,][]{2019MNRAS.483.3363H,2019MNRAS.482.1304E}. 
Global simulations of more massive galaxies typically have resolution below what is required to resolve the adiabatic stage of SNR evolution (and therefore to follow hot gas creation), instead adopting ``momentum'' feedback for most SN events.  This is likely  why the mass loading of hot gas is lower than that in our simulations; e.g., in the galactic center models of \citet{2019MNRAS.490.4401A} that
have similar  conditions to our model R2 but mass resolution $2\times10^3\Msun$, the hot gas mass loading factor was $<0.1$, even though for warm gas the fountain-like behavior and mass loading were similar to what we found. In  contrast, for the SPH models (MW and Sbc, with mass resolution $500\Msun$) that have similar conditions  to our model R4, \citet{2012MNRAS.421.3522H} found similar mass-loading, but in high-velocity escaping rather than moderate-velocity fountain-like warm outflows. This may have been a consequence of the particular implementation of momentum feedback in their particle-based method, which has since been replaced \citep{2018MNRAS.477.1578H}.

Although an apples-to-apples comparison with existing simulations is not immediately possible, we discuss our results for scaling relations in comparison with work by \citet[][C13 and C15, respectively, hereafter]{2013MNRAS.429.1922C,2015MNRAS.446.2125C}, \citet[][M16 hereafter]{2016MNRAS.459.2311M}, and \citet[][L17 hereafter]{2017ApJ...841..101L}, in which a parameter survey is conducted and scaling relations are presented. We also include results from \citet[][S18, hereafter]{2018MNRAS.478..302S}, \citet[][H19 hereafter]{2019MNRAS.483.3363H}, and \citet[][E19 hereafter]{2019MNRAS.482.1304E}.
For an in-depth comparison with other simulations (including  \citealt{2016MNRAS.456.3432G,2017MNRAS.466.1903G}) for Solar neighborhood  conditions (model R8), we refer the reader to the discussion in KO18.

Before presenting the results of our comparisons, we begin by summarizing details of the C13, M16, and L17 local simulations as well as high resolution global simulations of dwarfs (S18, H19, and E19).
\begin{itemize}
\item {\bf C13} conducted non-self-gravitating, unmagnetized local simulations of galactic disks covering a wide range of gas surface density ($2.5<\Sigma_{\rm gas}/\Surf<500$) and gravitational field (parameterized by gas fraction $f_g$ as $g_z\propto f_g^{1/2}$, which varies from 0.01 to 1).\footnote{By comparing the measured total weight with Eq. (23) in C13 for the hydrostatic pressure, we deduce $f_g=$0.1-0.3 for the R models and 0.3-0.5 for the LGR models (smaller value for higher surface density).} SFRs (and hence SN rates) are prescribed and stay constant over the duration of simulations. They adopt two relations for SFRs: (1) Kennicutt-Schmidt relation (C13-KS, hereafter) \citep{1998ApJ...498..541K}, 
\begin{equation}\label{eq:KS-sfr}
\Sigma_{\rm SFR} = 2.5\times10^{-4}\sfrunit\rbrackets{\frac{\Sigma_{\rm gas}}{\Surf}}^{1.4},
\end{equation} 
and (2) dynamical time prescription (C13-dyn, hereafter), 
\begin{equation}\label{eq:dyn-sfr}
\Sigma_{\rm SFR} = 8.2\times10^{-5}\sfrunit f_g^{-1}\rbrackets{\frac{\Sigma_{\rm gas}}{\Surf}}^{2}.
\end{equation} 
Note that the mean values of $\Ssfr$ in our simulation suite are similar to those of C13-dyn rather than C13-KS, especially at higher surface densities. SNe are placed randomly in the horizontal plane with the scale height identical to the initial gas scale height. The simulation box is smaller (especially, shorter), $L_x\times L_y\times L_z = 200\pc\times200\pc\times1\kpc$. The fiducial cooling function depends only on density, $n^2\Lambda$ with constant $\Lambda$, and cuts off at $T=10^4\Kel$ (some models include a $T$-dependent cooling function). No radiative heating is included. 

\item {\bf M16} ran non-self-gravitating, unmagnetized local simulations of galactic disks covering $\Sigma_{\rm gas} = 5,$ 50, and 500$\Surf$. The same scaling for SFR surface density is adopted as C13-KS, but the normalization is about a factor of two lower. They have two different SN seeding schemes, but we only compare with their FX models, in which SNe are randomly seeded in space and time within the initial disk scale height. Without a self-consistent treatment of star formation to determine realistic clustering of star formation and hence SNe, their SC models, in which SNe are preferentially seeded near density peaks, results in artificially enhanced cooling of SNe \citep[see also][]{2016MNRAS.456.3432G}. At their typical resolution of a few pc, in their SC models SNe are mostly realized via momentum injection following the prescription of \citet{2015MNRAS.450..504M}, which substantially changes outflow properties (energy loading factor and multiphase structure most significantly). A cubical box with $L=1\kpc$ is adopted. The cooling function depends on temperature, but cuts off at $T=10^4\Kel$. No radiative heating is included.

\item {\bf L17} performed non-self-gravitating, unmagnetized local simulations covering $1<\Sigma_{\rm gas}/\Surf<150$. SN rates and distributions are essentially the same as in C13, but additional exploration with independently varying SN scale heights was conducted. The adopted $\Ssfr$ were a bit higher than C13-KS, closer to C13-dyn and our mean $\Ssfr$ (but lower at higher surface densities and higher at lower surface densities). The cooling function depends on temperature and extends to $T\sim 300\Kel$ \citep{1993ApJ...413..137R}, and a constant photoelectric heating rate is adopted (the effect of the photoelectric heating is explored). The horizontal extent of  the simulation box is about 1/3 of ours, but larger (scales with gas surface density) than that of C13.\footnote{Note that a larger box is required in our simulations since natural temporal and spatial correlations of SNe arising from self-consistent modeling of star formation rates in our simulation produce larger superbubbles that would fill up the entire volume near the midplane if the box were not large enough. A smaller horizontal domain size makes overall evolution burstier and more synchronized and results in long-lasting imprints from an initial transient.}

\item {\bf S18, H19, and E19}: the high-resolution global dwarf simulations with which we compare have very low mass (S18: $M_{\rm vir}=10^{10}\Msun$ and $M_{\rm gas} = 1.8\times10^8\Msun$; H19: $M_{\rm vir}=10^{10}\Msun$ and $M_{\rm gas}=10^7\Msun$; and E19: $M_{\rm vir}=2.5\times10^{9}\Msun$ and $M_{\rm gas}=1.8\times10^6\Msun$). Unlike C13, M16, and L17, all of these simulations have self-consistent star formation and feedback. We note as a caveat in comparison with our results that $\Sgas$ and $\Ssfr$ can vary substantially within a global simulation, and even if one adopts a single value it will depend on the area. Here, we use the scale radius to define the area, but certainly outflows can emerge from locations beyond a scale radius. In the future, more direct comparison with homogeneous definitions would be desirable.

\end{itemize}

How and where outflow properties are measured matters a great deal, especially for mass loading (see \autoref{fig:R4loading}). C13 and C15 reported mass, energy, and metal loading factors of the total outflowing gas (without phase separation) by measuring the mean ejected mass, energy, and metals through the vertical boundaries $|z|=500\pc$. M16 reported the energy loading factor of total outflow gas measured at $1.5 z_{\rm eff}$, where $z_{\rm eff}$ is the initial scale height (with slightly different definition from \autoref{eq:H}), while they measure mass loading at both $1.5 z_{\rm eff}$ and $500\pc$. L17 reported mass, energy, and metal loading factors for total and hot phases separately by measuring the outflow fluxes averaged over space ($|z|=1-2.5\kpc$) and time (last 40\% of the simulation termination time). H19 and E19 measured outflow fluxes through spherical shells as a function of $r$ rather than $z$. For the purpose of our comparison here, we adopt the compilation of \citet{2020ApJ...890L..30L} at $r=1\kpc$. To make comparisons as fair as possible, we present plots for total outflow loading factors and match heights as closely as possible given the limitations of reported measurements.
We plot C13/C15 results in the panel for $|z|=500\pc$, M16 results in the panels for $|z|=H$ and $500\pc$, and L17, S18, H19, and E19 results in the panel for $|z|=1\kpc$.

\autoref{fig:mloading-surf-comp} plots $\eta_M$ vs $\Sigma_{\rm gas}$ measured at (a) $H$, (b) $2H$, (c) 500~pc, and (d) 1~kpc, in  comparison to the literature results. C13 reported two scaling relations between mass loading factor and gas surface density for two model series, C13-KS and C13-dyn:
\begin{eqnarray}
\eta_M^{\rm C13-KS} = 13\pm 10 \rbrackets{\frac{\Sigma_{\rm gas}}{\Surf}}^{-1.15\pm0.12}f_g^{0.16\pm0.14}
\label{eq:etaM-C13KS}\\
\eta_M^{\rm C13-dyn} = 20\pm 8 \rbrackets{\frac{\Sigma_{\rm gas}}{\Surf}}^{-0.82\pm0.07}f_g^{0.48\pm0.08}.
\label{eq:etaM-C13dyn}
\end{eqnarray}
In panel (c), we show these two scaling relations for the surface density range consistent with that used in C13.
In panel (c), the loading factors we have found are generally higher than in C13 and M16 at a given $\Sgas$. This is mainly because our self-regulated SFR surface densities are higher than their adopted values. 
From self-regulation, $\Sigma_{\rm SFR}\propto \Sigma_{\rm gas} g_z$ is expected, so that the vertical gravity should be taken into account in models assuming a prescribed SFR surface density.
Since the boxes in both C13 and M16 are shorter than ours, they generally adopted stronger vertical gravity to confine  the gas in the vertical domain.
However, their adopted $\Ssfr$ was not adjusted upward corresponding to the expectation from self-regulated star formation at higher $g_z$, and as a result their $\Ssfr$ values are lower than ours at a given $\Sgas$. In addition, stronger gravity would result in higher volume density at a given $\Sgas$, while there would be additional differences in volume filling factors of different gas phases due to their artificial cooling cutoff. 
The agreement of C13-dyn with  our results is better since the SFR prescription for this model set (implicitly) includes the effect of vertical gravity.
The overall better agreement of L17 with  our results shown in panel (d) is because the adopted $\Ssfr$ and vertical gravity are more consistent with our simulations (except at the lowest $\Sgas$).
We note that apparent better agreement in panel (a) with M16 is a coincidence, since $z_{\rm eff}$ is much smaller than the corresponding scale height in our models (see Column (5) in \autoref{tbl:time}). 

\autoref{fig:mloading-sfr-comp} plots the relation between $\eta_M$ and $\Sigma_{\rm SFR}$ in comparison to the literature results.
Using the C13 imposed relation between $\Sigma_{\rm gas}$ and $\Sigma_{\rm SFR}$ (\autoref{eq:KS-sfr} and \autoref{eq:dyn-sfr}), we can convert \autoref{eq:etaM-C13KS} and \autoref{eq:etaM-C13dyn} as a function of $\Sigma_{\rm SFR}$ to
\begin{eqnarray}
\eta_M^{\rm C13-KS} = 0.014 \rbrackets{\frac{\Sigma_{\rm SFR}}{\sfrunit}}^{-0.82}f_g^{0.16}
\label{eq:etaM-sfr-C13KS}\\
\eta_M^{\rm C13-dyn} = 0.42 \rbrackets{\frac{\Sigma_{\rm SFR}}{\sfrunit}}^{-0.41}f_g^{0.07}.
\label{eq:etaM-sfr-C13dyn}
\end{eqnarray}
In panel (c), we show these two scaling relations for the $\Ssfr$ range consistent with that used in C13. 
The difference in $\Ssfr$ ranges between the two C13 model series is clearly demonstrated.
Again, the C13-dyn results are in fairly good agreement with our results, while C13-KS gives substantially lower mass loading factors.
The M16 results in panel (c) also show much lower mass loading factors at a given $\Ssfr$ than our  results, as in C13-KS. 
This implies that $\Ssfr$ is not the only parameter that sets $\eta_M$, but the gravity and/or gas density both matter in setting $\eta_M$. 

At 1~kpc, our mass loading factors are again consistent with L17. Overall, we find that the slopes of our mass-loading relationships are similar at different heights, and these are in good agreement with the literature results  in cases where $\Sgas$, $g_z$, and $\Ssfr$ are consistent. In addition to local models, in  \autoref{fig:mloading-sfr-comp}~(d) we show total mass  loading from the global dwarf simulations of S18, H19, and E19.  These are all fairly consistent with  our results. The present agreement may imply that outflow loading factors are independent of global conditions. This is encouraging for the development of generalized cosmological subgrid wind launching models from local simulations, although it will be imperative to make further tests and comparisons in other regimes, including more extreme conditions.

We now turn to the energy loading factor. \autoref{fig:Eloading-sfr-comp} plots our relation between $\eta_E$ and $\Ssfr$ for all gas (dominated by the hot medium), in comparison with the literature results. Note that the energy loading factors from C13-dyn are not available, but both C13 and our results suggest that the energy loading factor is insensitive to $\Sigma_{\rm gas}$ and $\Sigma_{\rm SFR}$. $\eta_E\sim$ 0.05--0.5 encloses the result reported in C15, which also envelopes our results in panel (c) quite well. L17 obtained a rather narrower range between 0.1-0.3, again without strong dependence on $\Sigma_{\rm SFR}$. \autoref{fig:scaling-Eloading} also shows weak dependence of the hot energy loading factors as a function of all galactic parameters we consider. Interestingly, if the energy loading factor is measured at a fixed height, the dependence is even weaker: $\loading{E}{hot}\sim0.1$ at $|z|=500\pc$ and 0.05 at $|z|=1\kpc$. Overall, energy loading factors from our simulations are lower than the fiducial L17 results. The enhanced energy loading factors in L17 are the consequence of larger imposed SN scale heights ($\sim 150\pc$) in L17 compared to typical values in our simulations.  L17 showed that the energy loading factors increase as the scale height of SNe gets larger since SN explosions in the tenuous disk atmosphere more freely deliver injected energy to the extraplanar region without significant energy loss by cooling (see also the similar tests in Appendix B of KO18). In our simulations, the SN locations are determined by star formation, which in turn depend on the distribution of gas; with ``natural'' SN positioning with respect to the vertical gas profile, our lower energy loading factors are more consistent with superbubble breakout after shell formation \citep{2017ApJ...834...25K}.

The energy loading factor can in principle be increased by strongly correlated SNe, since early explosions create a low density cavity through which energy from subsequent events easily vents with minimal losses \citep{2018MNRAS.481.3325F}. Our simulations in fact have highly correlated SNe, with typical cluster particle masses in the range $10^{3}$--$10^{4}\Msun$ with maximum cluster mass up to $10^5$--$10^6\Msun$ (higher mass clusters for inner disk models). However, with long-term evolution and  self-consistent inflow/outflow, we find that the energy loading is much reduced compared to the idealized simulations of \citet{2018MNRAS.481.3325F}. This is because previous or neighboring events can both fill the atmosphere with fountain gas and can close off chimneys, both of which render energy venting more difficult (see \autoref{fig:slices} and \autoref{fig:massflux}). Global flow patterns driven by structures like spiral arms and/or bars may potentially reduce inflow/outflow interactions locally; if fountain flows originating in arm regions fall preferentially in low density, interarm regions, the energy loading factor may be enhanced in the arm region and reduced in the interarm region, while the global average stays similar. We are currently analyzing the outflow properties from local simulations with spiral arms utilizing the TIGRESS framework \citep{2020arXiv200605614K}. Higher energy loading factors in global dwarf simulations (as e.g. shown from H19, E19, and S18 in \autoref{fig:scaling-Eloading}) may also be related to the global geometry, but caution needs to be taken since the cooling rates in dwarf simulations are generally lower due to the lower metallicity.

In \autoref{fig:ZvsE}, we compare the correlation between our energy and SN-origin metal loading factors, in comparison with the literature results (see also \citealt{2020ApJ...890L..30L}). Since SNe drive outflows, the fluxes of energy and SN-origin metals have a common origin and are expected to correlate with each other, and C15 previously identified a tight correlation. Note that if we use total metal fluxes, this correlation gets weaker (almost disappears). In our simulations, even with $Z_{\rm SN}=10Z_{\rm ISM,0}$, more mass comes from the ISM than from SNe so the metal mass flux is dominated by the ISM-origin metals (see \autoref{fig:R4metalflux}). We consider both cumulative and instantaneous SN-origin metal fluxes. The former is for metals injected by SNe over the entire simulation duration and directly measurable from the metal tracer field employed in the simulation.
The latter is for metals injected by recent SN events and obtained from \autoref{eq:Zfluxsn} (see \autoref{sec:metal}). Note that although reported metal loading factors in other simulations technically correspond to cumulative SN-origin metals, their metal loadings can be interpreted as instantaneous ones since C13 and L17 run for a much shorter time than we do, and H19 and E19 consider low metallicity dwarfs.

\autoref{fig:ZvsE} plots energy loading factors as a function of cumulative ($\eta_{Z}^{{\rm SN,cum}}$; (a) and (c)) and instantaneous ($\eta_{Z}^{{\rm SN}}$; (b) and (d)) SN-origin metal loading factors measured at $|z|=1\kpc$. The top row ((a) and (b)) is for the total outflow and the bottom row ((c) and (d)) is for the hot outflow. The dotted line is $\eta_{E}=\eta_{Z}$ and the dashed line is $\eta_{E}=0.4\eta_{Z}$, as suggested by C15. In contrast to other scaling relations, we use all points from time evolution for fitting since temporal correlations within a model between the two loading factors are in fact physically meaningful; there is no temporal offset between two fluxes, and loading factors use the same denominator (up to a constant factor).

Without radiative energy loss, the energy loading factor would be equal to the (instantaneous, SN-origin) metal loading factor. As energy is lost by radiative cooling, $\eta_{E}<1$ and $\eta_E<\eta_{Z}$ (C15). In addition, due to ``recycled'' metals through fountain flows, the cumulative SN-origin metal flux is larger than the instantaneous one. However, energy in fountain flows is radiated away and not ``recycled,'' so that the ratio $\eta_E/\eta_Z^{\rm SN,cum}\sim 0.09$ is smaller than $\eta_E/\eta_{Z}^{\rm SN}\sim 0.3$, as shown in \autoref{fig:ZvsE}(a) and (b).\footnote{Given the approximate nature of the instantaneous ISM metallicity (see \autoref{sec:metal}), $\eta_Z^{\rm SN}$ can be erroneous if $\overline{Z}\approx \overline{Z}_{\rm ISM}$. This is most serious for R16 in \autoref{fig:ZvsE}(b) when cool outflows, which originate from the gas near the midplane used to define $\overline{Z}_{\rm ISM}$, dominate metal flux. Outliers for low $\eta_Z^{\rm SN}$ points are subject to the definition of $\overline{Z}_{\rm ISM}$ and should not be considered significant.} The relation reported in C15 for the total outflowing gas, $\eta_E=0.4\eta_{Z}$, is quite close to our result for the instantaneous metal flux measurement.

As expected, the correlation gets tighter when only hot outflows are considered ((c) and (d)) since cooling is minimal in hot outflows. The energy-to-metal loading factor ratio is also increased with the instantaneous metal loading factor. The slope in hot outflows is steeper than unity, implying less efficient cooling when there is more efficient loading of SN-origin metals in hot outflows. In other words, successful breakouts due to clustered SN events (indicated by high SN-origin metal loading factor) load SN-energy to outflows more efficiently \citep{2018MNRAS.481.3325F}. Using the fitting result in \autoref{fig:ZvsE}(d), 
\begin{equation}\label{eq:hot-EZ}
    \loading{E}{hot} =0.81{\eta_{Z, {\rm hot}}^{\rm SN}}^{1.15},
\end{equation}
and \autoref{eq:vB}, we obtain
\begin{eqnarray}\label{eq:hot-vB-fSN}
    \overline{v}_{\mathcal{B}, {\rm hot}} &=& \rbrackets{\frac{2E_{\rm SN}}{m_*}}^{1/2} \rbrackets{\frac{\loading{E}{hot}}{\loading{M}{hot}}}^{1/2}
    = 2.9\times10^3\kms
    {f_{M, {\rm hot}}^{\rm SN}}^{0.58} \loading{M}{hot}^{0.08}.
\end{eqnarray}
This says that the specific energy in hot outflows is most sensitive to the fraction of genuine SN material in outflows, 
\begin{equation}\label{eq:fMSNhot}
    f_{M, {\rm hot}}^{\rm SN} \equiv \frac{ \overline{Z}_{\rm hot}-\overline{Z}_{\rm ISM}}{Z_{\rm SN}-\overline{Z}_{\rm ISM}},
\end{equation}
which varies from event to event. To enhance $f_{M, {\rm hot}}^{\rm SN}$ and hence $\overline{v}_{\mathcal{B}, {\rm hot}}$ on average, SN feedback needs to occur either preferentially outside the main gas disk (L17) or inside a region in which a vertical cavity has been opened. The latter case is not easily realized in our simulations, but may be possible in central starbursts.

\subsection{Comparison with Observations}\label{sec:dis-obs}

Observations of galactic outflows (winds) are challenging because the outflow is much 
more tenuous than the underlying galactic disk, so that both emission and absorption lines are weaker. At the same time, outflows possess complex, multiphase structure, demanding high sensitivity observations of many gas tracers to quantify the mass (total and metal), momentum, and energy budget of the outflow. Currently, direct observational constraints for outflow characteristics and their scaling relations with galactic properties are neither strong nor comprehensive \citep[see][for a review]{2018Galax...6..138R}. 

Optical and UV absorption lines surveys provide the largest body of data to study correlations between the outflow characteristics and galaxy properties \citep{2005ApJ...621..227M,2005ApJS..160..115R,2014A&A...568A..14A,2015ApJ...811..149C,2015ApJ...809..147H,2016ApJ...822....9H,2016A&A...590A.125C}. From line profiles, it is relatively straightforward to derive the characteristic velocity of the outflow (modulo different definitions adopted in different studies). A shallow, positive correlation between outflow velocity and SFR is consistently observed in both neutral and ionized outflows; $\vout\propto \dot{M}_{*}^{0.15-0.35}$. \citet{2016ApJ...822....9H} presented a similar correlation between outflow velocity and SFR surface density, $\vout\propto\Sigma_{\rm SFR}^{0.34}$ (essentially the same correlation is seen in stacking analysis of galaxies at $z\sim2$ by \citealt{2019ApJ...873..122D}), while \citet{2015ApJ...811..149C} did not find a convincing correlation of $\vout$ with $\Sigma_{\rm SFR}$. 
Consistent with the observations, we find weak scalings, approximately $\overline{v}_{\rm out} \propto \Ssfr^{0.2}$, for both hot and cool gas, but an order of magnitude higher velocity for the former (\autoref{fig:scaling-v}(a)). We note that these observations treat galaxies as a whole, and are therefore not directly equivalent to our scaling relations (which would require observational resolution of $\lesssim \kpc$ and sufficient sensitivity to detect individual disk ``patches''). Also, the range of $\Sigma_{\rm SFR}$ in observations described above is generally on the high side, $\Sigma_{\rm SFR}>0.1\sfrunit$, which only marginally overlaps with our parameter space. 

The mass loading factor is a more difficult quantity to measure  empirically. In  estimating the mass loading from observed interstellar absorption lines, many assumptions are involved, including the covering area of the outflow (a combination of the opening angle, characteristic radius, and covering fraction of the outflow), the column density conversion from a specific species to total hydrogen, and the characteristic velocity \citep[e.g.,][]{2005ApJS..160..115R}.
The reported mass loading factors from observations of dwarf starbursts and LIRGs/ULIRGs are in  the range $\eta_M\sim0.1-10$, and have found either negative correlation \citep[e.g.,][]{2015ApJ...809..147H,2017MNRAS.469.4831C} or no correlation \citep[e.g.,][]{1999ApJ...513..156M,2019ApJ...886...74M} with galaxy mass (or circular velocity). Although the  full galaxy mass range in these studies is $\log M_*\sim7-11$, the low mass galaxy samples (at $\log M_*\sim7-8$) used in the study that found negative correlation are more extreme starbursts than those in the study reported no correlation \citep[see][]{2019ApJ...886...74M}. \citet{2014A&A...568A..14A} observed local LIRGs and ULIRGs ($\log M_*\sim 9.5-11$) with  integral field spectroscopy and obtained $\eta_M\propto M_*^{-0.43}$, similar to \citet{2017MNRAS.469.4831C}. 
A direct comparison with our results is not possible, since our work measures outflow rates and galactic properties locally, in contrast to the global outflow rates and galaxy mass reported in observations.  Still, it is encouraging that the observed estimates of $\eta_M$ are similar to what we find (\autoref{fig:mloading-sfr-comp}) at $\Ssfr \sim 0.1 - 1$, which overlaps with the observed range for these samples. 

Interestingly, \citet{2014A&A...568A..14A} reported a positive correlation between $\eta_M$ and $\Sigma_{\rm SFR}$ with a log-log slope of $0.17$, which is apparently in tension with our results (see \autoref{fig:scaling-Mloading}, with slopes $\sim -0.5$ for cool gas) and those from other numerical simulations, which all show negative scaling for  $\eta_M$ vs. $\Ssfr$ (\autoref{fig:mloading-sfr-comp}). However, in \citet{2014A&A...568A..14A} $\Sigma_{\rm SFR}\sim 0.1-100\sfrunit$, which only marginally overlaps with the high  end of our $\Ssfr$ range. 
Furthermore, the scatter in their mass loading factor is large and the significance of the fit is not high (Figure 14 of \citealt{2014A&A...568A..14A}). 
Nevertheless, there is overall agreement in the range of mass loading factor, $\eta_M\sim 0.1-1$.
Our results also suggest an intriguing possibility of a weakened correlation between $\eta_M$ and $\Ssfr$ at high $\Ssfr$, where hot outflows begin to  dominate the total mass (\autoref{fig:scaling-loading} (a) and (b)).

In the future, spatially-resolved outflow observations utilizing sensitive integral field unit observations offer the promise of enabling direct comparison with the kind of local scaling relations reported here. 
With future computational advances, it will also be possible to run global simulations with the current resolution and physics of our local simulations, to connect with observed global relationships.  

\subsection{Physical interpretation of Scaling Relations}\label{sec:dis-interpretation}

Multiphase outflow launching in our simulation suite is an outcome of intricate interactions between SN feedback and ISM dynamics, with complexity that precludes a purely analytic theory that can explain our quantitative findings. 
Nevertheless, we are able to obtain insight to the physics behind the  emergent scaling relations we have found using a simple theoretical model of superbubble evolution and breakout.
Given the simple assumptions we adopt (e.g., uniform background medium and spherical symmetry), we will mainly focus on parameter dependence rather than coefficients.

\citet{1977ApJ...218..377W} developed an analytic theory for the evolution of stellar wind-blown bubbles, and essentially the same theory has subsequently been applied to superbubbles driven by clustered SNe   \citep{1988ApJ...324..776M,1987ApJ...317..190M,2019MNRAS.490.1961E}. In the model, the evolution after the radiative shell formation is characterized by the energy injection rate $\dEin$ and the ambient medium density $\rho_0$. In \citet{1977ApJ...218..377W}, the injected energy is shared among kinetic energy of the cooled shell $\Esh$, thermal energy of the hot interior $\Ehot$, and radiative energy losses in the forward shocks $\Ecool$. The classical theory predicts $\dEsh=(15/77) \dEin$, $\dEhot=(5/11)\dEin$, and $\dEcool=(27/77)\dEin$.
Since $\dEin \propto$SFR, to zeroth order this explains why energy loading factors of both cool and hot outflows are nearly constant with SFR (see \autoref{fig:scaling-loading} (e) and (f)).

The classical theory neglects cooling at the interface between the hot interior and cool, dense shell, while in reality the interface cooling $\dEintcool$ is crucial for understanding the energy budgets in superbubbles \citep[e.g.,][]{2017ApJ...834...25K,2018MNRAS.481.3325F,2019MNRAS.483.3647G}. Mixing layers between hot and cool gas are mediated by both (M)HD instabilities and radiative cooling, best explored with very high-resolution simulations \citep[e.g.,][]{2020arXiv200308390F}. In the current simulations, the existence of intermediate temperature phase in outflows demonstrates that cooling in the mixing layers plays a role in reducing the injected energy.  

For present purposes, we employ a model used in 1D simulations of \citet{2019MNRAS.490.1961E}, in which interface mixing is parameterized via a diffusion coefficient $\lambda \delta v$. The resulting interface cooling rate is $\dEintcool = \theta \dEin$, with $\theta$ depending on $\lambda\delta v$ and ambient density $\rho_0$ as $\theta/(1-\theta)\propto (\lambda \delta v)^{1/2} \rho_0^{1/2}$. Inclusion of the interface cooling reduces $\dEin$ to $(1-\theta)\dEin$, and with less power the bubble expands less rapidly.  This results in $\dEhot=(5/11)(1-\theta)\dEin$ and $\dEcool=(27/77)(1-\theta)\dEin$, so that a constant $\theta$ would still imply energy loading of hot and cool outflows that are independent of SFR.
In reality, the ambient medium in the real ISM (and in the current simulations) is highly inhomogeneous and vertically stratified and the  diffusion coefficient representing details of mixing layer varies, so that $\theta$ is not constant.  The  weak scaling between $\eta_E$ and $\Ssfr$ (\autoref{fig:scaling-loading} (e) and (f)) presumably arises from weak dependencies in the averages over these variations.

Since a superbubble's interior temperature depends very weakly on the ambient medium density ($T_\mathrm{hot}\propto\rho_0^{2/35}$ for  conduction-mediated evaporation from \citealt{1977ApJ...218..377W,2019MNRAS.490.1961E}, and $T_\mathrm{hot}$ is also insensitive to $\rho_0$ from simulations of expansion in an inhomogeneous medium without conduction from \citealt{2017ApJ...834...25K}), the constant mass loading factor of hot outflows (\autoref{fig:scaling-loading} (b)) is easily understood from $\loading{M}{hot}\sim \loading{E}{hot}/T_\mathrm{hot}$ with weakly-varying $T_\mathrm{hot}$.

For the mass loading factor of cool outflows, $\loading{M}{cool}\sim \loading{E}{cool}/\voutcool^2$, we need to understand what determines the characteristic outflow velocity of the cool phase. To this end, we seek a scaling relation of the cooled shell velocity when a superbubble breaks out of the disk (roughly $R\sim H$).
Applying the theory of \citet{2019MNRAS.490.1961E}, the bubble radius follows
\begin{equation}
    R(t) = 83\pc (1-\theta)^{1/5} 
    \rbrackets{\frac{\dot{E}_{\rm in}}{10^{46}\ergyr}}^{1/5}
    \rbrackets{\frac{n_0}{\pcc}}^{-1/5}
    \rbrackets{\frac{t}{{\rm Myr}}}^{3/5}
\end{equation}
and the shell velocity is
\begin{equation}
    v_{\rm sh}(t) = dR/dt = 49\kms (1-\theta)^{1/5} 
    \rbrackets{\frac{\dot{E}_{\rm in}}{10^{46}\ergyr}}^{1/5}
    \rbrackets{\frac{n_0}{\pcc}}^{-1/5}
    \rbrackets{\frac{t}{{\rm Myr}}}^{-2/5}.
\end{equation}
The time at which the bubble radius reaches the disk scale height is
\begin{equation}
    t_H \equiv 8.5\Myr (1-\theta)^{-1/3}
    \rbrackets{\frac{\dot{E}_{\rm in}}{10^{46}\ergyr}}^{-1/3}
    \rbrackets{\frac{n_0}{\pcc}}^{1/3}
    \rbrackets{\frac{H}{300\pc}}^{5/3}
\end{equation}
so that
\begin{equation}
    v_{\rm sh}(t_H) = 21\kms  (1-\theta)^{1/3}
    \rbrackets{\frac{\dot{E}_{\rm in}}{10^{46}\ergyr}}^{1/3}
    \rbrackets{\frac{n_0}{\pcc}}^{-1/3}
    \rbrackets{\frac{H}{300\pc}}^{-2/3}.
\end{equation}
Assuming all SNe that explode within an area $\pi H^2$ contribute to superbubble breakout at $R=H$, 
\begin{equation}
    \dEin(<H) =\pi H^2  E_{\rm SN}\frac{\Sigma_{\rm SFR}}{m_*}    =3\times10^{47}\ergyr    \rbrackets{\frac{\Sigma_{\rm SFR}}{0.1\sfrunit}} 
    \rbrackets{\frac{H}{300\pc}}^{2}.
\end{equation}
We obtain
\begin{equation}
    v_{\rm sh}(t_H) = 63\kms  (1-\theta)^{1/3}
    \rbrackets{\frac{\Sigma_{\rm SFR}}{0.1\sfrunit}}^{1/3}
    \rbrackets{\frac{n_0}{\pcc}}^{-1/3}.
\end{equation}

Since SFRs in our simulations agree well with the pressure-regulated, feedback-modulated star formation theory (Ostriker \& Kim in prep.; see also \citealt{2010ApJ...721..975O,2011ApJ...731...41O,2011ApJ...743...25K}), we may use the relationships $\mathcal{W} = P_{\rm mid} = \Upsilon \Ssfr$
where $\Upsilon$ is the total feedback yield\footnote{The original notation used in \citet{2011ApJ...743...25K,2013ApJ...776....1K,2015ApJ...815...67K} for feedback yields was $\eta$, but  here we instead use $\Upsilon$ since $\eta$ in the present paper is used to denote  outflow loading factors.}  (allowing for thermal and magnetic as well as turbulent terms). 
We assume the characteristic ambient medium density to be the midplane density $\rho_{\rm mid}$, which is
\begin{equation}
    \rho_{\rm mid} = \frac{P_{\rm mid}}{\sigma_{\rm z,eff}^2} =
    \frac{\Upsilon \Sigma_{\rm {SFR}}}{\sigma_{\rm z,eff}^2}
\end{equation}
or
\begin{equation}
    n_{\rm mid} = 1.7\pcc
    \rbrackets{\frac{\Upsilon}{10^3\kms}}
    \rbrackets{\frac{\Sigma_{\rm SFR}}{0.1\sfrunit}}
    \rbrackets{\frac{\sigma_{\rm z,eff}}{40\kms}}^{-2}.
\end{equation}
We then finally obtain
\begin{equation}
    v_{\rm sh}(t_H) = 52\kms  (1-\theta)^{1/3}
    \rbrackets{\frac{\Upsilon}{10^3\kms}}^{-1/3}
    \rbrackets{\frac{\sigma_{\rm z,eff}}{40 \kms}}^{2/3},
\end{equation}
with no explicit dependence of $v_{\rm sh}(t_H)$ on $\Ssfr$. Note that in previous work seeking a physical interpretation of the observed weak scaling between the outflow velocity and $\Ssfr$, the empirical Kennicutt-Schmidt relation $\Ssfr\propto \Sigma^{1.4}$ \citep{1998ApJ...498..541K} was instead adopted to get $v_{\rm sh}(t_H)\propto \Ssfr^{0.1}H^{1/3}$ \citep[e.g.,][]{2004ApJ...606..829S,2010AJ....140..445C}.

In our simulation suite, we find a weak scaling of $\sigma_{\rm z,eff}\propto \Ssfr^{0.18}$ (see \autoref{tbl:time}) and $\Upsilon\propto \Ssfr^{-0.15}$ (Ostriker \& Kim in prep.; see also \citealt{2011ApJ...743...25K,2013ApJ...776....1K,2015ApJ...815...67K}), yielding $v_{\rm sh}(t_H)\propto \Ssfr^{0.17}$. Modulo a hidden dependence in $(1-\theta)$, this explains the weak, positive scaling $\voutcool\propto\Ssfr^{0.23}$ and hence $\loading{M}{cool}\propto \voutcool^{-2}\propto \Ssfr^{-0.46}$, similar to the results shown in \autoref{fig:scaling-v}(a) and \autoref{fig:scaling-loading}(a). 

We emphasize that $\voutcool$ is a characteristic velocity from a rather wide distribution of $\vout$ rather than a single ``shell'' velocity $v_{\rm sh}(t_H)$ as in the above simple theory.
Even in idealized simulations of multiple SNe in an inhomogeneous medium \citep{2017ApJ...834...25K}, the distribution of expanding velocities is broad, while the characteristic ``knee'' in the velocity distribution increases with the energy injection rate (parameterized by an interval between SNe), but is insensitive to density.

\subsection{TIGRESS Outflow Models in Context}\label{sec:dis-context}

The methods used in this work have clear pros and cons in the context of galactic wind research. Here we review the advantages and also discuss limitations of our methodology.

With the uniformly high resolution of our simulations ($2\pc$-$8\pc$; higher resolution for denser condition), the outflow characteristics studied in this work arise not from \emph{ad hoc} assumptions but from resolved key physical processes at every relevant step:
\begin{itemize}
    \item \emph{Star formation} -- Star formation occurs in gravitationally collapsing objects at high density and pressure that is distinct from the ambient ISM \citep[e.g.,][]{2019arXiv191105078M}. 
    \item \emph{SN injection} -- Self-regulated SFRs and a population synthesis model applied to star cluster particles provide 
    SN rates and positions that have realistic space-time correlations with respect to each other and the distribution  of ISM gas.
    \item \emph{Superbubble evolution} -- The Sedov-Taylor stage of SNR evolution is resolved for more than 90\% of individual SNe, 
    directly capturing hot gas creation and momentum injection.
    \item \emph{Multiphase outflow evolution} -- The evolution of low-density outflows in extraplanar regions is followed using the same spatial and time resolution as the higher-density ISM near the midplane \citep{2019arXiv191107872V}, without degrading the resolution as in (semi-)Lagrangian or adaptive mesh refinement schemes.
    \item \emph{Long-term evolution} -- Each model is run at least up to $1.5\torb$, covering a few star formation-feedback-wind launching-outflow/inflow cycles.
\end{itemize}
    
The main caveats  arise from the local approximation (adopted to achieve uniformly high resolution) and missing physics (adopted to enable long-term evolution and a  survey of parameters), e.g.,
\begin{itemize}
    \item \emph{Missing global geometry} -- Outflow evolution to scales large compared to the launch region cannot be captured in local models.  Without streamline opening, hot winds do not reach their asymptotic velocity \citep[e.g.][]{1985Natur.317...44C,2017MNRAS.470L..39F,2018MNRAS.478..302S}, and fountain flows that  travel large radial distances cannot be captured.  
    
    \item \emph{Missing radial and cosmic accretion} -- Our simulation adopts outflow boundary conditions in the vertical direction, and shearing-periodic boundary  conditions  in the horizontal directions.  There are therefore no sources of new gas to replace gas lost to star formation or winds. It is worth emphasizing that the galactic scale impact of outflows would not be solely determined by wind launching properties characterized in this paper, but also interaction with the CGM, which is in part shaped by cosmic flows that cannot be modeled in local simulations \citep{2020arXiv200616316F}. The relevant processes include cosmic accretion, gas flows driven by galaxy mergers, and outflows from satellite galaxies.

    \item \emph{Missing early feedback} -- We only include the two dominant channels of stellar feedback, SNe and radiative heating of warm-cold gas. 
    It has previously been argued that dynamics driven by ``early feedback'' in the form of radiation pressure, massive-star winds, and photoionization is needed to reduce densities  and make SNe effective \citep{2017MNRAS.466.1903G,2017MNRAS.466.3293P,2020MNRAS.491.2088K}.  In fact, the natural clustering of SNe in our  simulations means that we fully resolve radiative supernova remnant evolution $> 90\%$  of the time.  However, in environments where the free-fall times in dense clouds is short, the lack of early feedback means star clusters may significantly grow in the $\sim$ 3-4~Myr before the onset of the first SN; this may be responsible for unrealistically  high SFRs in our models R2 and R4. For lower density environments, SNe effectively disperse their parent clouds without excessive star formation. In  (short-term) simulations with conditions similar to model R8, \citet{2017MNRAS.466.1903G} found similar galactic outflow fluxes for models with and without stellar winds, while \citet{2020MNRAS.491.2088K} found similar  outflow fluxes for models with and without radiation pressure.

    \item \emph{Other missing physics} -- Thermal conduction and cosmic rays are two major missing physical processes that may have potentially significant impact on our results. Thermal conduction can load more hot gas during the superbubble evolution \citep[e.g.,][]{2019MNRAS.490.1961E}. Since superbubbles in our simulations expand in a highly inhomogeneous, turbulent ISM, there is a high level of mixing that can transfer gas between warm and hot phases \citep[see also][for evidence of this]{2020arXiv200210468S}.
    It remains unclear whether fully-realistic simulations that also include thermal conduction (which must be anisotropic to allow for the magnetic field) alter mass loading of hot outflows significantly.
    
    Cosmic rays are mainly accelerated in SN shocks and a provide a non-thermal pressure force with relatively low radiative losses. Cosmic rays advect with the gas and also diffuse along the magnetic field,  with flux limited by the Alfv\'en speed. Although there are large uncertainties in diffusion coefficients and numerical difficulties in modeling cosmic ray transport, cosmic-ray pressure gradients may be substantial and play a key role in driving cooler, smoother, and slower galactic winds \citep[e.g.,][]{2016ApJ...827L..29S,2018ApJ...854...89M,2018MNRAS.479.3042G}.
\end{itemize}

\section{Summary}\label{sec:summary}

This work quantifies characteristics of multiphase outflows emerging from self-consistent, high-resolution simulations of the star-forming ISM. Our suite of MHD simulations consists of 7 models covering a range of galactic conditions that appear within normal star-forming galaxies like the Milky Way. Each model represents a local, $\sim$kpc-scale region within a galactic disk. The ISM in each simulation is explicitly modeled by solving ideal MHD equations including the effect of galactic differential rotation, gas self-gravity, external gravity from the stellar disk and dark matter halo, optically thin cooling from 10~K to $10^9$~K, photoelectric heating onto small grains by FUV radiation, and 
energy and momentum input from SNe. Gas collapses to make star cluster particles, which produce in-situ and runaway SNe. The TIGRESS framework (see KO17 for numerical details) allows us to follow long-term evolution (more than an orbit time, at least a few feedback cycles after the initial transient) of the star-forming ISM, with self-regulated SFRs and ISM properties.
Self-regulation cycles of star formation and feedback modulate outflows and inflows self-consistently (\autoref{fig:massflux}). Galactic winds emanating from superbubble breakout possess multiphase structure with distinct characteristics (see KO18 and \citealt{2019arXiv191107872V} for in-depth analysis of the Solar neighborhood model R8). We measure fluxes of mass (total and metal), momentum, and energy of each thermal phase 
of the outflowing gas at four different locations: $|z| =  H,\, 2H,\, 500\pc,$ and $1\kpc$; results are given in \autoref{sec:windprop}.
We present scaling relations for wind loading factors, characteristic velocities, and metal properties as a function of a variety of local galactic properties.  These scaling relations are reported separately for cool and hot phases (\autoref{sec:scaling}), and we also compare scalings of total loading with results from other recent simulations (\autoref{sec:dis-sims}) and observations (\autoref{sec:dis-obs}). We provide a physical interpretation of scalings based on a simple theoretical model of superbubble breakout (\autoref{sec:dis-interpretation}).
We provide full information from our outflow analyses at \href{http://doi.org/10.5281/zenodo.3872049}{doi:10.5281/zenodo.3872049}, which we hope can serve as a benchmark for up-coming theoretical and observational studies.

Our key  findings for galactic outflows are  as follows:
\begin{enumerate}
    \item \emph{Overall evolution --} Star formation, SN feedback, and wind driving are all self-regulated and show clear cyclic behavior (\autoref{sec:evolution}). In low surface density models, the characteristic time scale for vertical oscillation $(\pi/G\rho_\mathrm{tot})^{1/2}$  is longer than the feedback time scale (or star cluster evolution time scale $\sim 40\Myr$), leading to a well-defined cyclic behavior for star formation and outflow fluxes governed by vertical oscillation. In high surface density models, in contrast, the natural vertical oscillation period is shorter than the duration of feedback from a burst, so that returning flows interfere with gas being launched by a burst.
    In these cases, evolution is more chaotic and no clear correspondence between midplane star bursts and outflows above the disk exists. We thus construct time-averaged outflow characteristics over a few feedback cycles ($0.5<t/\torb<1.5$) to quantify the overall behavior, rather than individual bursts. This is especially important in the measurement  of ``loading factors,'' for which a mismatch between time-dependent outflow fluxes and offset time-dependent reference fluxes (set by SN/star formation rates) can produce quite misleading instantaneous measurements for loading. 
    
    \item\emph{Emergent multiphase outflow ranges --}
    For the range $\Sigma_{\rm gas}\sim 1-100\Surf$ and $\Sigma_*/(2z_*)+\rho_{\rm dm} = 0.005-1\rhounit$ that are inputs to our simulations, the range of self-consistently regulated properties of the star-forming ISM disk are
    $\Sigma_{\rm SFR}\sim 10^{-4}-1\sfrunit$, 
    $P_{\rm mid}\approx \mathcal{W} \sim 10^3-10^6 k_B\pcc\Kel$, 
    $n_{\rm mid}\sim 0.05-50\pcc$, and 
    $\tdep\sim 10^2-10^4\Myr$. From fluxes
    measured at  $|z|=H$, the emergent loading factors of mass, momentum, and energy are $\eta_M\sim0.5-50$, $\eta_p\sim0.04-0.7$, $\eta_E\sim 0.005-0.02$, $\eta_{Z}^{\rm SN}\sim0.1$ for \emph{cool} outflows ($T<2\times10^{4}\Kel$) and $\eta_M\sim0.1-0.3$, $\eta_p\sim0.07-0.12$, $\eta_E\sim 0.05-0.25$, $\eta_{Z}^{\rm SN}\sim0.1-0.3$ for \emph{hot} outflows ($T>5\times10^{5}\Kel$). The intermediate phase ($2\times10^{4}\Kel<T<5\times10^{5}\Kel$) is subdominant for all loading factors. 
 
    We note that at fixed height, the hot outflow energy loading factor is essentially constant across simulations (e.g., \autoref{fig:Eloading-sfr-comp}(c) and (d)), $\loading{E}{hot} \approx 0.1$ for $|z|=500\pc$  and $\loading{E}{hot}=0.04$ at $1\kpc$. Similarly to the energy loading factor, at fixed heights $|z|=500\pc$ and $1\kpc$, the instantaneous SN-origin metal loading factor is more or less constant, $\loading{Z}{hot}^{\rm SN}= 0.16$ and $0.066$, respectively.
    Figures and the data at different heights are available at \href{http://doi.org/10.5281/zenodo.3872049}{doi:10.5281/zenodo.3872049}.

    \item \emph{Scaling of loading factors --}  We find that mass is primarily carried by cool outflows and energy is primarily carried by hot outflows, with the following scaling relations for loading factors at $|z|=H$:
    \begin{eqnarray}
    \log \loading{M}{cool} &=& -0.44^{+0.08}_{-0.08} \log \rbrackets{\frac{\Sigma_{\rm SFR,40}}{M_{\odot}{\rm \, kpc^{-2}\,yr^{-1}}}}-0.07^{+0.16}_{-0.15} \pm 0.27
    \quad \textrm{\autoref{fig:scaling-Mloading}(b)} \\
     &=& -0.54^{+0.08}_{-0.08} \log \rbrackets{\frac{\mathcal{W}/k_B}{{\rm cm^{-3}\,K}}}+3.23^{+0.39}_{-0.42} \pm 0.23
    \quad \textrm{\autoref{fig:scaling-Mloading}(g)} \\
     &=& 0.70^{+0.09}_{-0.10} \log \rbrackets{\frac{t_{\rm dep,40}}{{\rm Myr}}}-1.44^{+0.32}_{-0.29} \pm 0.23
    \quad \textrm{\autoref{fig:scaling-Mloading}(h)}
    \end{eqnarray}
    \begin{eqnarray}
    \log \loading{E}{hot} &=& 0.14^{+0.08}_{-0.08} \log \rbrackets{\frac{\Sigma_{\rm SFR,40}}{M_{\odot}{\rm \, kpc^{-2}\,yr^{-1}}}}-0.70^{+0.12}_{-0.14} \pm 0.22
    \quad \textrm{\autoref{fig:scaling-Eloading}(b)} \\
     &=& 0.17^{+0.09}_{-0.09} \log \rbrackets{\frac{\mathcal{W}/k_B}{{\rm cm^{-3}\,K}}}-1.73^{+0.49}_{-0.47} \pm 0.21
    \quad \textrm{\autoref{fig:scaling-Eloading}(g)} \\
     &=& -0.22^{+0.11}_{-0.10} \log \rbrackets{\frac{t_{\rm dep,40}}{{\rm Myr}}}-0.27^{+0.30}_{-0.32} \pm 0.20
    \quad \textrm{\autoref{fig:scaling-Eloading}(h)} 
    \end{eqnarray}
   
    The variation of mass loading factors with galaxy properties is strong in cool outflows and weak in hot outflows. In fact, \textit{all} loading factors of hot outflows only vary by a factor of 2-3 (right column of \autoref{fig:scaling-loading}), while galactic properties like 
    $\Ssfr$ vary more than 4 orders of magnitude.  
    
    For cool gas, the momentum loading also varies significantly across galaxy environments, while energy and metal loading do not (left column of \autoref{fig:scaling-loading}).  
    We find overall a similar level of correlations between loading factors and all local galactic properties we consider except $\Sgas$ (\autoref{fig:scaling-Mloading}, \autoref{fig:scaling-Eloading}). This is in part because the ``derived''  galactic properties ($\Ssfr$, $\Pmid$, $\mathcal{W}$, $n_{\rm mid}$, and $\tdep$) are self-regulated and connected with each other, and in part because our parameter choice assumes an implicit correlation between gas ($\Sgas$) and gravity ($\Sigma_*/(2z_*)$ and $\rho_\mathrm{dm}$) parameters (see \autoref{sec:scaling-input}).
    Subsequent work exploring a wider parameter space would be needed to cover conditions in nearby observable targets including dwarf starbursts and LIRGs/ULIRGs, and the full range of conditions that are relevant to theoretical galaxy formation models \citep{2020arXiv200616314M}. 
    
    \textit{We emphasize that the large mass loading of outflows at low SFR does not imply a massive cool wind because the cool gas outflow velocities are low.  Instead, at low SFR there is a heavily-loaded cool fountain.}
    
    \item \emph{Characteristic velocities --} We define two characteristic velocities, an outflow velocity $\overline{v}_{\rm out}$ (\autoref{eq:vout}), and a Bernoulli velocity $\overline{v}_{\mathcal{B}}$ (\autoref{eq:vB}). Since we include all gas that has positive outward velocity in computing outflow fluxes, the low 
    $\overline{v}_{\rm out} \sim 10$--$110
    \kms$ values we find for cool-phase outflows imply a large fraction of the gas will fall back as fountains, as indeed the simulations show. For cool outflows, $\overline{v}_\mathcal{B}\sim 20$--$140\kms$ is dominated by the kinetic term and is not much larger than $\overline{v}_{\rm out}$. For hot outflows, $\overline{v}_\mathcal{B}\sim 400$--$1400\kms$ is dominated by the thermal term, and is large enough that hot gas would escape far into halos. We find generally very weak scaling of the characteristic velocities with galactic properties.
    
    The velocity scaling relations at $|z|=H$ obtained in this work are:
    \begin{eqnarray}
    \log \rbrackets{\frac{\voutcool}{{\rm km\,s^{-1}}}} &=& 0.23^{+0.04}_{-0.04} \log \rbrackets{\frac{\Sigma_{\rm SFR,40}}{M_{\odot}{\rm \, kpc^{-2}\,yr^{-1}}}}+1.78^{+0.07}_{-0.07} \pm 0.14
    \quad \textrm{\autoref{fig:scaling-v}(a)} \\
     &=& 0.27^{+0.03}_{-0.03} \log \rbrackets{\frac{\mathcal{W}/k_B}{{\rm cm^{-3}\,K}}}+0.10^{+0.17}_{-0.17} \pm 0.10
    \quad \textrm{\autoref{tbl:scaling}} \\
     &=& -0.34^{+0.03}_{-0.04} \log \rbrackets{\frac{t_{\rm dep,40}}{{\rm Myr}}}+2.46^{+0.11}_{-0.11} \pm 0.08
    \quad \textrm{\autoref{tbl:scaling}}
    \end{eqnarray}
    \begin{eqnarray}
    \log \rbrackets{\frac{\vBhot}{{\rm km\,s^{-1}}}} &=& 0.11^{+0.04}_{-0.04} \log \rbrackets{\frac{\Sigma_{\rm SFR,40}}{M_{\odot}{\rm \, kpc^{-2}\,yr^{-1}}}}+3.04^{+0.08}_{-0.08} \pm 0.16
    \quad \textrm{\autoref{fig:scaling-v}(d)} \\
     &=& 0.13^{+0.04}_{-0.05} \log \rbrackets{\frac{\mathcal{W}/k_B}{{\rm cm^{-3}\,K}}}+2.25^{+0.23}_{-0.21} \pm 0.14
    \quad \textrm{\autoref{tbl:scaling}} \\
     &=& -0.17^{+0.06}_{-0.05} \log \rbrackets{\frac{t_{\rm dep,40}}{{\rm Myr}}}+3.37^{+0.16}_{-0.18} \pm 0.14
    \quad \textrm{\autoref{tbl:scaling}}
    \end{eqnarray}
    
    \item \emph{Metals --}
    Metals in outflows originate from both the ISM and SN. Recent SN-origin material in hot outflows amounts to typically 5--20\% of the mass and 30-60\% of the metal mass (these fractions generally increase with $\Ssfr$).
    
    The instantaneous SN-origin metal loading factor scales very weakly with 
    all galactic properties, e.g., at $|z|=H$,
    \begin{eqnarray}
    \log \loading{Z}{hot}^{\rm SN} &=& 0.11^{+0.07}_{-0.07} \log \rbrackets{\frac{\Sigma_{\rm SFR,40}}{M_{\odot}{\rm \, kpc^{-2}\,yr^{-1}}}}-0.61^{+0.11}_{-0.12} \pm 0.19
    \quad \textrm{\autoref{fig:scaling-loading}(h)}
    \end{eqnarray}
    The instantaneous SN-origin metal loading factor in cool outflows is nearly identical to that in hot outflows, slightly lower near the disk and higher farther away. 

    The metal enrichment factor $\zeta$ is nearly flat at low $\Ssfr$,
    $\zeta\approx 1$ and 1.5 for cool and hot outflows, respectively. $\zeta$ begins to increase with $\Ssfr$ above $\Sigma_{\rm SFR}>0.1\sfrunit$, reaching $\zeta\approx 1.1$ and 2 for cool and hot outflows, respectively. 
    
    There is a very tight, positive correlation between energy and SN-origin metal fluxes (and hence loading factors) in the hot outflow. A similar, but     looser correlation also exists for the total outflow. Taking all outflow time series into account, we find that the correlation is slightly super-linear with a log-log slope of $1.15$ at $|z|=1\kpc$ (\autoref{eq:hot-EZ}). This means that the energy loading in outflows is more efficient (radiative cooling is reduced) when more genuine SN material is loaded (= more successful breakout). This can be translated into a correlation between the Bernoulli velocity (or specific energy) and the SN-origin mass fraction in the outflow as in \autoref{eq:hot-vB-fSN}.

    \item \emph{Comparison with other simulations} -- Our results are overall consistent with previous local simulations as long as their adopted $\Ssfr$ are consistent with our predicted self-consistent values at a given $\Sgas$ and vertical gravity. However, mass and energy partitions between phases may still be quite sensitive to the adopted SN distribution and its mutual correlation with gas distribution. 
\end{enumerate}

Finally, we close this paper by putting the present work in the context of the general goal of the SMAUG project: the development of physical subgrid models for galaxy formation models.
Currently, in large-box cosmological simulations and semi-analytic models, galactic winds are often implemented via scaling relations, with velocities typically set by the halo potential and mass-loss rates tuned to match the resulting galaxy properties with observational constraints. 

With this kind of approach, the connection between galaxies and dark matter halos (e.g., the stellar mass-halo mass relation) is essentially \emph{imposed} rather than \emph{emergent}.
In particular, prescriptions of this kind do not account for the local ISM physics  involved in launching winds.  For example, our simulations show that the hot wind has a Bernoulli velocity that is nearly independent of local conditions (reflecting the characteristic temperature of hot ISM gas), which would lead to an asymptotic hot wind velocity that is independent of the halo potential, rather than scaling with the halo potential. Our simulations also show that most of the mass is carried by a low-velocity cool phase, with velocity relatively independent of local conditions but loading that decreases  with the local $\Ssfr$. 
Winds that are emergent from local galaxy properties (compared to previous globally-imposed wind scalings) are also likely to differ in their implications for global stellar mass-halo mass relationships, through the distribution of $\Ssfr$ in galaxies of different mass at varying redshift.

Potentially, large-box cosmological  simulations may require multiple layers of subgrid modeling to represent unresolved processes. The outflow characteristics quantified here provide the properties at the ``base'' of the outflow where it is launched, provided there is proper knowledge of the resolved conditions within the ISM on those scales. 
For cosmological zoom simulations, the resolution may be adequate (e.g. marginally resolving the ISM's scale height) to provide a  reasonable value for $\Ssfr$ (or  $P_\mathrm{mid}$ or other properties as shown in \autoref{fig:scaling-Mloading} and \autoref{fig:scaling-Eloading}), but insufficient to represent a multiphase outflow; our results could then be applied to model that outflow launching. For large-box cosmological simulations at lower resolution, a separate subgrid model would be required in order to predict $\Ssfr$ (or  $P_\mathrm{mid}$). 
A key conclusion is that to properly represent physically-realistic outflows, any subgrid model implementation must incorporate at least two distinct components, one for a hot, fast flow and the other for a cool, slow flow.

In this paper, we have provided scaling relations for certain properties of phase-separated outflows, focusing especially on the mass and energy loading  relations for cool and hot phases that enable comparisons with previous theoretical and observational work, and provide benchmarks for the future. However, we caution that the scaling relations provided here are insufficient to build a proper subgrid model for a cosmological simulation. In particular, while here we have provided information about ``typical'' (mass flux-weighted) velocities, the outflows in our simulations generally have a range of velocities (characterized as an exponential distribution for the warm gas in KO18) and temperatures. In a companion paper, we will quantify these distributions.  We will also provide a guide to combine with the phase-separated loading  relations of this paper to build a subgrid wind model for use in galaxy formation simulations and semi-analytic models.  

\acknowledgements
This work was carried out as part of the SMAUG project. SMAUG gratefully acknowledges support from the Center for Computational Astrophysics at the Flatiron Institute, which is supported by the Simons Foundation.
We are grateful to Amiel Sternberg for valuable discussions and  helpful comments on the manuscript.
The work of C.-G.K. was partly supported by a grant from the Simons Foundation (CCA 528307, E.C.O.). C.-G.K. and E.C.O. were supported in  part by NASA ATP grant No. NNX17AG26G. 
R.S.S., D.B.F., J.C.F., and C.C.H. were supported by the Simons Foundation through the Flatiron Institute.
G.L.B. acknowledges financial support from the NSF (grants AST-1615955, OAC-1835509, XSEDE).
Resources supporting this work were provided in part by the NASA High-End Computing (HEC) Program through the NASA Advanced Supercomputing (NAS) Division at Ames Research Center, in part by the Princeton Institute for Computational Science and Engineering (PICSciE) and the Office of Information Technology's High Performance Computing Center, and in part by the National Energy Research Scientific Computing Center, which is supported by the Office of Science of the U.S. Department of Energy under Contract No. DE-AC02-05CH11231.

\software{{\tt Athena} \citep{2008ApJS..178..137S,2009NewA...14..139S},
{\tt astropy} \citep{2013A&A...558A..33A,2018AJ....156..123T}, 
{\tt scipy} \citep{2020SciPy-NMeth},
{\tt numpy} \citep{vanderWalt2011}, 
{\tt IPython} \citep{Perez2007}, 
{\tt matplotlib} \citep{Hunter:2007},
{\tt linmix} \citep{2007ApJ...665.1489K},
{\tt xarray} \citep{hoyer2017xarray},
{\tt pandas} \citep{mckinney-proc-scipy-2010},
{\tt CMasher} \citep{CMasher},
{\tt corner} \citep{corner},
{\tt adstex} (\url{https://github.com/yymao/adstex})
}

\bibliographystyle{aasjournal}
\bibliography{ref}

\appendix

\section{Convergence with  Resolution and Box Size}\label{sec:app-conv}

The numerical convergence of the TIGRESS framework has been extensively demonstrated and discussed in KO17 (for general ISM properties, star formation rates, and outflow fluxes), and in KO18 (for multiphase characteristics of outflows). For the Solar neighborhood model (R8 in \autoref{tbl:model}), we showed that marginal convergence is achieved at 16~pc and more robust convergence at 8~pc. Due to the generally shorter dynamical time and length scales, we anticipate more stringent convergence conditions in higher density environments.  The simulation parameters shown in \autoref{tbl:model} indeed adopt finer spatial resolution for these models.  

\begin{figure*}
    \centering
    \includegraphics[width=\textwidth]{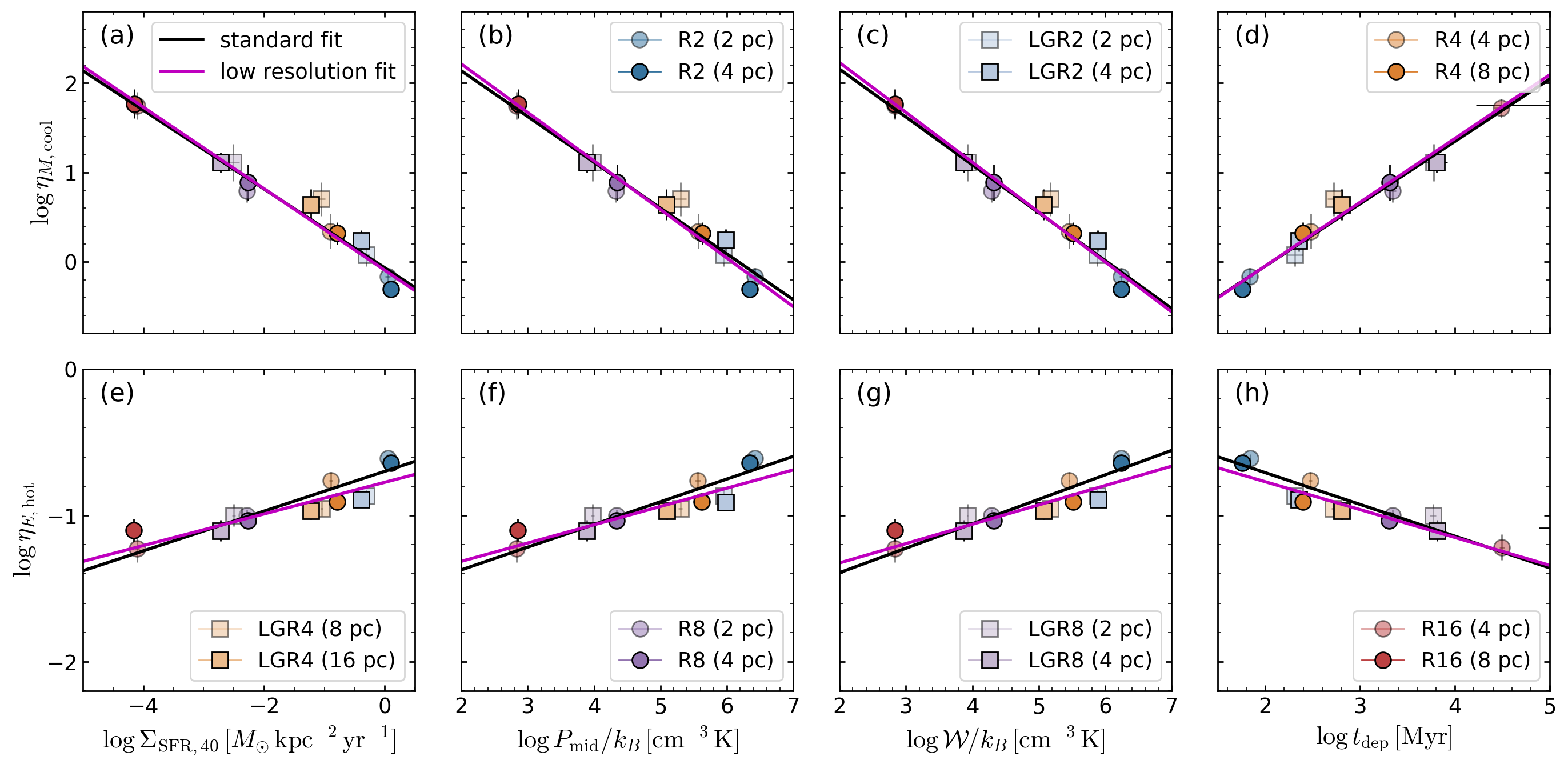}
    \caption{Resolution convergence for scaling relations of loading factors. {\bf Top:} scaling relations of cool mass loading factor. {\bf Bottom:} scaling relations of hot energy loading factor.
    The mass and energy fluxes are both measured at $|z|=H$. From left to right, the $x$ axes denote SFR surface density 
    with $\tbin=40\Myr$ ({\bf (a)/(e)}),
    midplane total pressure ({\bf (b)/(f)}), 
    total gas weight ({\bf (c)/(g)}), and 
    gas depletion time ({\bf (d)/(h)}).
    Simulation results from  standard and low resolution model suites are presented as lighter and darker symbols, respectively (see legends distributed over panels). The best fit lines for standard and low resolution models are shown as black and magenta solid lines, respectively.}
    \label{fig:scaling-conv}
\end{figure*}

To test resolution convergence, here we present results from a model suite with two times poorer spatial resolution than the standard model suite. Note that given the stochastic nature of each simulation's evolution, only statistical comparisons are possible between different resolutions. \autoref{fig:scaling-conv} plots the mass loading factor of cool outflows and energy loading factor of hot outflows for selected galactic properties (see \autoref{fig:scaling-Mloading} and \autoref{fig:scaling-Eloading}).
The lower resolution models are in good agreement with higher resolution models, falling on the reported scaling relations within one-sigma uncertainty levels. 

\begin{figure*}
    \centering
    \includegraphics[width=\textwidth]{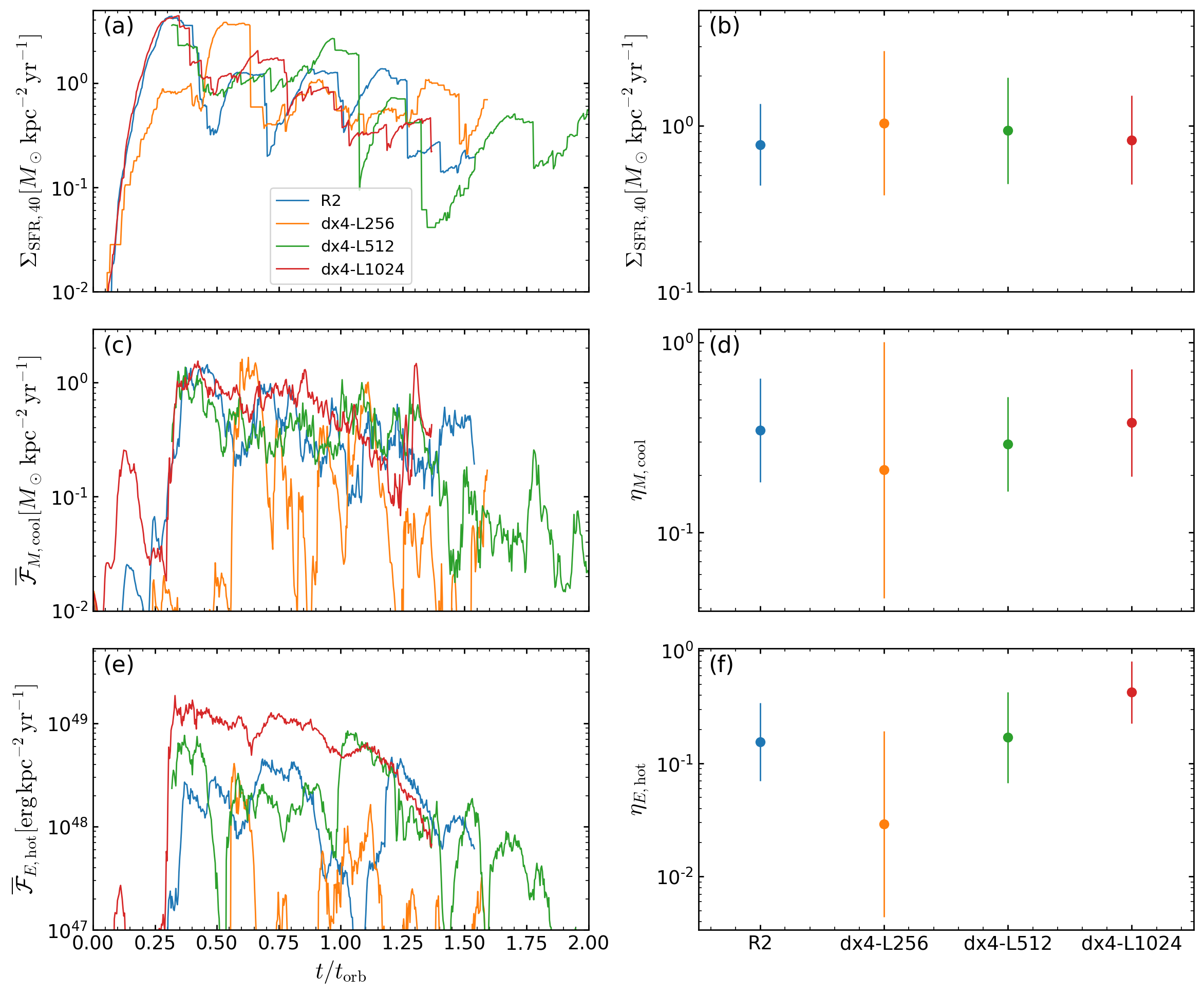}
    \caption{Box size convergence test for R2. The ``standard'' model adopts a spatial resolution $\Delta x=2\pc$ and horizontal domain size $L_x=L_y=512\pc$, while the other models adopt $\Delta x=4\pc$ with varying $L_x=L_y=L\pc$ shown in the model name. 
    {\bf Left:} Time evolution of (a) $\Sigma_{\rm SFR,40}$, (c) $\overline{\mathcal{F}}_{M, {\rm cool}}$, and (e) $\overline{\mathcal{F}}_{M, {\rm cool}}$ over $0.5<t/\torb<1.5$.
    {\bf Right:} Mean and standard deviation over $0.5<t/\torb<1.5$ for (b) $\Sigma_{\rm SFR,40}$, (d) $\loading{M}{cool}$, and (f) $\loading{E}{hot}$.}
    \label{fig:R2-conv}
\end{figure*}

To test box size convergence, we rerun model R2 at lower resolution $\Delta x=4\pc$ and varying horizontal domain sizes from smaller $L_x=L_y=256\pc$ to standard 512~pc to larger 1024~pc, while our standard choice is $\Delta x=2\pc$ and $L_x=L_y=512\pc$. We use model R2 because this model is expected to show the largest box size dependence due to its shortest gravitational time scale comparable to star cluster evolution time scale. \autoref{fig:R2-conv} compares time evolution (left) and mean/standard deviation (right) over $0.5<t/\torb<1.5$ for a few selected quantities, $\Ssfr$ in (a) and (b), mass flux and loading of cool outflows in (c) and (d), respectively, and energy flux and loading of hot outflows in (e) and (f), respectively. We confirm that the lower resolution model is in good agreement with the standard model as already demonstrated in \autoref{fig:scaling-conv}.

There are general increasing trends with box size in both mass and energy fluxes and loading factors. Our choice of box size is smaller or comparable to the Toomre length scale,
\begin{equation}
    \lambda_T\equiv\frac{4\pi^2 G \Sgas}{\kappa^2}=850\pc\rbrackets{\frac{\Sgas}{100\Surf}}\rbrackets{\frac{\Omega}{100\kms\kpc^{-1}}}^{-2},
\end{equation}
above which axisymmetric gravitational instability is suppressed by epicyclic motions. This means that if the large scale coherent structure is not destroyed by feedback within  the   gravitational time scale, the entire gas disk would collapse globally. For R2, $t_g\lesssim \tver, \tevol$, we anticipate large scale gravitational collapse from the initial conditions. In this case, star formation is more clustered with a larger box, resulting in stronger feedback and higher loading factors, especially, for the energy loading factor (\autoref{fig:R2-conv}(f)). Such strong bursts may indeed exist in galactic centers. As the validity of the local approximation is in question as $L$ gets closer to $R_0$, however, we limit our model to a moderate box size, but still large enough to capture spatial correlation of SNe to some extent. Global modeling is clearly necessary in this regime. 

\section{Instantaneous Loading Factors with Delayed Normalization}\label{sec:app-delay}

\begin{figure*}
    \centering
    \includegraphics[width=\textwidth]{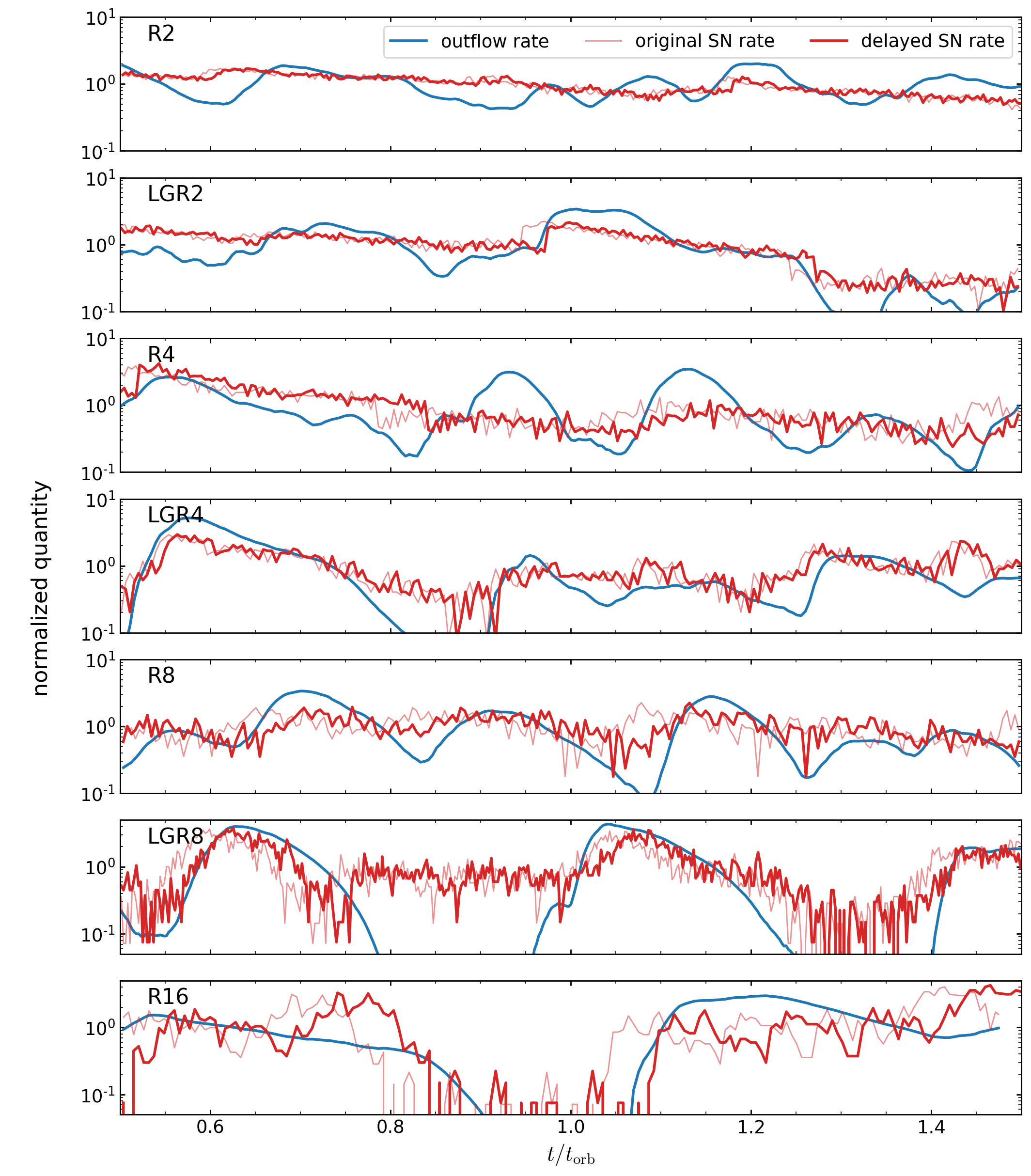}
    \caption{Comparison between outflow rate and SN rate. All quantities are normalized by their own mean over the time range shown in the plot. Mass outflow rates averaged over $|z|=H$--$2H$ are compared with original and delayed SN rates. The time delay maximizing the Pearson correlation coefficient (Column (3) in \autoref{tbl:tdelay}) is applied.}
    \label{fig:tdelay}
\end{figure*}

In our simulation suite (see Figure~\ref{fig:R4flux} for example), we observe more than an order of magnitude temporal fluctuations, and generally  a delay  between a peak in the SN rate and the enhancement in the outflow flux. As we discussed in \autoref{sec:loading}, the complicated quantitative behavior makes it difficult to define instantaneous loading factors in realistic simulations where both feedback injection rates and outflow rates are self-consistently modulated \citep[e.g.,][]{2015MNRAS.454.2691M}. For example, \autoref{fig:tdelay} plots the normalized mass outflow rate and SN rate for all models. Overall, there is stronger temporal fluctuation in outflow rates than SN rates. Often, a moderate level of continuous SN explosions does not create corresponding outflows (e.g., $t/\torb=0.8$--1 for LGR8), mainly due to strong inflows of material ejected by previous outflows. For this reason, attempting a one-to-one mapping of the peaks of outflow rate and SN rate (or SFR) with a constant time delay generally fails in our simulations. It is worse at higher SFRs and not particularly better for different physical quantities (momentum, energy, and metal) and phases.

Nevertheless, we have tested defining a delay time in two ways, in order to investigate the uncertainty in calculation of loading factors. (1) We calculate the Pearson correlation coefficient between the SN rate and outflow rates averaged over $|z|=H$--$2H$, and find the delay time that maximizes the correlation. We then compute the mean loading factors using the shifted reference flux. (2) We construct model fluxes, $\mathcal{F}_{q, {\rm model}} \equiv A_q\mathcal{F}_{q, {\rm ref}}(t-dt_q)$, with a grid of $dt\in (0,50\Myr)$ and $\log A_q\in(-2,2)$ for $q=M$ and $E$ to search $A_q$ and $dt_q$ that minimizes $\int_{0.5\torb}^{1.5\torb}(\mathcal{F}_q - \mathcal{F}_{q, {\rm model}})^2 dt$. Note that $A_q$ is equivalent to $\eta_q$. 

\autoref{tbl:tdelay} lists the delay times and loading factors obtained by two methods along with the loading factors without time delay. In \autoref{fig:tdelay}, we also show the result using the time delay of Column (3).  The delay times found in this way are longer in models with longer $\tver$. The derived loading factors are consistent within the intrinsic uncertainty arising from the temporal fluctuations and mismatch between outflow and SN rates. The mass loading factor estimated by model fitting gives generally smaller values, but not very different from other estimates.

\begin{deluxetable*}{ccCCCCC}
\tabletypesize{\footnotesize}
\tablecaption{Delay Times and Loading Factors\label{tbl:tdelay}}
\tablehead{
\colhead{Model} &
\colhead{($q$,ph)} &
\dcolhead{\eta_{q, {\rm ph}}} &
\dcolhead{dt^{\rm corr} [{\rm Myr}]} &
\dcolhead{\eta^{\rm corr}} &
\dcolhead{dt^{\rm model} [{\rm Myr}]} &
\dcolhead{\eta^{\rm model}} 
} 
\colnumbers
\startdata
R2     & ($E$,hot )  &     0.15 &      1.4 &     0.14 &      2.0 &     0.14\\
R2     & ($M$,cool)  &     0.84 &     0.98 &     0.83 &      3.1 &     0.36\\
R4     & ($E$,hot )  &     0.13 &      2.9 &     0.12 &      2.9 &     0.11\\
R4     & ($M$,cool)  &      1.8 &      6.4 &      1.6 &      6.4 &     0.79\\
R8     & ($E$,hot )  &    0.053 &      16 &    0.053 &      16 &    0.071\\
R8     & ($M$,cool)  &      4.5 &      12 &      4.5 &      12 &      4.0\\
R16    & ($E$,hot )  &    0.023 &      21 &    0.023 &      21 &    0.025\\
R16    & ($M$,cool)  &      31 &      26 &      32 &      35 &      25\\
LGR2   & ($E$,hot )  &    0.074 &      4.9 &    0.07 &      4.9 &    0.079\\
LGR2   & ($M$,cool)  &      1.2 &      3.4 &      1.1 &      2.9 &     0.89\\
LGR4   & ($E$,hot )  &    0.057 &      3.9 &    0.058 &      2.9 &    0.063\\
LGR4   & ($M$,cool)  &      3.4 &      2.0 &      3.4 &      2.0 &      3.2\\
LGR8   & ($E$,hot )  &    0.053 &      11 &    0.055 &      11 &    0.089\\
LGR8   & ($M$,cool)  &      7.3 &      11 &      7.6 &      11 &      8.9
\enddata
\tablecomments{
Column (2): combination of the outflow quantity `$q$' and phase `ph' used to measure the delay time and loading factor.
Column (3): loading factors reported in \autoref{sec:scaling} without any time delay.
Columns (4) and (5): delay time and loading factor maximizing the Pearson correlation coefficient between SN rate and outflow rate.
Columns (6) and (7): delay time and loading factor minimizing the difference between the delayed and measured model fluxes.
}
\end{deluxetable*}

\section{Scaling Relations with Input Parameters}\label{sec:scaling-input}

\begin{figure}
    \centering
    \includegraphics[width=\textwidth]{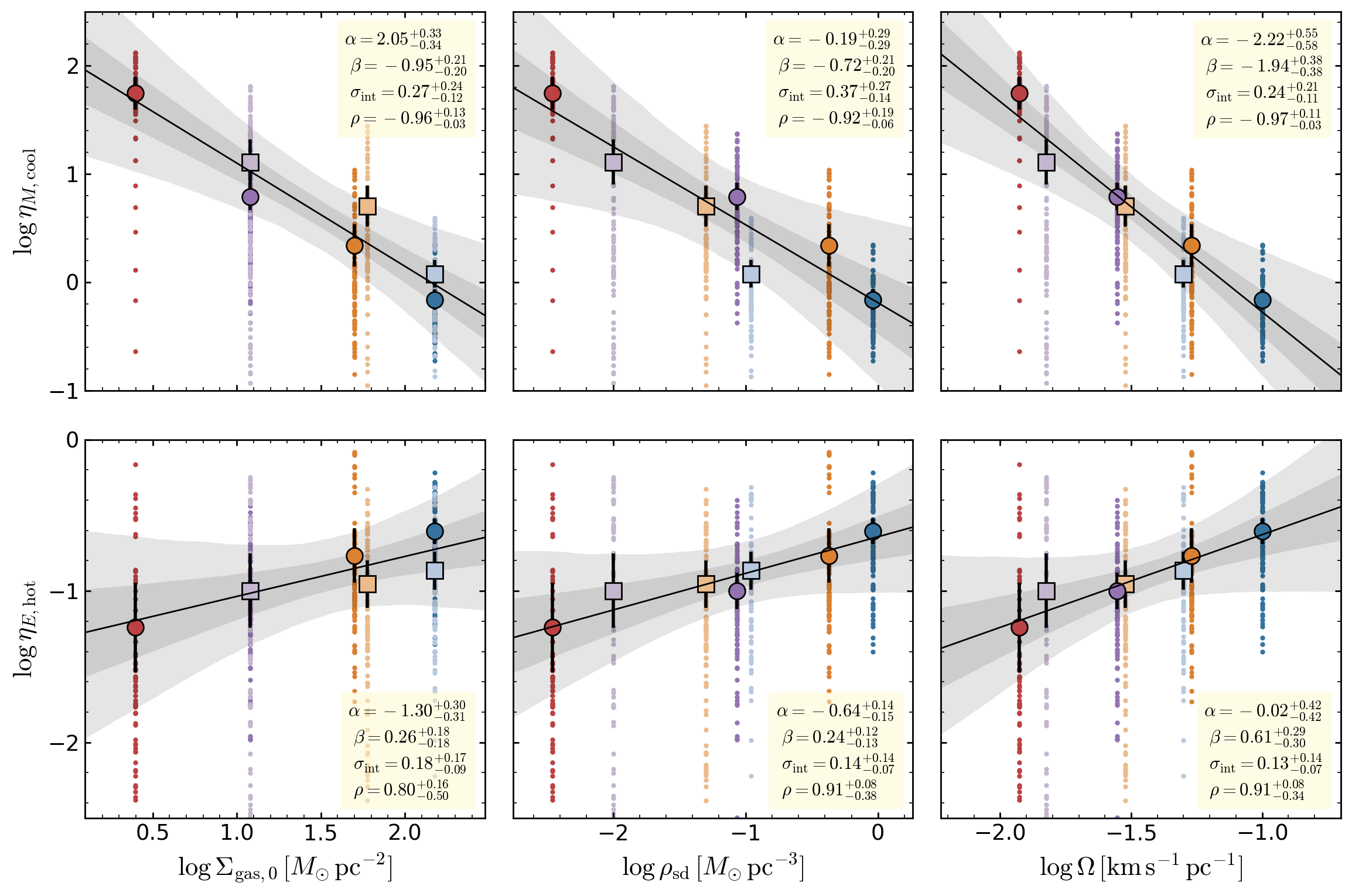}
    \caption{Scaling relations of cool mass loading and hot energy loading factors with simulation input parameters. The mass and energy fluxes are measured at $|z|=H$. Figures at different heights are available at \url{https://changgoo.github.io/tigress-wind-figureset/figureset.html}. The simulation results and fitting results are presented as in \autoref{fig:scaling-loading}. }
    \label{fig:scaling-xpar}
\end{figure}

In the main portion of the paper, we provided scaling relations for loading factors with respect to the self-regulated ISM properties such as $\Ssfr$ and $P_\mathrm{mid}$ (see \autoref{sec:scaling}).  Here, we additionally present scaling relations with  respect to the input model parameters in \autoref{tbl:model}. \autoref{fig:scaling-xpar} shows the mass loading factor of the cool outflow (top row) and the energy loading factor of the hot outflow (bottom row) at $|z|=H$ as a function of initial gas surface density ($\Sigma_{\rm gas,0}$), stellar+dark matter midplane volume density ($\rho_{\rm sd}\equiv \Sigma_*/(2z_*)+\rho_{\rm dm}$), and angular velocity of galactic rotation ($\Omega$). 
Based on the intrinsic scatter ($\sigma_{\rm int}$ in each panel of \autoref{fig:scaling-xpar}), we find that the energy loading $\loading{E}{hot}$ correlates better with ``gravity-parameter'' $\rho_{\rm sd}$, while mass loading $\loading{M}{cool}$ correlates better with ``gas-parameter'' $\Sigma_{\rm gas,0}$. Both loading factors show good correlation with $\Omega$. Overall, $\loading{M}{cool}$ better correlates with the self-regulated ISM properties shown in \autoref{fig:scaling-Mloading} than with the input model parameters shown in \autoref{fig:scaling-xpar}.
We note that the input parameters are not chosen to be fully independent of each other: roughly, $\Sigma_{\rm gas,0}\propto\rho_{\rm sd}$ with two different normalizations for the R and LGR series, and $\rho_{\rm sd}\propto \Omega^2$.

\end{document}